\documentclass[a4paper,11pt]{article}
\pdfoutput=1 

\usepackage{jheppub} 

\usepackage[T1]{fontenc} 
\usepackage{tikz}
\usetikzlibrary{arrows.meta, decorations.markings}

\title{Exact solution of the Seven-Vertex Model on a dynamical lattice}

\author[a]{Mateus  da Silva Junca,}
\author[b,c]{Ivan Kostov,}
\author[d]{and Andre Alves Lima} 

 \affiliation[a]{Departamento de F\'isica, Universidade Federal do
 Esp\'irito Santo, 29075-900, Vit\'oria, Brazil}
 \affiliation[b]{Universit\'e Paris-Saclay, CNRS, CEA, Institut de
 physique th\'eorique \\
	 91191 Gif-sur-Yvette, France} \affiliation[c]{Beijing Institute
	 of Mathematical Sciences and Applications \\
  Huairou 101408,Beijing,China } \affiliation[d]{Departamento de
  F\'isica, Universidade Federal de Pernambuco\\ 50670-90, Recife,
  Brazil}
  
\emailAdd{mateus.junca@edu.ufes.br}
\emailAdd{ivan.kostov@ipht.fr}
\emailAdd{alves.lima@ufpe.br}

  \abstract{ We give the complete solution of the one-parameter
  deformation of the six-vertex model on dynamical lattice introduced
  in \cite{Kostov:2025awi} and dubbed gravitational seven-vertex
  model.  This statistical model is mapped to a gas of self- 
  and mutually avoiding loops on dynamical triangulations, with a
  temperature coupling controlling the volume not occupied by loops.
  The phase diagram is characterised by massive, dilute and dense
  critical phases, similarly to the gravitational $O(n)$ loop model.
  There is however an important difference -- in our model the weights
  of the loops are not topological but depend on the form of the loop
  and on the curvature defects of the lattice via lattice spin
  connection.  The seven-vertex model on dynamical lattice is
  nevertheless exactly solvable after being reformulated as a
  large-$N$ matrix model, which we will refer to as 7vMM, and the
  solution in the scaling limit was found in \cite{ Kostov:2025awi}.
  Here we derive the full solution in terms of Jacobi theta functions
  and present the (non-algebraic) spectral curve of 7vMM in a
  parametric form.  We obtain the phase diagram in the space of the
  two coupling constants -- the cosmological constant and the
  temperature  -- and identify the critical phases along the
  boundary of the physical domain.  We derive the scaling solution of
  \cite{ Kostov:2025awi} as the asymptotic of the full solution in the
  vicinity of the tricritical point separating the phases of dense and
  massive loops.
  }

 \def\Det{ \text{Det}} 
  \newcommand\re[1]{({\ref{#1}})}

  \def\be{\begin{eqnarray} } 
  \def\ee{\end{eqnarray}}
  \def\Re{\mathrm{Re}}
\def\Im{\mathrm{Im}}

   \def\CA{{\mathcal{A}}}
 
  \def\X{\mathbf{X}}
  \def\uv{{_{\text{UV}}}}
  \def\ir{{_{\text{IR}}}}
\def\k{ \kappa} 

 \def\bI{{\mathbb I}} 
    \def\IR{{\mathbb{R}}}
    \def\IC{{\mathbb{C}}} 
\def\l {\lambda}
\def\CC{ {\mathcal C}}
 \def\CG{\mathcal{G}} 
  \def\p{\partial}
   \def\a{\alpha} 
   \def\b{\beta} 
  
 \def\e{\epsilon} 
 \def\s{\sigma}
  \def\t{\tau} 
  \def\vp{\varphi}
 \def\d{\delta}
  
 \def\RSGc{ R^{_{{SG}}}_{_{\rm{UV}}}}
   \def\RSGd{ R^{_{{SG}}}_{_{ \rm{IR}}}}
 \def\RSLc{ R^{_{{SL}}}_{_{\rm{UV}}}}
  \def\SL{{{SL}}}
  \def\SG{{{SG}}}
    \def\bw{\bar{ w}}
\def\U{\mathbf{U}}
\def\vt{\vartheta}
 \def\tvt{\tilde\vartheta}
\def\CF{{\cal F}} 
 \def\CS{\mathcal{S}}
 \def\CB{{\mathcal{B}}}
 \def\CO{{\mathcal{ O} }} 
\def\CL{{ \mathcal{ L} }}
  \def\CM{\mathcal{M}}
 \def\CZ{{ \mathcal{ Z} }}
 \def\CU{{\mathcal U}}
 \def\Z{{\mathbf Z}}
  \DeclareMathOperator{\Tr}{{\mathrm Tr}} 
 \def\sc{\mathrm{sc}}
\def\bb{\mathrm{b}}
\def\ba{\mathrm{a}}
  \def\Tstar{ T_*}
 \def\bq{\bar{q}}
   \def\cT{{\mathfrak c}}
 \def\rI{\mathrm{I}}
 \def\rII{\mathrm{II}}
 \def\rIII{\mathrm{III}}
 \def\tilka{{\tilde \kappa}}
\def\tscu{{\tilde u}}
\def\mgc{{\tilde \mu}}

\begin{document} 
\maketitle
\flushbottom

\section{Introduction}

 Statistical models on dynamical lattices
 are known as discrete models of 2D gravity%
 \footnote{For reviews, see
 \cite{1991hep.th....8019K,DiFrancesco:1993nw,Ginsparg:1993is}.}
 since they give a discretisation of the path integral over Riemann
 metrics \cite{Ambjorn:1985az,David:1985nj,Boulatov:1986jd}.
 The enhanced symmetry resulting from the sum over geometries makes it
 possible to solve, analytically, problems whose
 exact solutions are inaccessible on a flat lattice.  At the same
 time, there is a precise correspondence between the universal
 behaviours of statistical models on flat and on dynamical lattices, a
 remarkable and non-trivial fact which has been confirmed by studies
 of bulk and boundary geometrical critical phenomena
 \cite{Duplantier:1988wc,Duplantier:1989sx,Duplantier:1998fg,Duplantier:2003vx},
 massless flows connecting different critical points
 \cite{Boulatov:1986sb,AleshaZam-3pf,Kostov:2006ry,Ishimoto:2005ag},
 and by phenomena as subtle as the decay of metastable vacua through
 nucleation \cite{Zamolodchikov:2006xs,YL-On}.  
 Thus the discrete models of 2D gravity are a valuable tool for studying fundamental
 aspects of string theories and low-dimensional quantum gravity.
 In particular, these models are solvable not only at the fixed points where the 
 matter field is described by a CFT with $c\le 1$,  but also along the massive and
  massless renormalisation-group (RG) flows connecting different fixed points.
In the models of 2D gravity, the role of an IR cutoff is played by
the cosmological constant which controls the size of the two-dimensional universe.
 The best known example of a massless RG  flow in a non-rational
theory is the thermal flow in the $O(n)$ loop model whose
gravitational counterpart solves a simple transcendental equation \cite{Kostov:2006ry}.

 Coupling a statistical model to gravity by its transferring from a flat to a dynamical
  lattice is a two-step procedure. First, the model is defined on a lattice with
 curvature defects represented by a planar graph from a sufficiently
 large ensemble, and then the sum over all graphs in the ensemble is
 performed. 
The first step is straightforward and leads to graphical
representations in terms of Fortuin-Kasteleyn clusters
\cite{Fortuin:1971dw}, which is essentially the same as on a flat
lattice.  
The simplest cluster expansion on a dynamical lattice
is a gas of self- and mutually-avoiding loops, in terms of which 
the Ising \cite{Kazakov:1986hu,Boulatov:1986sb}, $O(n)$
\cite{Kostov:1988fy}, SOS and ADE height models \cite{Kostov:1989eg}
can all be formulated.
In all of them the fugacities are \emph{topological}, in the sense that they
do not depend on the geometry of the loops; for example, all
contractible loops have the same fugacity $n$.

It is well known that, on the flat lattice, the loop models can be 
given a local formulation as \emph{vertex} models,
for which the fluctuating degrees of freedom are defined on the links of the lattice, 
and Boltzmann weights
are associated with pairs of adjacent links.
 For example, the $O(n)$ loop
model can be described in this way by factorising the total weight of
a closed loop into local weights for the left- and right turns \cite{Nienhuis:1984wm}.
The complete equivalence between loop-gas and local vertex formulations 
holds only on the infinite flat lattice, however, and is spoiled in non-trivial topologies
(cylinder or torus) or in the presence of conical curvature defects, in
the sense that the non-contractible loops may have different fugacities.
On a dynamical lattice, where curvature defects proliferate, the distinction becomes radical.
One can still give a geometric representation of the vertex model in terms
 of a gas of oriented loops, but now one must assign \emph{non-topological} 
 phase factors dependent on the local curvature enclosed by the loops. 
 It is natural to call statistical
systems of this type \emph{gravitational vertex models}.  Remarkably, they
 are also integrable.  The simplest example is Baxter's six-vertex model
on the dynamical lattice, which was
formulated by Ginsparg \cite{Ginsparg:1991bi} as a large-$N$ matrix
model and to which we will refer as 6vMM. It was solved exactly in the works
\cite{Kostov:1999qx,ZinnJustin:1999wt,ELVEYPRICE2023105739}, where it was
found that the universal behaviour near the critical point is
described by Liouville gravity with Gaussian matter compactified at
 a radius determined by the vertex weights.
Qualitatively this is what one would expect, except for the fact that the 
compactification lengths on flat and on dynamical lattices are \emph{different}.

Apart from the solution of the 6vMM, there has been almost no research
concerning vertex models on dynamical lattices.  The phase diagram of
the gravitational vertex models (the critical and multicritical points
and the flows connecting them) remained largely unexplored, although 
it might reveal a wealth of rich and interesting new physics.  

Only very recently \cite{Kostov:2025awi} the flow in a one-parameter deformation of the
gravitational six-vertex model called \emph{gravitational seven-vertex model}
was analysed, and claimed to provide an integrable
discretisation of sine-Liouville gravity.  
While the six-vertex model can be
formulated as a gas of dense oriented loops,
the seven-vertex model allows for vacancies which lead to different
critical phases.  
In \cite{Kostov:2025awi}, the model on the dynamical lattice was
reformulated as a large-$N$ matrix model 
which we will call the 7vMM, and the solution encoded in its spectral curve was derived in the
scaling limit. Based on the scaling solution,
it was argued that the
flow connecting the dilute with the dense phases of the loop gas is
the gravitational analogue of the massless thermal flow in the
sine-Gordon theory with imaginary mass coupling discovered by
Nienhuis~\cite{Nienhuis:1984wm} in the context of the Coulomb gas, and
studied in depth by Fendley, Saleur and Al.~Zamolodchikov
\cite{Fendley:1993wq,Fendley:1993xa,Zamolodchikov:1994za}.  
In flat space, the endpoints of the massless flow are Gaussian fields
compactified at two different lengths. After coupling to gravity
this qualitative picture persists, but the compactification lengths
for the Gaussian fields at the endpoints of the flow both
change.

In   the present  paper, we expand and articulate some of the points discussed
in \cite{ Kostov:2025awi}.  Our main result is the complete solution
for the classical spectral curve of the 7vMM for any values of 
its two parameters -- the lattice cosmological constant $\k$ and the lattice 
`temperature' coupling constant $T$.  Based on the full solution, we
determine the critical line on which the partition function diverges.
The critical line consists of three analytic branches which host two
different critical phases: the phase of massive loops, which is
described by pure gravity, and the phase of dense loops, which is in the
same class of equivalence with the gravitational six-vertex model.  We
derive the scaling solution claimed in \cite{ Kostov:2025awi} by
zooming at the tricritical point separating the dense from one of the
massive phases.  The full solution thus allows us to relate the
arbitrary renormalisation constants in \cite{ Kostov:2025awi} to the
lattice parameters.

The paper is organised as follows. In section \ref{sect:7vMPG} we
give the precise definition of the $7$-vertex model on 
 a dynamical lattice, starting by reminding the basics about its well-known
prototype on a honeycomb lattice. In section \ref{Sect:7VMM}
we introduce the holographic dual: the seven-vertex Matrix Model (7vMM).
In section \ref{section:saddlepoyntanalysis}
we derive the parametric
form of the classical spectral curve of 7vMM in terms of Jacobi
theta-functions.  
In section \ref{SectCriticalCurves} we give the
statistical interpretation of the three analytic branches of the critical
line in the space of couplings. In section \ref{SectSolNearCritical} 
we derive the scaling limit by zooming at the tricritical point.

\section{The seven-vertex model on planar graphs}	
\label{sect:7vMPG}

\subsection{Vertex models and loop ensembles on infinite  regular  lattice }
 \label{sect:7vmWS}

\begin{figure}[h!]
        \centering
\begin{minipage}[t]{0.8\linewidth}
            \centering
             \includegraphics[width= 11 cm]{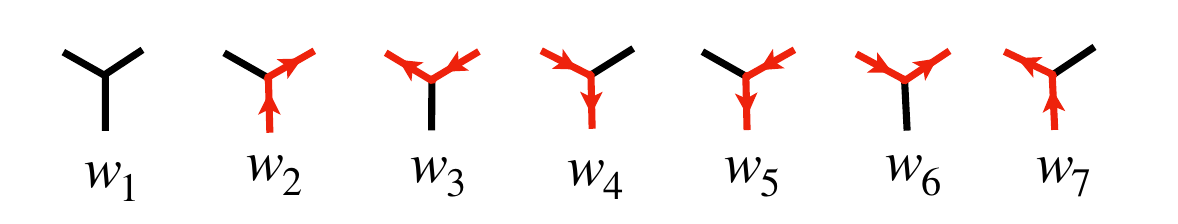}
 \caption{\small  The   vertices of the 7-vertex model.}
     \label{fig:vertices}
          \end{minipage} 
\end{figure}

Let us first recall the relation between vertex models and loop
ensembles on a regular lattice.  More specifically, we consider the
vertex model on a honeycomb lattice, first studied by Baxter
\cite{baxter1986q}, which gives a local formulation of the $O(n)$ loop
model \cite{Nienhuis:1984wm}.  This is the discrete model we are going
to focus on in this paper.  The local degrees of freedom of this
vertex model are arrows assigned to the bonds of the honeycomb lattice
obeying the ``ice rule'' that the numbers of incoming and outgoing
arrows at each site must be equal.  The ice rule allows seven
configurations at each vertex of the graph (assuming reflection
symmetry) as shown in fig.  \ref{fig:vertices} and we thus refer to it
as the seven-vertex model.

The local Boltzmann weights are associated with these seven vertex
configurations.  Assuming also invariance under $\pi/3$ rotations, there
are only 3 distinct vertices which can be parametrised up to a common
factor as
\be\begin{aligned}  \label{7vflat} w_1= T, \quad w_2=w_3=w_4 = w , \quad w_5=w_6=w_7 =
w^{-1} .  \end{aligned} \ee 
The  weight $T $ of the vacant vertex will be referred to as temperature.

Each allowed vertex configuration defines a set of closed contours on
the lattice.  The vertex model is thus mapped onto a gas of
self-avoiding and mutually avoiding oriented loops with partition
function
\be\begin{aligned} 
\label{loopexpansion}
Z_{7v} \sim   
 \sum_{\text{oriented loops}} 
 (1/T)^{
 V_{\text{loops}}}\ 
 w^{ (\text{left turns})-(\text{right
turns})}.
\end{aligned} \ee 
 The temperature $T$ controls the volume $V_{\text{loops}}$ occupied
 by loops (= the number of the vertices visited by loops).  
 
 A fundamental topological property of this system on an infinite
 lattice is that the number of right and left turns for any closed
 loop differs by exactly $\pm 6$.  Consequently, each oriented loop
 acquires a fugacity of $w ^6$ or $ w^{-6}$ depending on its
 orientation.  The system is equivalent to the $O(n)$ loop model,
 where the unoriented loop fugacity is defined as $n= w^6+w^{-6}$,
\be\begin{aligned}  \label{loopexpansion1} Z_{O(n)}\Big|_{\mathrm{honeycomb}}
=\sum_{\text{unoriented loops}} (1/T)^{ V_{\text{loops}}}\ \
n^{\text{loops}}, \qquad n= w^6+w^{-6}.  \end{aligned} \ee 
In this representation, $1/T$ serves as the monomer fugacity or the
chemical potential governing the presence of loop segments.

When $w$ is a unimodular complex number such that $|n|\le 2$, the
phase structure of the loop gas was conjectured by Nienhuis
\cite{Nienhuis:1984wm} based on Coulomb gas arguments.  The Coulomb
gas appears through the mapping of the loop model onto a SOS
(solid-on-solid) height model, the non-intersecting loops representing
the domain walls separating different heights.  The SOS height
variable renormalises at large distances as a Gaussian field and the
observables can be constructed as electric and magnetic operators for
this Gaussian field.

 The loop gas has finite correlation length above the
critical temperature
\be\begin{aligned}  \label{crT} T _c = (2 + \sqrt{2-n})^{1/2} .  
\end{aligned} \ee 
For $T< T_c$ the loops  almost fill 
the lattice and form
a critical phase called \emph{dense phase} of the loop gas. 
At $T=T_c$, where  the vertex model is exactly solvable,
the loop gas is in the  \emph{dilute phase} which is also critical.
When the temperature is tuned to $T_c$, both the typical length of the
loops and the volume not occupied by loops diverge.

The effective field theory for the large-distance behaviour of the
vertex/loop model in the vicinity of the critical temperature is the
sine-Gordon model with imaginary mass coupling.  The dilute and the
dense phases correspond to the ultraviolet (UV) and infrared (IR)
renormalisation-group fixed points, respectively.

  \subsection{The vertex model on dynamical lattice: Definition and
  loop expansion }
 \label{sect:7vmWS1}

\begin{figure}[h!]
		\centering
\begin{minipage}[t]{0.78\linewidth}
            \centering
             \includegraphics[width= 12 cm]{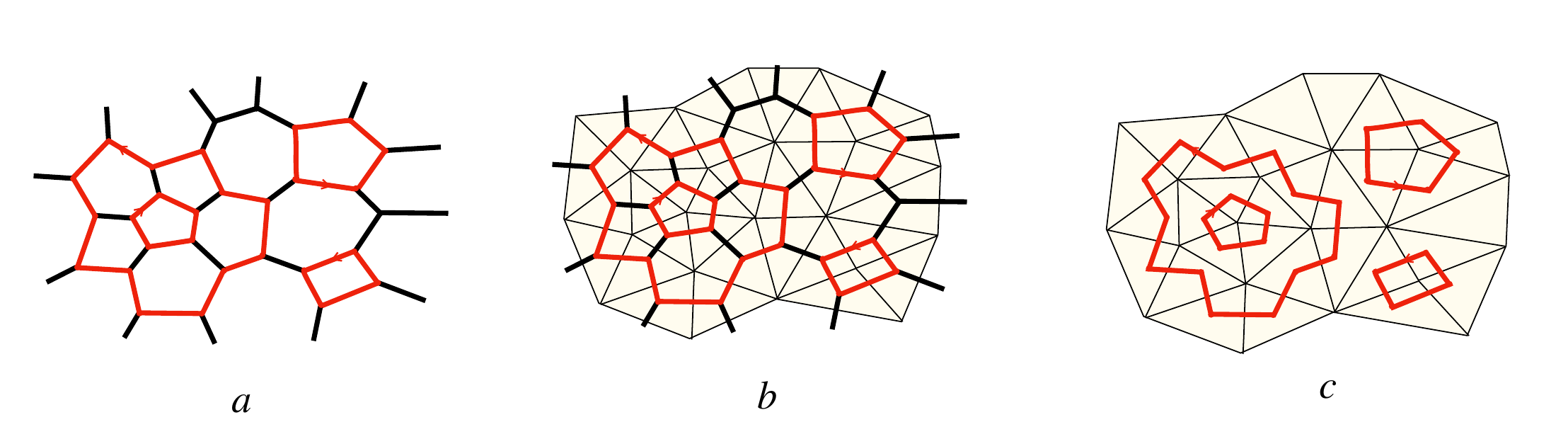}
\caption{\small A loop configuration on a trivalent graph $\CG$ with
the topology of the disk (a) becomes, after mapping the trivalent
graph to its dual (b), a loop configuration on a triangulation of the
disk (c).  To preserve the information about the topology the lines of
the trivalent graph should be thickened.  }
    \label{latticess}     \end{minipage} 
\end{figure}

Now let us see how coupling to gravity transforms the vertex models.
The partition function of the 7-vertex model is trivially generalised
to any tri-coordinated planar graph.  The vertex degrees of freedom
are again introduced by assigning orientations to some of the bonds in
such a way that the numbers of the incoming and the outgoing arrows at
each vertex are equal.  An example of an allowed vertex configuration
is given in fig.  \ref{latticess}.  Any vertex configuration obeying
the ice rule gives rise to a set of oriented loops and the partition
function takes the same form as \eqref{loopexpansion}.  The partition
function of the gravitational vertex model on the sphere is defined as
an additional sum over all trivalent graphs $\CG$ with the topology of
a sphere.   
 The volume $ V_\CG$ of the planar graph is controlled by a parameter $\k$,
which we refer to as lattice cosmological constant,
while the temperature $T$ controls the volume $V_{\text{loops}}$
occupied by loops:
 \be\begin{aligned}  \CF_{0} (\k, T)&=\sum_{\text{ graphs\ } \CG} \kappa^{V_\CG} \
 Z_{O(n)}\Big|_{\CG}
  \\
 Z_{O(n)}\Big|_{\CG}&= \sum_{\text{loops}\ \CL \text{ on \ } \CG} (1/T)^{
V_{\text{loops}}}\ \prod_{\CL} w^{ (\text{left turns})-(\text{right
turns})}.
 \label{loopexp7v} \end{aligned} 
\ee 
where the parameter $\k$ controls the volume $ V_\CG$ of the planar
graph and the temperature $T$ controls the volume $V_{\text{loops}}$
occupied by loops.

The principal difference with the flat lattice is that the fugacities
of the loops cease to be topological: the dynamics of the loops is now
entangled with the local geometry of the lattice.

We can think of a graph $\CG$ as the dual object to a triangulation $
{ \tilde{\mathcal{G}}}$ of the sphere or of the disk as  illustrated  in 
fig.\,\ref{latticess}. The vertex model can be mapped to a
statistical SOS type {\it height model} in which the local degrees of
freedom are integer-valued heights $h_i$ assigned to the sites $i \in
{ \tilde{\mathcal{G}}}$.  Each vertex configuration determines, up to
a constant, an integer-valued height function, with heights assigned
to the faces of the trivalent graph or, equivalently, to the sites of
$ { \tilde{\mathcal{G}}}$.  The oriented loops divide the planar graph
into domains of constant heights, with the oriented loops appearing as
domain boundaries.  Since the difference of the heights of two
neighbouring domains is $\pm 1$, all the heights but one can be
reconstructed taking into account the orientations of the loops.  The
Boltzmann weight of a height configuration is a product of local
factors associated with the elementary triangles $\Delta_{ijk} $,
 \be\begin{aligned}  \label{BweightsSOS} W_{\Delta} (h_1,h_2,h_3)&= 
  \delta_{h_1h_2}
 \delta_{h_2h_3}\delta_{h_3h_1}   +{1\over T} \delta_{h_1h_2}
 \delta_{h_2h_3} A_{h_3h_1} \, 
 w^ { h_3-h_1   } \ +{\rm cyclic },
 \\
A_{h,h'}&\equiv  \d_{h, h'+1}+\d_{h, h'-1} .  
\end{aligned} \ee 

\begin{figure}[h!]
		\centering
\begin{minipage}[t]{0.7\linewidth}
            \centering
           \includegraphics[width= 10 cm]{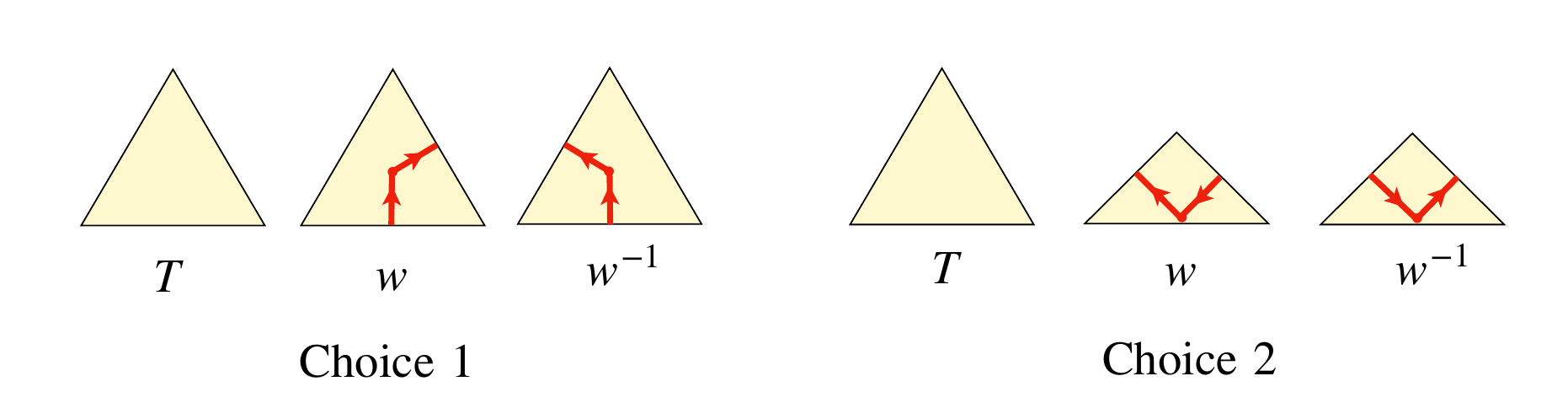}
\caption{\small Two choices for the form of the triangular tiles. }
    \label{choices12}     \end{minipage} 
\end{figure}

Choosing the precise shape of the triangular tiles that compose the discrete surface is not necessary to define and solve the model, whose partition function is unambiguously defined by \eqref{loopexp7v}. 
We can define the triangles to have edges of different lengths, with edges of the same length being allowed to be glued together; two obvious choices are depicted in fig.\,\ref{choices12}.
A distinction between the two choices arises, however, when 
we make a connection with the effective CFT in the continuum limit.
It turns out that they lead to different couplings to the local curvature of the surface
because the local curvature on the triangulation dual to a
trivalent graph $\CG$ is not uniquely determined by  the graph's connectivity.
 For a  triangulation $ {\tilde{\mathcal{G}}}$, the local curvature at the site $i\in {\tilde{\mathcal{G}}}$ is defined to be the deficit angle
\begin{equation} 	\label{defrhat}
\hat R_i = 2\left( 2\pi -  \a_i \right)
\end{equation} 
where $\a_i$ is the sum of the angles of all corners of triangles meeting at $i$.
Let us consider the two choices in fig.~\ref{choices12}.

 \begin{itemize}
 
\item \emph{ Choice 1.} The most symmetric choice is that of
equilateral triangles with angles $ \pi/3$ associated with their
vertices.  There are three distinct types of triangles:
one empty triangle with Boltzmann weight $T$ and two triangles visited
by a loop that makes a right- or left-turn at angle $\pi/3$, with Boltzmann weights
$w^{\pm 1} $. 
For a flat triangulation,
there must be
exactly six triangles meeting at each vertex.
A different number makes a conical curvature defect, with curvature given
by the missing angle.

The phase factor of the
loops in the expansion \eqref{loopexp7v} can be thought of as a
holonomy factor for a particle with spin.  Each left/right turn
contribute an elementary holonomy $ w^{\pm 1}$, a lattice analogue of
the spin connection defined on pairs of adjacent links.  The phase of
the holonomy factor, known also as Berry phase, is the total angle
swept of the normal vector.  The geometrical phase of a loop $\CL$ is
\be
w^{ (\text{left turns})-(\text{right turns})}
=
w ^{3 \hat K_\CL/\pi }
\ee
 where $ \hat K_\CL$ is the total geodesic curvature along the loop $\CL$.
  By the Gauss-Bonnet formula, the phase factor can be expressed in
 terms of the local Gaussian curvature \eqref{defrhat}.
 Since all angles are equal to $\pi/3$, the Gaussian
curvature at site $i$ is determined by the coordination number $C_i$
(the number of triangles having $i$ as a vertex), namely
$\hat R_i = {2\pi\over 3}(6-C_i) $,
 so that the flat lattice is the
regular hexagonal lattice with $C_i=6$ at all sites.  Let $ {
\tilde{\mathcal{G}}}_\CL$ be the disk with boundary given by the loop
$\CL$.  Then the geodesic curvature $\hat K_\CL$ of the loop is
related to the integrated Gaussian curvature $\hat R_\CL$ of $ {
\tilde{\mathcal{G}}}_\CL$ by
 \be\begin{aligned} 
 \label{GBf}
  \hat R_\CL+ 2 \hat K_\CL = 4\pi, \qquad 
   \hat R_\CL\equiv \sum_{i\in { \tilde{\mathcal{G}}}_\CL} \hat R_i .
 \end{aligned} \ee 

 \smallskip
 
  \item  
\emph{Choice 2.} The other possible choice for the geometry of the
triangular tiles is that they are either equilateral triangles with one side 
of unit length opposite the angle $\pi/2$  and two shorter sides opposite 
two angles $\pi/4$. By construction, the
loops are not allowed to cross the longer side.  Each segment of a
loop now represents a left/right turn at $\pm \pi/2$.  The geometrical
phase of a loop $\CL$ is
\be
w^{ (\text{left turns})-(\text{right turns})}
=
w ^{2 \hat K_\CL/\pi }.
\ee
With this choice for the form of the tiles, the Gauss-Bonnet theorem
\eqref{GBf} holds, but the Gaussian curvature \re{defrhat} is not the
same as for the choice 1.  In particular, sites with zero curvature
can be a meeting point of 4, 6, 7 or 8 triangles, and there are
infinitely many non-equivalent flat triangulations.  Flat surfaces for
choice 2 have curvature defects for choice 1 and vice versa.
 \end{itemize}

 \smallskip

To summarise, the loop expansions the 7-vertex model and the $O(n)$
model are identical on the infinite honeycomb lattice, but react
differently to curvature defects and therefore define different
discrete models of 2D gravity.  In the gravitational $O(n)$ model, the
loops have fugacity $n$.  In the gravitational vertex model, the
oriented loops are weighted by a geometrical phase which is
proportional to the geodesic curvature of the loop.  The coefficient
of proportionality depends on the form chosen for the triangular tiles
the world surface is made of.
 
 Anticipating the exact solution of the vertex model we are going to
 present later, the definition of the discrete curvature matches the
 continuum theory if we take the choice 2 for the gravitational vertex
 model while the choice 1 is the natural one for the $O(n)$ model.   
 Assuming the choice 2, we
 parametrise the vertex weight $w$ in a way compatible with the
 solution of the six-vertex model ($T=0$) obtained in
 \cite{Kostov:1999qx,ZinnJustin:1999wt,ELVEYPRICE2023105739}.  When
 $T=0$, the isoceles triangles can be glued pairwise along their long sides to
 make squares.  This brings us to the formulation of the six-vertex
 model on random quadrangulations used in
 \cite{Kostov:1999qx,ZinnJustin:1999wt,ELVEYPRICE2023105739}.   
 
  To match with the notations
used in \cite{Kostov:1999qx}, we will mostly use throughout this paper
 the parameter $\l$ such that
\begin{equation} 
\label{defw}
 w=e^{ 
i\pi \l/2} .
\end{equation} 
But we will also use often, at different places depending on convenience, three other parameters related to $\l\in [0,1]$, which we register here, namely
\begin{equation}\begin{aligned}  \label{parameterss}
  \b &= \pi (1-\lambda), 
  \\
    q&= - w^{-2 }=  e^{i \beta} ,  
    \,
    \\
  b&= \sqrt{1-\l\over 1+\l}  .
\end{aligned} 
\end{equation} 

\section{The 7-vertex matrix model}	\label{Sect:7VMM}

The sum over planar graphs decorated by vertices is generated by the
perturbative expansion of the following matrix integral, which we
refer to as `7-vertex matrix model' or shortly `7vMM' with partition
function
\be	\label{SLMM}
\begin{aligned}
\CZ _N &= \int d\X d\mathbf{Z} d\mathbf{Z}^\dag \ e^{-{1\over \hbar}
\CS },
\\
	\CS &= \Tr \left[ \tfrac{1}{2}\X^2 + \Z^{\dagger} \Z -
	\tfrac{1}{3} T \; \X^3 - \bw \X \Z\Z^{\dagger} - w \X \Z^{\dagger}
	\Z \right] , \qquad w\equiv e^{i \pi \lambda / 2} .
\end{aligned}
\ee
The integration variables $\X$ and $\Z$ are respectively Hermitian and
complex $N\times N$ matrices with flat integration measure
 \begin{equation}
	d\X d\Z d\Z^\dagger \equiv \left[ \prod_{i=1}^N dX_{ii}
	\prod_{j>i} d (\Re X_{ij}) \ d (\Im X_{ij}) \right] \left[
	\prod_{i,j=1}^N d ( \Re Z_{ij}) \ d (\Im Z_{ij}) \right].
\end{equation}
The perturbative free energy of the matrix model is a sum over all
connected planar (i.e.~with thickened lines) trivalent Feynman graphs,
obtained by expanding the non-gaussian part and performing Wick
contractions.  The perturbative partition function is exponential of
the sum of all connected graphs built according to the following
Feynman rules: \be\begin{aligned}
        \includegraphics[width= 12 cm]{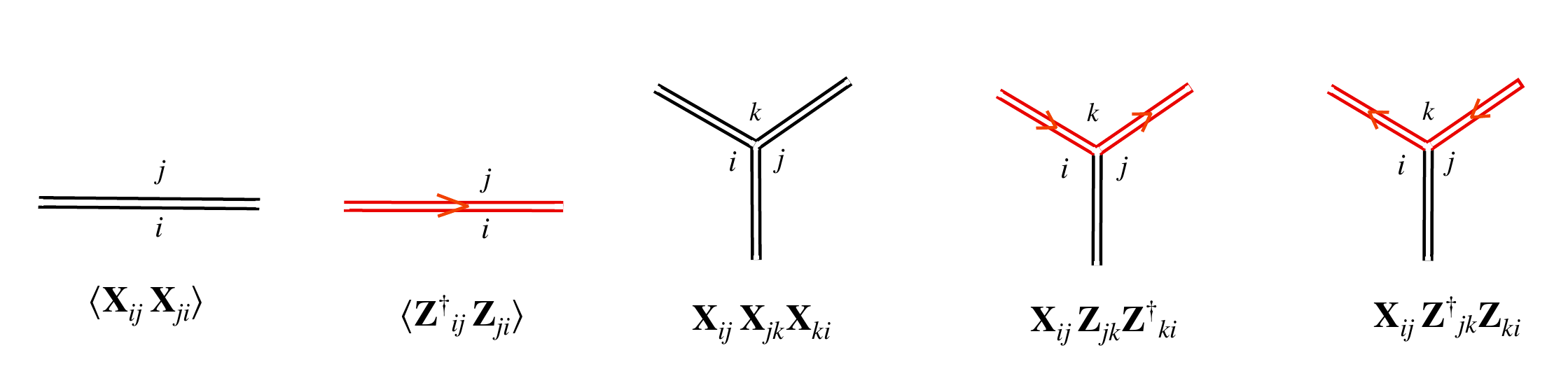}
            .
            \end{aligned} \ee 
Here we used the 't Hooft's double-line notations \cite{t1974planar},
with the `index lines' standing for Kronecker delta symbols.

 The topology of the connected planar graphs is controlled by the
 `Planck constant' $\hbar \sim 1/N$, and the asymptotic semiclassical
 expansion is also a topological expansion,
\be \label{thooftexp} 
\hbar^2 \log \CZ_N 
	=
	  \sum_{g\ge 0} \hbar^{2g} \CF_g (\kappa, \lambda),
	\quad 
	\kappa \equiv  \hbar N, 
\ee
 with $\CF_g( \kappa, \l)$ being the partition function of the vertex
 model, eq.  \eqref{loopexp7v}, in the ensemble of trivalent planar
 graphs with $g$ handles.

The basic observables, which serve as  building blocks for most  of the
observables in the ensemble, are the moments of the matrix $\X$,
\begin{equation}
W_L= \hbar\left\langle  \Tr [\X^L] \right\rangle   \, ,  \qquad  L=0,1,2,...\  .
\end{equation}
They are the expansion coefficients for the resolvent of the random matrix $\X$
\begin{equation}	 \label{resolvent} 
W(x) = \sum _{L=0}^\infty x^{-L-1} W_L
	=\hbar \left\langle \Tr \frac{1}{ x-\X} \right\rangle .
\end{equation}
Alternatively, if the resolvent is known, its series expansion can be
used to extract the moments.  The resolvent is itself an important
observable, the disk amplitude with a marked boundary and boundary
cosmological constant $x$.  It is the derivative of the disk partition
function without unmarked boundary,
\begin{equation}	\label{defOm} 
\Omega(x) = \hbar \, \Big\langle \Tr \log( x- \X) \Big\rangle .
\end{equation}

The Gaussian integral over the complex matrix produces the inverse
power of the determinant of the quadratic form
	\be Z^\dagger_{ij} \left( \delta_{ik} \delta_{lj} - w\,
	\delta_{ik} X_{lj} - \bw\, X_{ik} \delta_{lj} \right) Z_{jk} \ee
and, ignoring the unimportant multiplicative factor, we express the
partition function of the matrix model as a one-matrix integral
\begin{equation}
	\CZ_N \sim \int \!  d\X \ \frac{ e^{- \frac1\hbar \Tr (
	\frac{1}{2}\X^2 - \frac{1}{3} T \; \X^3 ) } }{ \Det \left( \bI
	\otimes \bI - w\, \bI \otimes \X - \bw\, \X \otimes \bI \right) }.
\end{equation}
Finally, the Hermitian measure is invariant under $\U(N)$
transformations that can be used to diagonalize $\X$, reducing $\CZ_N$
to an $N$-fold integral over the real eigenvalues $x_1, \cdots, x_N$,
\begin{equation} \label{eigvalinx1bis}
	\CZ _N \sim \int_{-\infty}^{\infty} \prod_{i =1}^N {dx_i \ e^{-
	\frac1\hbar (\frac12 x_i^2 - \frac13 T \, x_i^3)} \over 1- (w+\bw)
	x_i } \ \frac{ \Delta^2(x) }{\prod_{j\ne k} \big[ 1- w\, x_k -
	\bw\, x_j\big]}
\end{equation}
where $\Delta(x) = \prod_{j > i} (x_i - x_j) $ is the Vandermonde
determinant.

The integrand has the symmetries
\begin{equation}	\label{dualitysm1}
\{\l, T, x\}
	\equiv \{\l+4, T, x\}  
	\equiv \{ - \l, T, x\}
	\equiv \{ 2-\l, -T, -x\}, 
\end{equation}
which are shared also by the loop expansion \eqref{loopexp7v}.  They
allow us to restrict $\l$, so in the following we assume that
\begin{equation}
\lambda \in [0,1) .
\end{equation}

Because of the cubic potential, for any $T\ne 0$ the integral
\eqref{eigvalinx1bis} is divergent.  In addition, the integrand have
simple pole at
\be\begin{aligned} 
x_\text{pole} = {1\over w+\bw}  >  0, 
\end{aligned} \ee 
in each of the variables $x_j$, $j=1,..., N$.  As usual, the integral
can be made convergent by deforming the integration contour away from
the pole and so that it goes to infinity in one of the Stokes sectors
for the cubic potential.  The choice of the contour does not affect
the perturbative expansion and only brings exponentially small
non-perturbative effects necessary to repair the asymptotic
semiclassical expansion.  We can thus restrict the integration to the
classically accessible values which are distributed in a connected
interval around the minimum of the external potential.

It is convenient to make a change of variables in the integral
\eqref{eigvalinx1bis} such that the pole occurs at the origin.  We do
this by the redefinition
\begin{equation}	\label{RESCALING} 
T \to (w+\bw ) T , \qquad x \to \frac{x + 1}{ w+\bw }, \qquad \hbar
\to \frac{\hbar}{ (w+\bw ) ^2}
\end{equation}
 and the eigenvalue integral takes the form, up to a constant factor,
\begin{equation}	 \label{eigvalinxter}
	\CZ_N \sim \int \prod_{j=1}^N dx_j \ e^{- \frac{1}{\hbar} V(x_j)}
	\ \frac{\prod_{k< j} (x _k- x _j )^2 }{ \prod_{k,j} ( w x_k+ \bw
	x_j ) }
\end{equation}
with the cubic potential
\begin{equation}	\label{newpotential}
V(x)=  (1-T ) \, x + ( \tfrac12 -T ) \, x^2 - \tfrac13 T \, x^3.
\end{equation}
Now the pole and the minimum are fixed respectively at
$x = 0$
and
$x = -1$,
while the position of the maximum depends on the value of the
temperature coupling, producing different critical regimes, to be
discussed in section \ref{SectCriticalCurves}.

\section{Exact solution for the 7vMM}	
	\label{section:saddlepoyntanalysis}

In this section, we derive this solution of the 7vMM in the large-$N$
limit which is encoded in its classical spectral curve.  Although the
spectral curve is not algebraic as in the minimal models of 2D
gravity, it can be reconstructed from its behaviour at infinity and a
pair of quasi-periodicity conditions.  The spectral curve can be
obtained rigorously from the Virasoro constraint formulated in \cite{
Kostov:2025awi} but we prefer to follow here the traditional approach
of \cite{Brezin:1977sv,Brezin:1979ba}.

\subsection{Saddle-point analysis}	 

It is useful to think about the integral \eqref{eigvalinxter} in terms
of a Dyson gas of $N$ charged particles at positions $x_i$, with a
density distribution
\begin{equation}
\rho(x) = \hbar \sum_{i=1}^N \delta(x -x_i) .
\end{equation}
In the thermodynamic limit for the Dyson gas, $N \to \infty$, the
infinite number of eigenvalues are expected to fill a single
continuous interval.  Since the (only) minimum of $V(x)$ is at
$x_\text{min} = -1$, and since the partition function diverges if the
eigenvalues reach the origin, the support of $\rho(x)$ in this limit
is an interval $[-{\mathrm{a} } , -{\mathrm{b} } ]$ on the negative
axis, ${\mathrm{a} } >{\mathrm{b} } >0$.  Meanwhile, $\rho(x)$ becomes
a continuous function normalised by the condition that it counts the
$N$ eigenvalues:
\begin{equation} \label{normrho} 
\int_{-{\mathrm{a} } }^{-{\mathrm{b} } } dx \; \rho(x) = \k, \qquad
\k\equiv N\hbar .
\end{equation}
Thus, in the thermodynamical limit, the sum over eigenvalues becomes
an integral weighted by the macroscopical spectral density, and knowing
$\rho(x)$ allows us to compute any macroscopic statistical average.  Among the
basic observables, the disk partition function with unmarked boundary
\eqref{defOm} is given by
\begin{equation} \label{defdiskam} 
\Omega(x) = \int_{-{\mathrm{a} } }^{-{\mathrm{b} } } \!  dx' \;
\rho(x') \log(x-x') ,
\end{equation}
and the resolvent \eqref{resolvent} by
\begin{equation}	\label{defdiskamW} 
W(x) = {\p \Omega(x)\over \p x}
	= \int_{-{\mathrm{a} } }^{-{\mathrm{b} } }dx' \frac{\rho(x')}{x-x'} .
\end{equation}
Conversely, if we know the resolvent in the thermodynamical limit, the
eigenvalue density $\rho(x)$ can be extracted from the discontinuity
of $W(x)$ on its support $[-{\mathrm{a} } , -{\mathrm{b} } ]$,
\begin{equation} \label{spctrlds}
W(x -i0) - W(x + i0) = 2\pi i \rho(x ) \qquad \text{for}\quad x
\in[-{\mathrm{a} } ,-{\mathrm{b} } ] .
\end{equation}

In the thermodynamical limit $\hbar\to 0$, the eigenvalue integral
\eqref{eigvalinxter} is saturated by a saddle-point.  One way of
formulating the problem is to rewrite the partition function as the
functional integral
\begin{equation}	\label{CZNfunctinterhomu}
	\CZ_N = \int \!  [D\rho] \int \!  d\tilka \ e^{- {1\over \hbar^2}
	\CA[\rho,\tilka] } ,
\end{equation}
where
\begin{equation} \label{ActIrhomu}
\begin{aligned}
	\CA[\rho,\tilka, \k] &= - \frac{1}{2} -\!\!\!\!\!\!\!  \int \!  dx
	- \!\!\!\!\!\!\!  \int dx' \ \rho(x) \, \log\left( \frac{ ( x-x'
	)^2 }{ (w x+ \bw x')(\bw x +w x')} \right) \rho(x') \\
&\quad
	+\int  dx \rho(x) V(x)
	+  \tilka \, \Big( \kappa- \int \! dx \, \rho(x)  \Big) .
\end{aligned}
\end{equation}
Here we have introduced a Lagrange multiplier $\tilka$ to enforce the
normalisation \eqref{normrho}.  From now on we will express $w$ in
terms of $q=-w^{-2}$, cf eq.\,\eqref{parameterss}.  The stationarity
conditions for $\tilde\k$ and $\rho$ give respectively
\begin{equation}	\label{SadlledeltI}
\int dx \rho(x) =\k \quad \text{and}\quad \CS(x) =\tilde\k \quad
\text{for} \quad x \in [-{\mathrm{a} } ,-{\mathrm{b} } ] ,
\end{equation}
where
\begin{equation}	\label{eigvalinxterAct}
\CS(x) = V(x) - 2 \Omega(x) + \Omega(q x) +\Omega( \bar  q x) 
\end{equation}
can be interpreted as the effective potential for a probe eigenvalue
particle at $x$ interacting with the Dyson gas.  Besides the external
potential $V(x)$, the probe particle feels the Coulomb repulsion from
the eigenvalues $x_i$ due to $\Omega(x)$, and the Coulomb attraction
to their images rotated at $ q x_i$ and $\bar q x_i$.  For $x\in
[-{\mathrm{a} } ,-{\mathrm{b} } ]$ the integral in the expression for
$\CS$ must be understood as the principal value.

Once the saddle point density is found, the free energy $\CF_0
=\hbar^2\log \CZ$ is evaluated to
\begin{equation}
\begin{aligned}
	\CF_0 [\rho , \k] &= - \CA[\rho,\tilka, \k] _{\text{s.p.}} = -
	\tfrac12 -\!\!\!\!\!\!\!  \int _{-{\mathrm{a} } }^{-{\mathrm{b} }
	} dx \rho(x) \big( \CS(x) - V(x) \big) \, - \int dx \rho(x) V(x)
	\\
	&= - {\textstyle{1\over 2}} \k\tilka - {\textstyle{1\over 2}} \int
	_{-{\mathrm{a} } }^{-{\mathrm{b} } } dx \rho(x) V(x).
\end{aligned}
\end{equation}
Note that the saddle-point equation for $\tilka$  can be written as 
\begin{equation}
\frac{\p\tilde  \CF_0  (\tilka) }{ \p \tilka} = \k, 
\qquad 
\tilde  \CF_0  (\tilka) \equiv -   \CF_0   (\k) + \k \tilka.
\end{equation}
The function $\tilde \CF_0 (\tilka)$ can be interpreted as the grand
canonical free energy where $\tilka$ is a free parameter and $\k$ is a
dynamical variable.  Differentiating the second equation we find
\begin{equation}	\label{dkapF}
\frac{\p  \CF_0  (\k) }{ \p \k} = \tilka.
\end{equation}

\subsection{The saddle-point equation as a boundary value problem}
\label{section:saddlepoynteqs}

We can solve the saddle point equation for $\rho(x)$, the second of
eqs.\eqref{SadlledeltI}, by reformulating it as a boundary value
problem for the meromorphic function $\CS(x)$.  The Riemann surface of
this function has three cuts, on $[-{\mathrm{a} } , -{\mathrm{b} } ]$,
$[-q{\mathrm{a} } , -q{\mathrm{b} } ]$ and $[-\bq {\mathrm{a} } , -\bq
{\mathrm{b} } ]$.  The saddle-point equation relates the values of $S$
at the edges of these cuts:
\begin{equation} \label{eqsS} 
\begin{aligned}
\tilka 
&=   {\textstyle{1\over 2}} \big[ \CS (x+i0) + \CS (x-i0) \big]
\\
	&= 
	V(x) - \Omega(x+i0) - \Omega(x-i0) + \Omega(q x) + \Omega(\bar q x) 
\qquad
	\text{for $x \in [-{\mathrm{a} } , -{\mathrm{b} } ]$.}
\end{aligned}
\end{equation}
Recall that $\Omega(q x)$ and $\Omega(\bar q x)$ are analytic on
$[-{\mathrm{a} } ,-{\mathrm{b} } ]$.

It is convenient to formulate this condition for a function with
simpler analytic structure, namely
\begin{equation}	\label{DefPhixf} 
\Phi (x) = \frac{ \Omega(q^{1/2}x) - \Omega(\bar q^{1/2} x) }{ i }
	- U(x) ,
\end{equation}
where $U(x)$ is a polynomial in $x$ to be fixed.  The meromorphic
function $\Phi(x)$ has only two cuts on $[-q^{1/2}{\mathrm{a} } ,
-q^{1/2}{\mathrm{b} } ]$ and $[-\bq^{1/2}{\mathrm{a} } , -\bq
^{1/2}{\mathrm{b} } ]$.  Then, for $x \in [-{\mathrm{a} }
,-{\mathrm{b} } ]$,
\begin{equation}	\label{TwoTErmYfA}
\begin{aligned}
\Phi(q^{\frac12  } e^{\mp i0}x  )  - \Phi (q^{\frac12  } e^{\pm i0}x  ) 
&	=
	 i \tilka
	 + i V(x)
	- U(q^{1/2} x \mp i0) + U(\bar q^{1/2} x \pm i0).
\end{aligned}
\end{equation}
We now choose $U(x)$ in such a way as to cancel $V(x)$ in the last
line, that is, we require that
\begin{equation}	\label{MapUtoV}
V(x) = \frac{U(q^{1/2} x)- U(q^{-1/2} x) }{i} 
\end{equation}
which determines $U$ as the cubic polynomial
\begin{equation} \label{defUpot}
U(x) = t_1 x + \tfrac12 t_2 \; x^2 + \tfrac13 t_3 \;  x^3 
\end{equation}
with coefficients
\begin{equation}	   \label{t1t2t3}
t_1=  \tfrac{1}{2} (1-T ) \sec( \tfrac12 \pi \lambda ) ,
\qquad
t_2 =  ( \tfrac{1}{2} - T ) \csc ( \pi \lambda ),
\qquad
t_3 = \tfrac{1}{2} T \, \sec ( \tfrac32 \pi  \lambda ).
\end{equation}
Note that the map \eqref{MapUtoV} between $V(x)$ and $U(x)$ cannot be
inverted for $\l = \frac13$ which leads to $t_3 $ having a simple pole
there.  The observables are not singular at this point, however.
 
Now the saddle-point condition \eqref{eqsS} reduces to the two-term
functional relation
\begin{equation}	\label{TwoTermPhi}
\Phi (q^{\frac12 \mp 0  }    ) - \Phi (  q^{-\frac12 \pm  0}x )
	=i \tilka .
\end{equation}
which can be regarded as boundary conditions on the edges of the two
cuts of $\Phi(x)$.  Solving the saddle-point equation amounts to
solving this boundary value problem.

\subsection{Spectral curve}

The solution of the saddle-point condition for the matrix integral is
more conveniently described in terms of the derivative
\begin{equation}	\label{defPhi1} 
Y(x) \equiv \p_x \Phi(x) .
\end{equation}
The derivative cancels the constant in the r.h.s.~of
eq.\,\eqref{TwoTermPhi}, leading to an homogeneous functional
relation,
\begin{equation}	\label{eqspa}
q^{1/2} Y(q^{{1\over 2}-0} x) - q^{ -1/2} Y(  q^{-{1\over 2}+0} x) = 0 , 
	\qquad  
	x \in [-{\mathrm{a} }  ,-{\mathrm{b} } ]. 
\end{equation}
This homogenous equation is what we will solve explicitly below.  The
meromorphic function $Y(x)$ is related to the resolvent as
\begin{equation}	\label{defPhi} 
Y(x) = \frac{ q^{1/2} W(q^{1/2}x) - \bq^{1/2} W(\bar q^{1/2} x) }{i} -
U'(x) ,
\end{equation}
with $U(x)$ given by \eqref{defUpot}.  Its asymptotics at $x\to\infty$
is fixed by the expansion coefficients of $W(x)$ at infinity,
\begin{equation}	\label{asymJ}
Y(x) = - U'(x) - \sum_{n=1}^\infty \frac{2 W_n \sin ( \frac12 n\beta)
}{ x^{n+1}} ,
\end{equation}
thus finding $Y(x)$ is equivalent to finding $W(x)$, hence to solving
the model.  We will refer, slightly abusively, to the map $x \mapsto
y=Y(x)$ as the spectral curve of our matrix model.

\begin{figure}[h!]
		\centering
\begin{minipage}[t]{0.7\linewidth}
            \centering
       \includegraphics[width=0.4\textwidth]{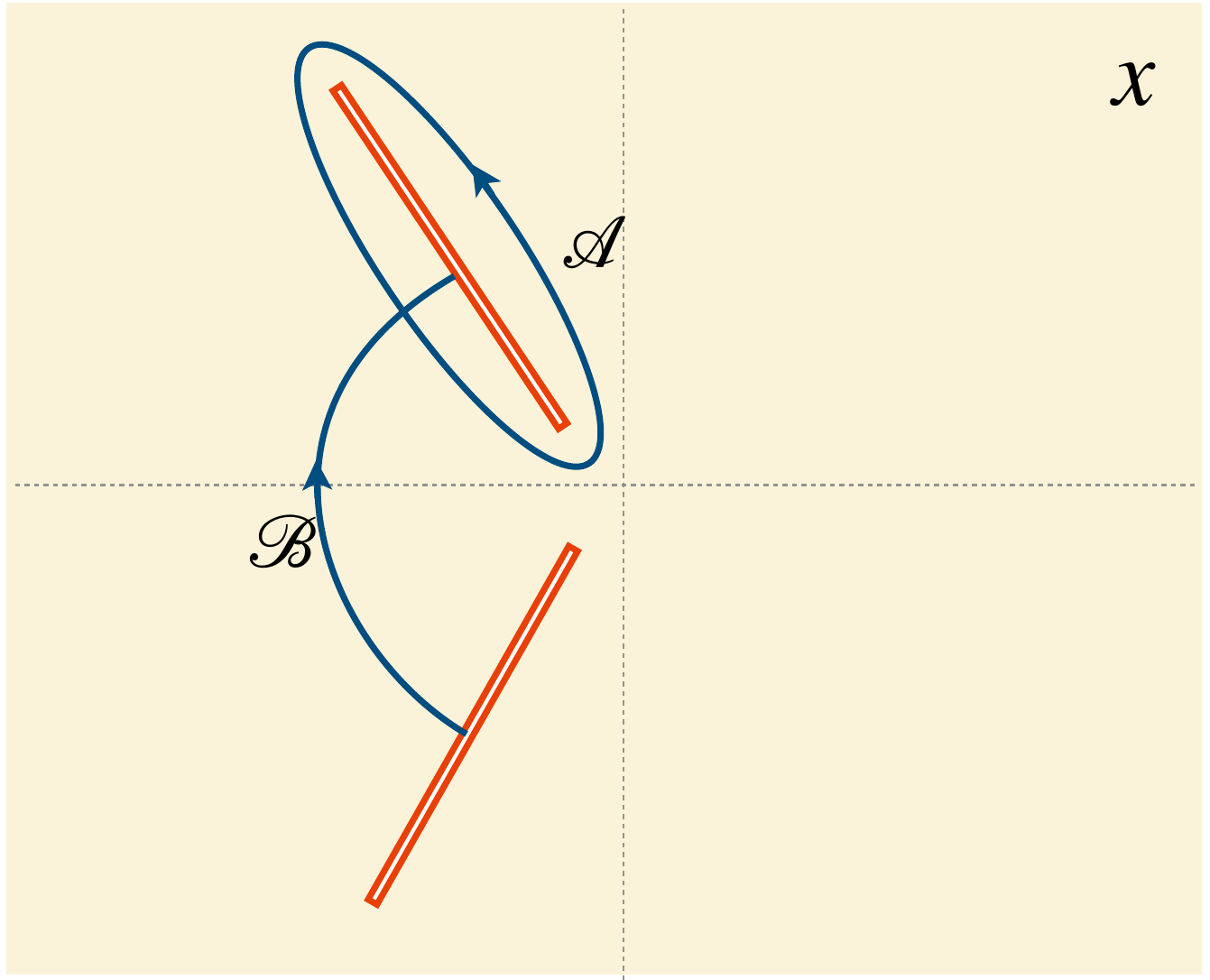} 
\caption{\small Spectral plane with cuts of the spectral function
$Y(x)$ and its cycles.  The upper and lower branch cuts are at $[-
q^{1/2 }{\mathrm{a} } , - q^{1/2} {\mathrm{b} } ]$ and $[- \bq^{1/2
}{\mathrm{a} } , - \bq^{1/2} {\mathrm{b} } ]$, respectively.  }
    \label{Curve_cycles}     \end{minipage} 
\end{figure}

The physical sheet of the Riemann surface of $Y(x)$ is the
complex $x$-plane with two cuts,
\begin{equation}	\label{CutsofYx}
[- \bq^{1/2 }{\mathrm{a} } ,  - \bq^{1/2} {\mathrm{b} } ]
\quad \text{and}\quad
[- q^{1/2 }{\mathrm{a} } ,  - q^{1/2} {\mathrm{b} } ] ,
\end{equation}
which correspond to the cuts of $W(x)$ rotated away from the real
axis, as shown in fig.\ref{Curve_cycles}.  Eqs.(\ref{eqspa}) can be
interpreted as a requirement of quasi-periodicity along one of the
main cycles of a torus obtained by the appropriate identification of
the edges of the two cuts.  After the identification, the main cycles
of the torus are homotopic to the contours $\CA$ and $\CB$, the latter
enclosing the upper branch cut $[-{\mathrm{a} } q^{ \frac12} ,
-{\mathrm{b} } q^{ \frac12}]$, and the former connecting the two cuts.

The integrals of $Y(x)$ along $\CA$ and $\CB$ are determined by the
parameters $\k$ and $T$.  eq.\,\eqref{defdiskamW} implies that
\begin{equation}	\label{normnBef} 
\begin{aligned}
 \oint_{\CA} \frac{dx}{2\pi} \, Y(x) &= \frac{q^{1/2}}{2\pi i}
 \oint_\CA dx \, W(q^{1/2} x) = \frac{1}{2\pi i}
 \oint\limits_{[-{\mathrm{a} } , -{\mathrm{b} } ]} \!\!  dx \, W(x) =
 \kappa
\end{aligned}
\end{equation}
where the last integral is over a contour enclosing the original
branch cut of $W(x)$.  As for the other cycle,
\begin{equation}
\begin{aligned}
\int_{\CB} \ dx \, Y(x) 
	&= 
\int_{\CB}  {dx}  \, \p_x \Phi (x) 
	=
	 {\Phi (  q^{\frac12 -0}  x) - \Phi (q^{-\frac12 +0} x )} 
	= i{ \tilka   } 
\end{aligned}
\end{equation}
where $ x \in [-{\mathrm{a} } , -{\mathrm{b} } ]$, and we used
eq.\,\eqref{TwoTermPhi}.  In summary, the integrals over the cycles
compute the thermodynamic properties of the model
\begin{equation}	\label{normn} 
\oint_{\CA} \frac{dx}{2\pi} \, Y(x) = \kappa, \qquad \int_{\CB}
 {dx} \, Y(x) =  i{\tilka}  =  i
{\p\CF \over \partial\kappa }.
\end{equation}

As mentioned above, knowing $Y(x)$ is essentially equivalent to
knowing the resolvent, hence to solving the model.  This can be put
quite clearly by noting that, as shown by a computation analogous to
\eqref{normnBef}, any average can be expressed as a contour integral
with $Y(x)$, namely
\begin{equation}	\label{spectrd}
\int _{-\mathrm{a}}^{-{\mathrm{b} } } \!  dx \; \rho(x) f(x) =
\oint\limits_{[-{\mathrm{a} } , -{\mathrm{b} } ]} \frac{dx}{2\pi} \;
q^{-\frac12 } Y( q^{-\frac12 } x) f(x)
\end{equation}
for any function $f(x)$ analytic in the vicinity of the interval
$[-{\mathrm{a} } , -{\mathrm{b} } ]$.  Alternatively, we can obtain
the spectral density as the discontinuity of $Y(x)$ across one of its
branch cuts:
\begin{equation}	\label{rhofromYdisc}
\rho(x) = \frac1{2\pi} \, \bq^{1/2} \Big[ Y( q^{\frac12 -0} x) - Y(
q^{-\frac12 +0} x ) \Big]
\end{equation}
Therefore $Y(x)$ fully encodes the solution of the matrix model in the
thermodynamical limit.

\subsection{Uniformisation map and parametric form  of the spectral curve }
\label{sec:unif}

 We will be interested mainly in the generic case of irrational
 $\lambda$, in which case the Riemann surface of the function $y=Y(x)$
 has infinitely many sheets, with two cuts in each sheet.  We are
 going to transform the boundary condition \eqref{eqspa} into a
 quasi-periodicity condition by introducing a global uniformisation
 parameter $s$,
\begin{equation} \label{canonu}
x = x(s), \qquad y=y(s) .
\end{equation}
Obviously, what we need is an elliptic parametrisation such that the
main sheet is mapped to a two-dimensional torus with a modular
parameter $\tau$ determined by the positions of the two cuts, which
are mapped to one of the main cycles of the torus.
It will be convenient to take $\tau \in \IR$ and choose the
uniformisation domain of the physical sheet of $Y(x)$ to be the
rectangle
\begin{equation}	\label{ParamRetangles}
- {\textstyle{1\over 2}} \pi \tau < \Re s < {\textstyle{1\over 2}} \pi
\tau ,
\qquad
-  {\textstyle{1\over 2}} \pi  < \Im s <   {\textstyle{1\over 2}} \pi ,
\end{equation}
with the branch cuts mapped to the horizontal edges of the rectangle,
see fig.\ref{Uniformization_Map_s}.  An equivalent parametrisation
with complex modular parameter is given in Appendix \ref{app_dualpar}.

Te infinitely foliated Riemann surface of $y=Y(x)$ is parameterised by
the infinite vertical strip $|\Re \, u| < \frac12 \pi$.  The
saddle-point equations \eqref{eqspa} are then equivalent to the
requirement that the functions \eqref{canonu} are periodic around the
cycle $\CA $ (the horizontal sides of the rectangle) and
quasi-periodic around the cycle $\CB $ (the vertical sides),
\begin{equation} \label{periodis}
\begin{aligned}
 x(s + \pi \tau) &= x(s), &\qquad x(s+ i\pi) &= q x(s) ,
\\
y (s + \pi  \tau) &= y(s), &\qquad y (s + i\pi) &= \bar q y(s).
\end{aligned}
\end{equation}
These conditions are solved by an elliptic theta function:
\begin{subequations}		\label{solutionS} 
\begin{align}
x(s) &= x_0 \frac{ \vt (\frac{\beta\tau}{2} - s)}{\vt(s)} ,
\label{solutionSX} 
\\
y(s)&
	\begin{aligned}[t]
 	&= 
 	 -  \frac{ y_0 }{ x_0 }  x(-s)
	+ 
	\frac{ y_1 }{  x_0 }  x'(-s)
	\\
	&= y_0 \frac{\vt( \frac{\beta \tau}{2} +s)}{\vt(s)} + y_1 \left(
	\frac{\vt'( \frac{\tau \b}{2} +s) }{ \vt(s) } - \frac{ \vt'(s)
	\vt( \frac{\tau\b}{2} +s) }{ \vt(s)^2 } \right)
\end{aligned}
\label{solutionSY} 
\end{align}
\end{subequations}
where $x_0$, $y_0$ and $y_1$ are constants, we recall that $q=
e^{i\beta}$ with $\b = \pi(1-\l)$, and we have defined\footnote{Our
conventions for the Jacobi theta functions $\theta_a$ are the same as
in \cite{GR}.  In particular, $\theta _1(u,e^{i\pi\tau}) \equiv 2
\sum_{n = 0}^\infty (-1)^n e^{i(n + \frac12)^2 \pi \tau} \sin [(2n+1)
u] $.  In Wolfram Mathematica the functions are implemented as
$\theta_a(u, q)= {\tt EllipticTheta[a, u, q]}$.}
\begin{equation}	\label{Defofvt}
\vt(s) \equiv i \sqrt{\tau} \ \exp \left( - s^2 / \pi \tau \right)
\theta_1(- is, e^{-\pi \tau}) .
\end{equation}
This choice of the  theta function is adapted to the continuum
limit at $\t\to\infty$.  The parameterisation most convenient for
computing the expansion of the partition function at $\k=0$ is the
dual one, which we give in Appendix \ref{app_dualpar}.
The quasi-periodicity conditions \eqref{periodis} follow from the
properties  
\begin{equation} \label{periodisThetas}
\vt(s + i\pi) = - e^{\pi/\tau} e^{-2is/\tau} \vt(s),
\qquad
\vt(s + \pi\tau) =  - \vt(s) = \vt(-s) .
\end{equation}
\begin{figure}[t]
\centering
\begin{minipage}[t]{0.8\linewidth}
\centering
\includegraphics[width= 12.5 cm]{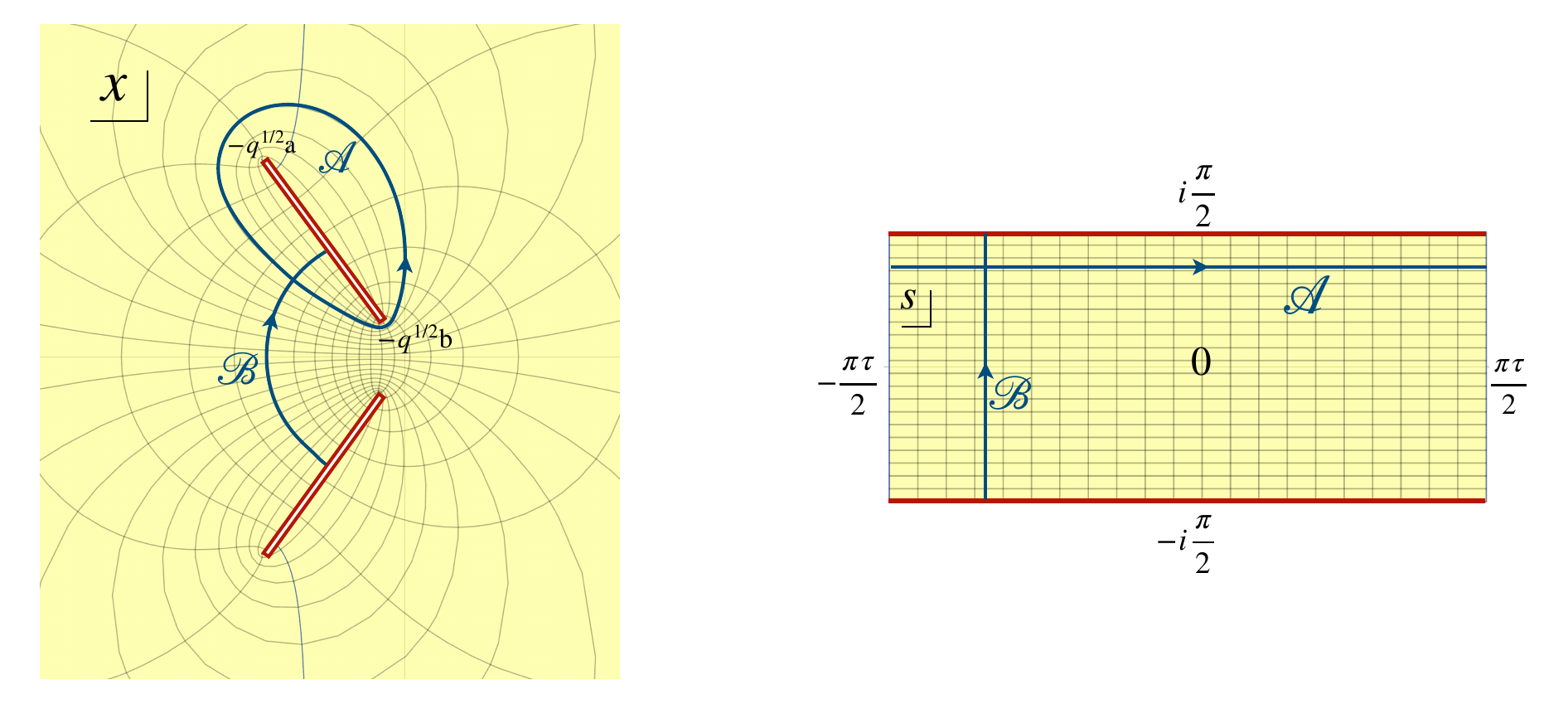}
\caption{The uniformisation map restricted to the main sheet of the
Riemann surface of $Y(x)$.}
\label{Uniformization_Map_s}
\end{minipage} 
\end{figure}

\begin{figure}[t]
\centering
\begin{minipage}[t]{0.6\linewidth}
\centering
\includegraphics[width= 7 cm]{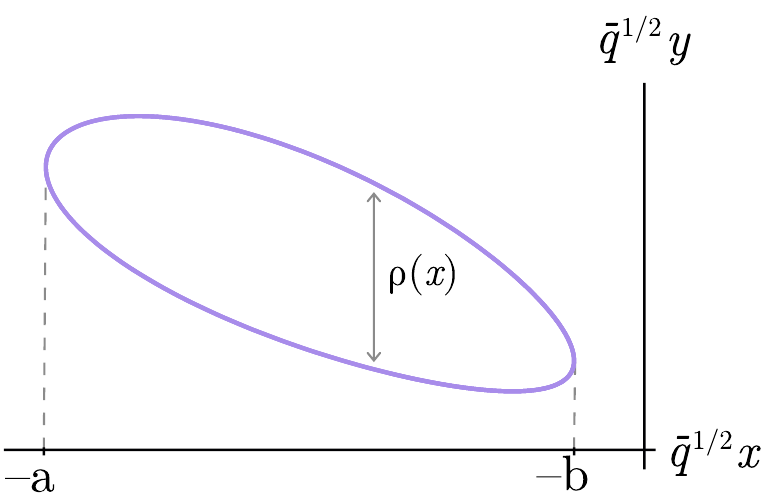}
\caption{ Section of the spectral curve around a branch cut computed
from eq.\,\eqref{rhofromYdisc}.  The spectral density equals the
vertical distance between the two branches }
\label{Spectral_Density_Profile}
\end{minipage} 
\end{figure}

The particular combination of theta functions in the solution
\eqref{solutionS} is determined by the analytic properties of $Y(x)$.
We have chosen the parametrisation so as to place the point $x=\infty$
at the origin of the $s$-plane.  Then the function $x(s)$ must have a
single pole at $s = 0$, which (combined with the quasi-periodic
conditions) fixes the ratio of theta functions in \eqref{solutionS}.
The expansion \eqref{asymJ} of $Y(x)$ implies that $y(s)$ must have a
\emph{second order} pole at $s=0$, which is produced by the derivative
in the expression for $y(s)$ in \eqref{solutionS}
\cite{Kostov:1999qx,ELVEYPRICE2023105739}.
The parameters $x_0$, $y_0$, $y_1$ in eqs.\eqref{solutionS} are
completely fixed by the asymptotic condition \eqref{asymJ},
\begin{equation}	\label{x0exact}
\begin{aligned}
y_1( T,\tau) &= \tfrac12 \sec( \tfrac{3\pi \l}{2} ) \, \cT_1 \,
T \, x_0^2 , \\
y_0( T,\tau) &= - (\tfrac12 - T) \csc(\pi\l) \, x_0 - \sec(
\tfrac{3\pi \l}{2} ) \cT_2 \, T \, x_0^2 , \\
x_0( T,\tau) 
		&= 
		- \frac{1}{\cT_2} 
		\frac{2(1-T) \sin ( \tfrac{\pi \l}{2} ) 
			}{ 
		(1-2T) + \sqrt{(T-T_+) (T-T_-) / T_+ T_-}
		} ,
\end{aligned}
\end{equation}
where 
\begin{equation}	\label{Tpmlatau}
		T_\pm( \tau) = \frac12 \left( 1 \pm \sqrt{ \frac{ \cT_2^2 +
		\cT_1 \cT_3 }{ \frac13 \cot^2 ( \frac{\pi\l}{2}) \, \cT_2^2 +
		\cT_1 \cT_3 } } \right) ,
\end{equation}
and
\begin{equation}	\label{defCCC}
\cT_1 \equiv \frac{\vt(\frac{\b\t}{2}) }{ \vt'(0)}, \qquad \cT_2
\equiv - \frac{\vt'(\frac{\b \tau}{2})}{\vt'(0)}, \qquad \cT_3 \equiv
\frac{\vt''(\frac{\b \tau}{2})}{2\vt'(0)} - \frac{\vt(\frac{\b
\tau}{2}) \vt'''(0)}{6\vt'(0)^2} .
\end{equation}
We present the derivation in Appendix \ref{app_solution}.  The
spectral density can be computed by eq.\,\eqref{rhofromYdisc} as the
discontinuity of $Y(x)$ across one of the branch cuts.  This is
illustrated by fig.  \ref{Spectral_Density_Profile}, where we plot the
corresponding sections of the spectral curve as a parametric curve
from our exact solution.  We have closed curves with turning points at
the two branch points.  The spectral density is (apart from a factor
of $2\pi$) the distance between the upper and the lower portion of the
curves.  The profile of the curve changes with $\tau$ and with $T$.

The branch points of $Y(x)$ are determined by the condition that
$(dY/dx)^{-1} = 0$, hence $x'(s) / y'(s) = 0$ for finite $y(s)$,
$x(s)$.  Since $y(s)$ is analytic for $s \neq 0$, one must solve
$x'(s) = 0$, which imposed the condition
\begin{equation}	\label{sbp}
\frac{\vt '(s) }{ \vt(s)} + \frac{\vt '( \frac{\b \t}{2} - s) }{ \vt(
\frac{\b \t}{2} -s)} = 0 .
\end{equation}
This condition is satisfied at two pairs of points
$s^\pm_{\mathrm{a}}$ and $s_{\mathrm{b} } ^\pm$ in the parametrisation
rectangle, which are the parameters of the branch points:
\begin{equation}	\label{brncpointparm}
x (s_{{\mathrm{a}}}^\pm) = - q^{\mp 1/2} {\mathrm{a} } ,
\qquad
x (s_{{\mathrm{b} } }^\pm) = - q^{\mp 1/2} {\mathrm{b} }  .
\end{equation}
The solutions
$s^\pm_{\mathrm{a}}$ and $s_{\mathrm{b} } ^\pm$
can be written as
\begin{equation}
s_{\mathrm{a}}^\pm = s_{\mathrm{a} } \pm \tfrac12 i \pi, 
\qquad 
s_{\mathrm{b} } ^\pm = s_{\mathrm{b} }  \pm \tfrac12 i \pi,
\qquad
s_{\mathrm{a} } < 0 < s_{\mathrm{b} } . 
\end{equation}
(Note that $s _{\mathrm{a}} ^\pm$ and $s_{{\mathrm{b} } }^\pm$ lie at
the upper and lower edges of the rectangle \eqref{ParamRetangles}.)
The real parts $s_{\mathrm{a}} $ and $s_{\mathrm{b} } $ are not
independent but satisfy
\begin{equation}	\label{sbbsbasum}
s_{\mathrm{b} }  + s_{\mathrm{a} } = \tfrac12 \beta \tau,
\end{equation}
which follows from the symmetries of eq.\,\eqref{sbp}.
eq.\,\eqref{sbbsbasum} yields a useful identity relating the points
${\mathrm{a}} $ and ${\mathrm{b} } $.  It is easy to check that the
function \eqref{solutionSX} satisfies
\begin{equation}	\label{xxinv}
x(s) \, x(\tfrac12 \beta \tau - s) = x_0^2 ,
\end{equation}
then, from \eqref{sbbsbasum}, it follows that
\begin{equation}	\label{babbisx0}
{\mathrm{a} } {\mathrm{b} }  = x_0^2 .
\end{equation}
Recall that the uniformisation map has been constructed in such a way
that $x = 0$ is at $s = \frac12 \beta \tau$, while $|x| = \infty$ is
at $s = 0$.  The combination of eqs.\eqref{babbisx0} and
\eqref{sbbsbasum} then means that
\begin{equation}	\label{bbnear0}
{\mathrm{b} }  \to 0
\quad \Rightarrow \quad
s_{\mathrm{b} }  \approx \tfrac12 \beta \tau \to \infty .
\end{equation}

To complete the solution, we need to evaluate the integrals of the
one-form $Y(x)dx $ over the $\CA$- and $\CB$-cycles which give
respectively $\k$ and $\p_\k\CF_0 $, according to eqs.\eqref{normn}.
These two integrals translate to the $s$-plane as
\begin{align}
\kappa \ \ &= \frac{1}{2 \pi } \int_{-\frac{\pi \tau}{2}}^{\frac{\pi
\tau}{2}} ds \ x'(s+ {\textstyle{1\over 2}} i \pi-i0 ) \, y(s +
{\textstyle{1\over 2}} i \pi-i0 ) \nonumber \\
& =  \frac{1}{2 \pi } \int_{-\frac{\pi \tau}{2}}^{\frac{\pi \tau}{2}}
ds \ x'(s+ i0 ) \, y(s + i0 ) ,
\label{normX}
\\
\frac{\partial \CF_0 }{\partial \kappa} & = \frac{1}{i} \int_{-
\frac{i \pi}{2} }^{ \frac{i\pi}{2} } ds \ x'(s ) \, y(s ) .
\label{eqC2}
\end{align}
The first integral determines implicitly $\t=\t(\k, T)$ thus fixing
completely the solution.  (In the second line the contour is deformed
close to the real axis but above the origin where $x(s)$ and $y(s)$
are singular.)  Both integrals can be computed explicitly in terms of
elliptic functions but the result is too long and not very
illuminating, so we do not report it here.  We will return to these
integrals when considering the scaling limit.

 The \emph{second} derivative of $\CF_0$ (at fixed $T$),
 or the susceptibility under changes of the lattice cosmological constant,
has a
remarkably simple relation to the modular parameter.  From
the second equation \eqref{normn}, we have
\begin{equation}	\label{p2CF0intY}
\p^2_\k  \CF_0 = \frac{1}{i} \int_\CB \! dx \; \p_\kappa Y(x) .
\end{equation}
The function $\p_\kappa Y(x)$ obviously satisfies the functional
equations \eqref{eqspa}, hence it can be resolved by the same
uniformisation map, with $x$ given by eq.\,\eqref{solutionSX}.  The
difference from $Y(x)$ is the asymptotics at $x\to\infty$.  Since the
external potential $U(x)$ does not depend on $\kappa$,
eq.\,\eqref{asymJ} gives $\p_\k Y(x) = C / x^2 + O(x^{-3})$, where $C$
is a constant.  In the $s$-plane, we therefore must have a double zero
at the origin, $\p_\kappa y(s) \sim 1 / x^2(s) \sim s^2$.  This,
together with the monodromy conditions \eqref{periodis} determines $
\p_\k y(s) = {\mathfrak c} / x'(s) $ for some constant ${\mathfrak
c}$.  The constant $ {\mathfrak c} $ is fixed by the integral over the
$\CA$-cycle
\begin{equation}	\label{normnk} 
\oint_\CA \frac{dx}{2 \pi } \; \p_\kappa Y(x) = 1 .
\end{equation}
 In the   $s$-parametrisation, the integral becomes trivial,  
\begin{equation} 	\label{normX1}
1 = \frac{1}{2\pi} \int_{-\frac{\pi \tau}{2}}^{\frac{\pi \tau}{2}} ds
\ x'(s+ \tfrac12 i \pi) \, \p_\kappa y(s + \tfrac12 i \pi ) =
\frac{{\mathfrak c}}{2\pi} \int_{-\frac{\pi \tau}{2}}^{\frac{\pi
\tau}{2}} ds = \frac{{\mathfrak c} \tau}{2} ,
\end{equation}
hence ${\mathfrak c} = 2/\t$ and
\begin{equation}	\label{dkappay}
\p_\k  y(s) = \frac{2 / \tau}{x'(s)}  .
\end{equation}
 The integral over the $\CB$-cycle in eq.\,\eqref{p2CF0intY} is then equally trivial, and we find 
\begin{equation}	\label{frenb}
\p_\kappa^2 \CF_0 = \frac{2\pi   }{\tau }
\end{equation}
with $\t=\t(\k,T)$ determined by \eqref{normX}.

\section{Phase diagram and critical behaviour} 
\label{SectCriticalCurves}

\def\Tstarm{  \tilde{T}_*}

\subsection{Critical curve}

Now we are ready to reconstruct the phase diagram of the 7-vertex
model in the space of the couplings $\kappa$ and $T$.  The observables
are analytic functions of $\kappa$ and $T$ in the physical domain
where the partition function converges:
\begin{equation}
0<\k<\k_c(T), \qquad -\infty<T<\infty.
\end{equation}
When the cosmological constant $\k$ is tuned to its critical value
$\k_c(T)$, the volume of the typical planar graph diverges and the sum
over graphs can be replaced by a path integral over Riemann metrics
with a generalised Polyakov measure \cite{Polyakov:1981rd}.  This is
why the critical points on the flat lattice are tricritical points on
the dynamical lattice.

 As in any discrete model of 2D gravity, the phase transitions are of third order,
 which means that the \emph{third} derivative of the free energy with respect 
to $\k$  becomes singular at the critical line,
\begin{equation}	\label{p3F0}
\left( {\p^3 \CF_0 \over \p \kappa^3 } \right)_T\to \infty \qquad
\Rightarrow\qquad \k=\k_c(T).
\end{equation}
 Using the relation between the modular parameter and the
 susceptibility, eq.\,\eqref{frenb}, we can cast the equation for the
 critical curve in the form
\begin{equation}	\label{pkappptau}
\left({\p \k \over \p \tau }\right)_T= 0 \qquad \Rightarrow\qquad \k=\k_c(T).
\end{equation}
 Here the function $\k=\k(\t, T)$
is given by the normalisation integral \re{normX}
and defines a surface in the
 three-dimensional space with coordinates $(\k, T, \t)$.
 When parametrised by $\k $ and $T$, the same surface is represented by a
 function $\t =\t(\k, T)$, not necessarily single-valued.  The
 condition \re{pkappptau} determines the boundary of the physical
 domain where the function in question is single-valued, so the critical
 curve is the one-dimensional analogue of the turning point of a
 curve.


 The equation of the critical curve can be
 decomposed into three conditions for the branch points $\ba$ and
 $\bb$, each of which implies \eqref{pkappptau}.   From
 eq.~\eqref{babbisx0} we have
\begin{equation}	\label{2x0px0papb}
2 x_0 \frac{( \p x_0 / \p \tau)_T }{(\p \k / \p \tau)_T} 
	= 
	\mathrm{a}  \left( \frac{\partial  {\mathrm{b}}}{\partial \kappa} \right)_T
	+
	\mathrm{b}  \left( \frac{\partial \mathrm{a} }{\partial \kappa} \right)_T .
\end{equation}
Since $x_0$ is an analytic function of $\tau$, as long as $x_0 \neq
0$, the l.h.s.~diverges on the critical curve \eqref{pkappptau}.  This
means that either $(\p \k / \p\mathrm{b} )_T = 0$ or $(\p \k / \p
\mathrm{a} )_T = 0$.  The other way to achieve the critical line is by
taking the limit $x_0\to 0$, as it can be shown that $( \p x_0 / \p
\tau )_T$ vanishes along with $x_0$.  Since only the right branch
point can reach the origin, $x_0=0$ is equivalent to $\bb=0$.  Thus we
have three analytic branches of the critical line:
\begin{subequations}	
\label{critkappaIIII}	
 \begin{flalign} 
 \label{kappabrI}
\left(  \partial \kappa /\partial \ba  \right)_T &= 0\ \ \Rightarrow \ \k^\rI(T),
\\
 \bb &= 0\ \ \Rightarrow \ \k^\rII(T),
 \label{kappabrII}
\\
\left(  \partial \kappa/\partial \bb  \right)_T &= 0 \ \ \Rightarrow \ \k^\rIII(T)
\, .
\label{kappabrIII} 
\end{flalign}
\end{subequations}
 Since $\bb\to 0$ is the hyperbolic limit $\t\to\infty$ of our solution,
 the equation for the  branch II   can be written also as  
\begin{equation}
\k^\rII (T)=\lim _{\t\to\infty} \k(\t, T)
\end{equation}
where the function $\k=\k(\t, T)$
is defined by the normalisation integral \re{normX}.

\subsection{Massive, dense and dilute loops}

To interpret the three branches of the critical line from the
perspective of statistical mechanics of loops on planar graphs, notice
that each branch contains a point where the solution is already known.
These are points: $T \to -\infty$, $T=0$ and $T\to +\infty$.  At
$T\to\pm\infty$ the loops disappear and the partition function is that
of the ensemble of empty trivalent planar graphs with effective
coupling constant $T^2\k$.  This problem has been solved in the
classical paper \cite{Brezin:1977sv} and the critical value of the
coupling is known to be $1/12\sqrt{3}$.  Thus the critical behaviour
for large positive or negative $T$ is that of pure Liouville gravity
(no matter fields), and also the asymptotics of the critical curve at
$T\to \pm\infty$ is
\begin{equation}	\label{asymTbig}
\lim_{T\to\pm\infty} \k_c(T) ={\textstyle {1 \over  12\sqrt{3}} } \, {  T^{-2} } .
\end{equation}
For finite but sufficiently large $T$, positive or negative,  the loops 
are small and their only effect is to renormalise the cosmological constant. 
The critical behaviour on the  branches I and III is that of pure gravity. 
Note that the symmetry $T\to -T$ is lost for finite $T$.

The branch II contains the point $T=0$ where the gas of fully packed
loops is mapped to the six-vertex model whose critical coupling $\k_c
^{\text{6v}}= \k_c(0)$ has been computed in
\cite{Kostov:1999qx,ELVEYPRICE2023105739} and whose critical behaviour
is described by Liouville gravity with Gaussian ($c=1$) matter field
compactified at radius \cite{Kostov:1999qx}
\begin{equation}	\label{Rdense}
R\big|_{ T=0}={1-\l\over 2}.
\end{equation}
 When $T\ne 0$, the loops are not fully packed any more but the
 vacancies are scarce and the universal critical behaviour remains the
 same along the branch II. This is the critical phase of \emph{dense
 loops}.

The three branches of the critical curve are parametrised by three temperature
 intervals:
\begin{equation}
\rI = ( -\infty, \Tstarm) ,
\qquad
\rII = ( \Tstarm , \Tstar) ,
\qquad
\rIII =  ( \Tstar , \infty) .
\end{equation}
  The transition temperatures $\Tstarm $ and $ \Tstar$, which must
  satisfy $ \Tstarm<0< \Tstar $, will be computed in Section
  \ref{SectSolNearCritical}.  As we shall see later, a new physics
  emerges near the critical temperature $T=T_*$ separating regimes II
  and III. The tricritical point $(T_*, \k_*)$ with $\k_* \equiv
  \k_c(T_*)$ is the phase of \emph{dilute} loops.  In the vicinity of
  this point, the free energy and the observables are given by scaling
  functions of the two renormalised coupling constants.  The scaling
  solution will be discussed in detail in section
  \ref{SectSolNearCritical}.

\subsection{Edge behaviour of the spectral density  along the critical line}

\begin{figure}[b]
\centering
	\begin{minipage}{0.8\textwidth}
	\centering
	\includegraphics[scale=0.23]{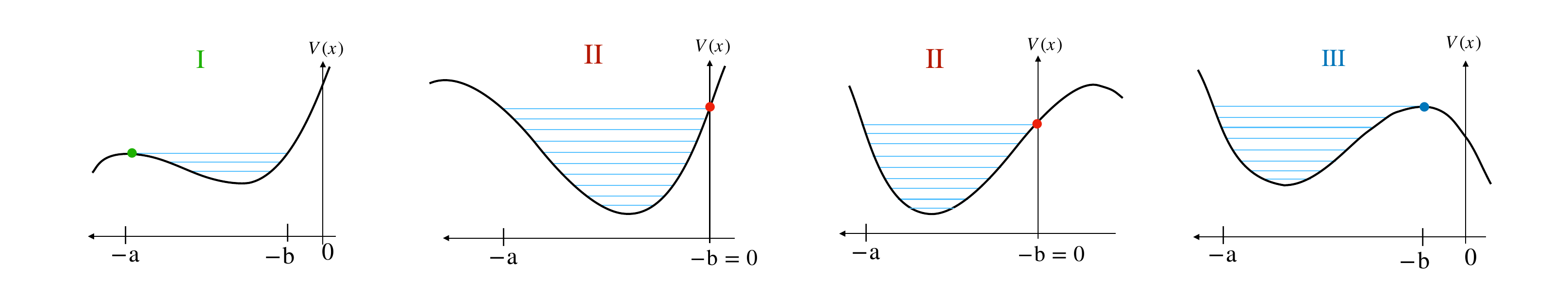}
		\caption{%
Qualitative shapes of the effective potential for different ranges of
temperature.  The support of the spectral density is $[-\mathrm{a} ,-\mathrm{b} ]$. 
The Fermi sea is represented by blue lines.  A critical configuration 
	occurs when the support approaches either the top
of the potential or the origin $x = 0$.  }
	\label{Fig_Cubic_V_after_change}
	\end{minipage}
\end{figure}

Let us now see how the spectral density behaves in the different
situations leading to criticality.  There are four possible
configurations, depicted in fig.\,\ref{Fig_Cubic_V_after_change}.
When $\k=\hbar N$ grows, the support of the eigenvalue density also
grows until it reaches either $x=0$ or the top of the potential.  (For
simplicity we ignore the interaction between the eigenvalues, this
does not change this qualitative argument.)

 \emph{Regime I.} In the leftmost panel criticality happens when the
 left branch point, $- \mathrm{a} $, reaches the maximum, whence $(\p
 \k/\p \mathrm{a} )_T=0$.  The edge behaviour of the density is in
 this case
\begin{equation}	\label{nearedgInearedgIII}
\rho_c^{\rm I}(x) 
	\sim
	\begin{cases}
	(-\mathrm{a} -x)^{3/2} &\qquad \text{for $x \to -\ba$}\, ; \\
	(- {\mathrm{b}}-x)^{1/2}      &\qquad \text{for $x \to - {\mathrm{b}}$}\, .
	\end{cases}
\ %
\end{equation}

  \emph{Regime II.} The second and the third panels of
  fig.\,\ref{Fig_Cubic_V_after_change} depict, respectively, positive
  and negative temperatures such that the minimum and the maximum are
  situated on different sides from the origin.  The critical behaviour
  corresponds to the eigenvalue distribution hitting the origin before
  reaching the maximum, and the critical curve is given by $
  {\mathrm{b}}=0$.  In the segment II, the critical behaviour of the
  dense loops is qualitatively the same as that at $T = 0$, with edge
  behaviour \cite{Kostov:1999qx}
\begin{equation}	\label{nearedgII}
\rho _c^{\rm II} (x) 
	\sim 
	\begin{cases}
	(-{\mathrm{a}}-x)^{1/2} &\qquad \text{for $x \to -{\mathrm{a}}$}
	\\
	(-x)^{\frac{1-\l}{1+\l}}	&\qquad \text{for $x \to -{\mathrm{b}}= 0$}.
	\end{cases}
\end{equation}

 \emph{Regime III.} For $T$ sufficiently large and positive, both the
 bottom and the top of the potential are on the left of the origin
 as shown in the rightmost panel of
 fig.\,\ref{Fig_Cubic_V_after_change}.  The pure-gravity singular
 behaviour occurs when the right end of the interval, $-
 {\mathrm{b}}$, reaches the top of the potential for some
 $\k=\k^\rIII(T)$.  When this happens, $\p {\mathrm{b}}/\p \k$
 diverges and so $(\p \k/\p {\mathrm{b}})_T=0$.  The edge behaviour of
 the spectral density in the regime $\rIII$ is
  \begin{equation}
\rho_c^{\rm III}(x) 
	\sim
	\begin{cases}
	(-\mathrm{a} -x)^{1/2}     &\qquad \text{for $x \to -\ba$}\,;
	\\
	(- {\mathrm{b}}-x)^{3/2}     &\qquad \text{for $x \to - {\mathrm{b}}$}\,.
	\end{cases}
\end{equation}
 
\emph{The phase boundary I/II.} At the point $\tilde T_*$ separating
regimes $\rI$ and $\rII$, we observe a superposition of the critical
singularities associated with the edge behaviour at both ends of the
eigenvalue interval.  The singular behaviour of $\rho_c(x)$ at
$T=\tilde T_*$ is
\begin{equation}	\label{nearedgIIIandII}
\rho_c^{\rm I/II}(x) 
	\sim
	\begin{cases}
	(-{\mathrm{a}}-x)^{3/2} &\qquad \text{for $x \to - {\mathrm{a}}$}
	\\
	(-x)^{\frac{1-\l}{1+\l}} &\qquad \text{for $x \to -{\mathrm{b}}=
	0$}.  \end{cases}
\end{equation}

\emph{The phase boundary II/III.} The point $T_*$ separating regimes
$\rII$ and $\rIII$ corresponds to the maximum occurring exactly at the
origin.  At this point the two mechanisms creating a singularity act
on the same edge $- {\mathrm{b}}$ and we observe a genuine tricritical
behaviour resulting from the interplay between the diverging size of
the lattice and the diverging size of the loops.  The result is that
the typical length of the loops diverges but so does the volume that
is not occupied by loops, producing a dilute critical phase of the
gravitational loop gas.  We have \cite{Kostov:2025awi}
\begin{equation}	\label{rhocrit}
\rho_c^{\rm{II/III}}(x) 
	\sim 
	\begin{cases}
	(- {\mathrm{a}}-x)^{1/2}     &\qquad \text{for $x \to - {\mathrm{a}}$}
	\\
	(-x)^{\frac{1+\l}{1-\l}}	&\qquad \text{for $x \to - {\mathrm{b}} = 0$}.
	\end{cases}
\end{equation}
Note that the exponent near $x = 0$ is \emph{different} from the one
in \eqref{nearedgII}.  This scaling behaviour, which will be derived
in Section \ref{SectSolNearCritical} below, characterises the
\emph{tricritical point} of the gravitational 7v-model.  The values of
$T_*$ and $\tilde T_*$ will be computed in the next section.

\subsection{The phase diagram}

\begin{figure}[h]
\centering
	\begin{minipage}{0.9\textwidth}
	\centering
	\includegraphics[width=0.93\textwidth]{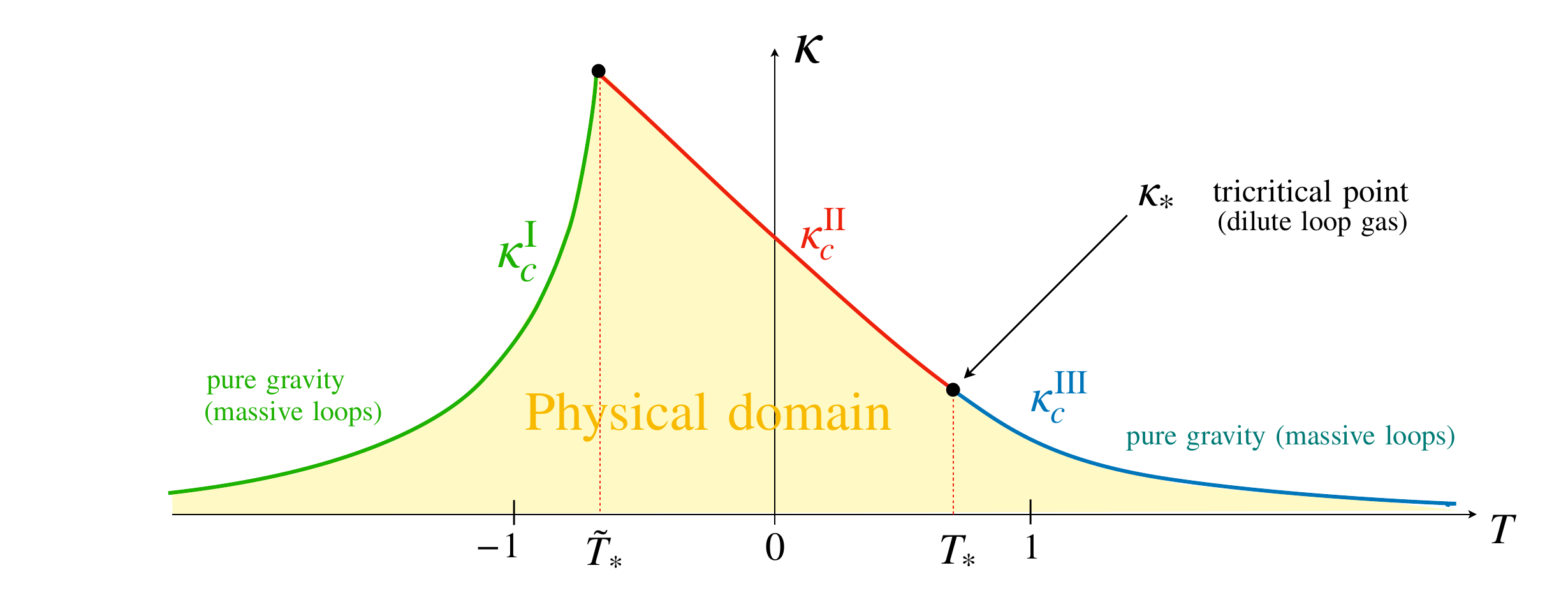} 
	\caption{%
Phase diagram of the gravitational seven-vertex model on the $T$-$\k$
plane.  The partition function converges on the physical region
(yellow).  Its boundary, the critical line $\k_c(T)$, consists of
three different analytic branches $\k_c^{\rm I}$, $\k_c^{\rm II} $ and
$\k_c^{\rm III}$, defined respectively in the segments
$\mathrm{I}=(-\infty, \Tstarm)$, $\mathrm{II}=(\Tstarm, \Tstar )$ and
$\mathrm{III}=( \Tstar , \infty)$.  In segments $ {\rm I}$ and $ {\rm
III}$ the loops are massive (pure gravity) while in segment $ {\rm
II}$ the loops are massless and form a dense filling phase.  At the
tricritical point $T=T_*$, the loop gas is in the dilute critical
phase in which both the length of the loops and the space not occupied
by loops diverge.  }
	\label{Phase_diagram_Simple}
	\end{minipage}
\end{figure}

Summarising, in fig.  \ref{Phase_diagram_Simple} we draw the phase
diagram of our statistical model.  The form of the critical curve
corresponds to the choice $\l=0.2$.  The phase diagram of the
gravitational loop gas is qualitatively the same as the phase diagram
of the loop gas on a flat lattice we briefly sketched in section
\ref{sect:7vmWS}.

\bigskip

The phase diagram is qualitatively the same as that for the 7v
model on a regular lattice \cite{Nienhuis:1984wm}, including the RG
flow from the critical point $T=T_*$ to the phase II or III. In the
model coupled to gravity, the lattice cosmological constant $\k$
controls the volume of the world sheet and thus plays the role of the
IR cutoff parameter.

\section{The solution at and  near  criticality}	
\label{SectSolNearCritical}

In this section we examine the universal behaviour in the interesting
parts of the phase diagram, namely the dense phase and the vicinity of
the critical point $\{\Tstar ,\k_*\}$, which is expected to be
described by sine-Liouville gravity.  In appendix
\ref{AppTauInftyLimit} we extract the universal behaviour by taking
the hyperbolic limit $\tau \to \infty$ of the general solution
\eqref{solutionS}.  It is however technically easier, and perhaps more
instructive, to obtain the solution in the continuum limit directly by
solving the suitably adjusted monodromy conditions \eqref{eqspa}.

\subsection{Solution in the dense critical phase}
\label{section:criticalsolution}

\def\Tcp{T^c_+}\def\Tcm{T^c_-}\def\xoc   {x^c_0} \def\yoc  {y^c_0} \def\yonec{y^c_1}

At ${\mathrm{b} } =0$, two of the branch points of the Riemann surface
of $y=Y(x)$ merge at the origin and the solution can be obtained in
elementary functions.  Introducing a hyperbolic uniformisation
parameter $s$ to resolve the remaining pair of branch points
$\mathrm{a} _\pm$ we write
\begin{subequations}		\label{SolCritCurv}
\begin{align}
x_c(s)& = \xoc     \ \frac{ e^{-\lambda s} }{ 2\sinh(s) }
\label{SolCritCurvA}
\\ 
y_c(s) &
\begin{aligned}[t]
	&= \left( \yoc + \yonec \ \p_s \right) \frac{ e^{\lambda s} }{
	2\sinh(s)}
\\
	&= \left( \yoc + \l \yonec \right) \frac{ e^{\lambda s} }{
	2\sinh(s) } - \yonec \, \frac{ \coth(s)\, e^{\lambda s} }{
	2\sinh(s) }.
\end{aligned}
\label{SolCritCurvB}
\end{align}
\end{subequations}
The second term in the expression for $y(s)$ reflects the quadratic
term in the asymptotic of $Y(x)$ at $x\to\infty$ which translates into
a second-order pole at $s=0$.  Since we are on the critical curve,
$\k=\k_c(T)$ and the constants $\xoc    , \yoc   $ and $y_{1c}$ are
functions only of the temperature $T$ and the parameter $\l$,
\begin{equation}	\label{exprx}
\begin{aligned}
\yonec   
	&= \tfrac14 \sec( \tfrac{3\pi \l}{2} ) \, T \,  (\xoc   ) ^2 ,
\\
	\yoc &= - (\tfrac12 - T) \csc(\pi\l) \, \xoc - \tfrac12 \lambda \,
	\sec( \tfrac{3\pi \l}{2} ) \, T \, (\xoc )^2 , \\
		\xoc &= \frac{4 (1-T) \sin ( \tfrac{\pi \l}{2} ) }{ \lambda
		\left( 1-2T + \sqrt{(T-\Tcp ) (T-\Tcm) / \Tcp \Tcm} \right) }
		,
\end{aligned}
\end{equation}
with 
\begin{equation}	\label{Tcpmla}
	T_c^\pm = \frac12 \pm \frac{\sin(\tfrac{\pi\l}{2})}{2} \sqrt{
	\frac{ 1 - 9 \lambda^2 }{ (1 - \lambda^2) \sin^2(
	\tfrac{\pi\l}{2}) - 2 \lambda^2 } }.
\end{equation}
Eqs.  \eqref{exprx} and \eqref{Tcpmla} are derived in Appendix
\ref{app_solution} by matching the asymptotics \eqref{asymJ}.  They
can also be found by taking $\tau \to \infty$ in the exact solutions
\eqref{x0y0t0ExactC}, see Appendix \ref{AppTauInftyLimit}.  It is not
hard to check that $\Tcp (\l)$ and $\Tcm(\l)$ are, respectively,
always positive and always negative for $\l \in [0,1]$, hence
$\xoc    $, $\yoc   $, $y_{1c}$ are all real for 
$T \in (\Tcm , \Tcp )$.

\subsection{Critical curve for the dense phase and tricritical point}
\label{SectTricritPoint}
 
The critical curve in the dense phase can be obtained explicitly by
evaluating the integral \eqref{normX} for the solution
\eqref{SolCritCurv},
\begin{equation}	\label{kappacc}
\begin{aligned}
	\kappa^\rII_{c} (T) &= - \frac{ \xoc \Big[ 3 \lambda \, \yoc + (3
	\lambda^2 - 1)\, \yonec \Big] }{12\pi} ,
\end{aligned}
\end{equation}
 where the dependence on $T$ is only through the parameters $\xoc $,
 $\yoc $ and $ \yonec $ defined in \eqref{exprx}.  Since for $\Tcm < T
 < \Tcp $, these parameters are all real and $\kappa^\rII_{c}(T)$ is
 real as well.

\begin{figure}[t]
\centering
	\begin{minipage}{0.5\textwidth}
	\centering
	\includegraphics[scale=0.5]{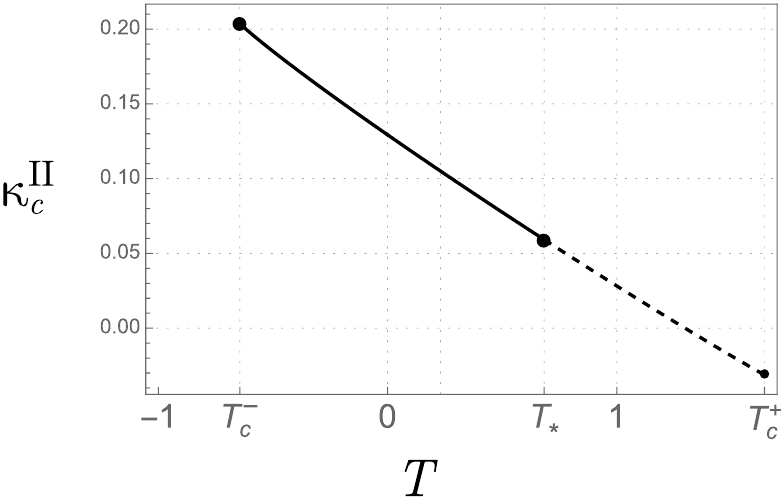}
	\caption{%
	The dense critical line, $\kappa^{\rm II}_c (T)$.  Here $\lambda =
	0.2$.  The dashed segment with $\Tstar < T < \Tcp $ is not
	physical, as shown in Section \ref{SectCriticalCurves}.  }
	\label{Kappa_c_II_Tpm}
	\end{minipage}
\end{figure}

In this way we reconstructed the segment $\k^\rII_c(T)$ of the
critical curve in fig.\,\ref{Phase_diagram_Simple}.  As an
illustration, in fig.\,\ref{Kappa_c_II_Tpm} we plot $\k_c^\rII$
against $T$ for $\l=0.2$.  The critical line in the dense phase starts
at $\Tcm$ but does not extend all the way to $\Tcp $.  Instead, it
ends at the tricritical point with temperature $\Tstar$ which we now
are going to compute.

Instead of looking for the common solution of the conditions
\eqref{kappabrI} and \eqref{kappabrII}, we will deduce the critical
temperature from the behaviour of spectral density.  We can compute
$\rho(x)$ from the discontinuity of $Y(x)$ across one of its cuts, see
eq.\,\eqref{rhofromYdisc}.  The upper cut is on the upper edge of the
parametrisation strip, so the parametric form of $q^{1/2} Y(q^{1/2}
x)$ is $q^{1/2} y_{c}(s + \frac{ i \pi}2)$, $q^{-1/2} x_c(s+ \frac{ i
\pi}2)$.  The critical behaviour happens near $x = 0$, which maps to
$s\to\pm \infty$, with the two signs corresponding to the two sides of
the cut.  The functions \eqref{SolCritCurv} behave on the two sides of
the cut as
\begin{equation}	\label{nearcrit}
\begin{aligned}
		x_c(s) &= \xoc \ e^{-(1+\l)s}, &\quad y_c(s) &= \big[ \yoc -
		(1-\l) \yonec \big] \ e^{-(1-\l)s}, &\quad \text{for $s\to
		+\infty$} , \\
		x_c(s) &= - \xoc     \ e^{(1-\l)s}, &\quad y_c(s) &= - \big[
		\yoc    + (1+\l) \yonec   \big] \ e^{(1+\l)s}, &\quad \text{for
		$s\to -\infty$} ,
\end{aligned}
\end{equation}
up to corrections of order $e^{-2|s|}$.  Therefore the spectral
density is given for $0 < x / \xoc \ll 1$ by
\begin{equation}    \label{eqdensitydensephase}
		\rho_{c }(x) \approx - \frac{1}{2\pi} \Big[ y_{0c } -
		(1-\lambda)\yonec \Big] \left|{x \over \xoc
		}\right|^{\frac{1-\lambda}{1+\lambda}} + \frac{1}{2\pi} \Big[
		\yoc + (1+\lambda) \yonec \Big] \left|{x \over \xoc
		}\right|^{\frac{1+\lambda}{1-\lambda}} .
\end{equation}
The leading contribution comes from the smaller  power,
\begin{equation}	        \label{scalingrhodense}
	\rho_{c }(x) \approx - \frac{1}{2\pi} \Big[ y_{0c } -
	(1-\lambda)\yonec \Big] \left|{x \over \xoc}\right|
	^{\frac{1-\lambda}{1+\lambda}} + \text{subleading terms}.
\end{equation}
The coefficient in front of the leading power is positive for
sufficiently small temperatures and decreases monotonically with $T$.
The critical line ends at the point $\Tstar $ where this coefficient
vanishes.  For  $T>\Tstar $ it becomes negative, and so does
$\rho(x)$.  Thus the tricritical temperature $\Tstar$ is determined by
\begin{equation}	        \label{DefSolfTstarpm}
\yoc (\Tstar) - (1-\lambda)\yonec (\Tstar) = 0
\end{equation}
which yields
\begin{equation} \label{crittemp}
\begin{aligned}
		\Tstar = \frac12 + \frac12 \sqrt{ \frac{2(1-3\lambda)
		\sin^2(\tfrac12 \pi \l) }{ (1+\lambda) \cos(\pi \lambda) -
		2\lambda } } \qquad \qquad (0 < \l < 1).
\end{aligned}
\end{equation}
Exactly \emph{at} the tricritical point, the leading term in
\eqref{eqdensitydensephase} vanishes by construction, and we are left
with the second, subdominant term.  The spectral density changes from
\eqref{scalingrhodense} to
\begin{equation} \label{scalingrhocrit}
	\rho_*(x) \equiv \rho_c(x) \big|_{ T=\Tstar} = \frac{1}{2\pi}
	\Big[ y_{0c } + (1+ \lambda)\yonec \Big] \left|{x \over
	\xoc}\right|^{\frac{1+\lambda}{1- \lambda}} + \text{subleading
	terms}
\end{equation}
and the coefficient can be shown to be positive for $T = \Tstar$.
This proves the behaviour predicted in \cite{ Kostov:2025awi} and
discussed above in \eqref{rhocrit}.

\subsection{Scaling limit}

We can study the vicinity of the critical line in phase space by
blowing up the spectral curve near the origin of the spectral plane
while the branch points of $Y(x)$ come close to zero but do \emph{not}
vanish.  This means that in the elliptic parametrisation
\re{solutionSX}, the modular parameter is large, $\tau \gg 1$, but
still finite.  We control the position of the branch point
${\mathrm{b} } $ by introducing a small parameter $\e(\tau)$ such that
$\lim_{\tau \to \infty} \e(\tau) = 0$ and
\begin{equation}	\label{DefofM}
{\mathrm{b} }  \equiv \e(\tau) \times \text{finite coefficient}  .
\end{equation}
Then we blow up the vicinity at the origin $x=0$ is done by rescaling
the spectral coordinate,
\begin{equation}	\label{xofsxscofssc}
x(s) / \xoc      \equiv 
	 \e(\tau) \; x_\sc (s_\sc) ,
\end{equation}
 where we shifted the uniformisation parameter as
 \be\begin{aligned} s_\sc \equiv s - \frac12 \beta \tau \qquad
 \text{with}\ \b = \pi (1-\l) \label{defssc} \end{aligned} \ee
 before taking the limit  $\tau \to \infty$, cf. \eqref{bbnear0}.

The function $x_\sc(s_\sc)$ is finite as $\tau \to \infty$.  Taking
the $\tau \to \infty$ limit, cf  eq.\,\eqref{xlimittauinfScaling}, we
have
\begin{equation}	\label{Mvss}
\e(\tau) = \frac1M \exp \big[ - \tfrac12 \pi  \tau  (1 - \l^2) \big] ,
\end{equation}
where $M$ is a finite parameter taking into account the freedom in
choosing the normalisation factor in the definition of $\e(\tau)$.
Then eqs.\eqref{xlimittauinfScaling} and \eqref{xofsxscofssc} give
\begin{equation}	\label{sclx}
x_\sc(s) = -2 M e^{- \l s} \sinh(s) 
\end{equation}
where from now on we will denote by $s$ the shifted parameter
\eqref{defssc}.  Notice that eq.\,\eqref{sclx} can also be obtained from
the solution  in the dense phase  \eqref{SolCritCurv}
 by applying  the   reflection symmetry \eqref{xxinv}   to the definition
\eqref{xofsxscofssc}:
\begin{equation}	\label{sclx2}
\begin{aligned}
	x_\sc(s) \equiv \lim_{\tau \to \infty} \frac1{\e(\tau)} \frac{x(s
	+ \tfrac12 \beta \tau)}{\xoc }
	= \lim_{\tau \to \infty} \frac1{\e(\tau)} \frac{x_0^2}{ \xoc \,
	x(-s)} .
\end{aligned}
\end{equation}
%

Geometrically, the constant $M$ measures the distance of the branch
points ${\mathrm{b} } _\pm $ to the origin in the scaling limit where
the branch points $\mathrm{a} _\pm$ are sent to infinity.  The
parameters $s^\pm_{\mathrm{b} } = s_{\mathrm{b} } \pm \frac12 i \pi$
of the branch points ${\mathrm{b} } _\pm$ satisfy $x_\sc'(s \pm
\frac12 i \pi) = 0$ which yields
\begin{equation}	\label{solforsbb}
s_{\mathrm{b} }  = \frac12 \log \Big( \frac{1+\l}{1-\l} \Big) .
\end{equation}
This gives the exact relation between  the branch point and $M$,
\begin{equation}	\label{solforsbb1}
\begin{aligned}
{\mathrm{b} }  / \e
	= 
	 \frac{2M
	}{
	(1+\l)^{\frac{1+\l}{2}}
	(1-\l)^{\frac{1-\l}{2}}
	}  .
\end{aligned}
\end{equation}
The r.h.s.~is finite, which checks with our assumed scaling in
\eqref{DefofM}.  Moreover, apart from the $\tau$-independent
multiplicative factor, we see that $M \sim {\mathrm{b} }  / \e$ is the rescaled
position of the branch point.  After a Laplace transform, ${\mathrm{b} } $
determines the exponential decay of the disk amplitude with given
boundary length, and therefore gives the renormalised boundary entropy
produced by the fluctuations of the bulk in the continuum limit, see
section \ref{SectBoundLengthDistro}.  This is why we will refer to $M$
in what follows as `boundary entropy'.

\subsubsection{Blowing up  the  vicinity of the tricritical point}

The limit of $y(s)$ is more subtle.  After the shift of $s$, the
function $y(s)$ maps to $y(s + \tfrac12 \beta \tau)$; we can expand
the theta functions and, at leading order we find a finite function
times $\e^{(1-\l) / (1+\l)}$, as shown in Appendix
\ref{AppTauInftyLimit}, cf.~eq.\,\eqref{ylimittauinfScaling}; this
corresponds to the solution for $s \to + \infty$ in \eqref{nearcrit}
but, as we have seen in section \ref{SectTricritPoint}, the
\emph{sub}leading term is also important near the tricritical point.
We can obtain both leading and subleading contributions more easily by
noting that we can always determine $y(s)$ from $x(s)$ by
eq.\,\eqref{solutionSY}.  Using the periodicity of the exact solution,
$x(s - \pi \tau) = x(s)$, we can write
\begin{equation}
	\lim_{\tau \to \infty}
	x(-s - \tfrac12 \beta\tau )
	=
	 \lim_{\e \to 0} 
	\frac1{\xoc    }
	\Big[
	(M \e)^{\frac{1+\l}{1-\l}} e^{(1+\l)s}
	- (M \e)^{\frac{1-\l}{1+\l}} e^{-(1-\l)s} 
	\Big]
\end{equation}
and then $y(s)$ can be computed easily by taking derivatives of this
expression,
\begin{equation}	\label{yscb}
\begin{aligned}
\!\!\!\! &\lim_{\tau \to \infty} y(s + \tfrac12 \beta \tau)
\\
	& = - \frac{ \yoc + (1+\lambda) \yonec }{\xoc }
	\e^{\frac{1+\l}{1-\l}} \left[ M^{\frac{1+\l}{1-\l}} e^{(1+\l)s} -
	\frac{\yoc - (1-\lambda) y_{1c}}{\yoc + (1+\lambda) y_{1c}} \
	\e^{- \frac{4\l}{1-\l^2}} M^{\frac{1-\l}{1+\l}} e^{-(1-\l)s}
	\right].
\end{aligned}
\end{equation}

Now we focus on the vicinity of the tricritical point, where the
coefficient of the leading term in the r.h.s.~of \eqref{yscb}
vanishes, see eq.\,\eqref{DefSolfTstarpm}; there, at leading order,
\begin{equation} \label{vanishy0clasa}
	\yoc - (1-\lambda) y_{1c} = \text{constant} \times (T - \Tstar) +
	O(T-\Tstar)^2 .
\end{equation}
This leads us to define the \emph{renormalised temperature coupling}
\begin{equation} \label{Deftscal}
t	=
	\text{constant} \times 
	\e^{- \theta}
	(T - \Tstar) ,
\qquad
\theta = \frac{4 \l}{1-\l^2} .
\end{equation}
The constant in \eqref{Deftscal} is chosen such that
\eqref{vanishy0clasa} is conveniently simplified.  Then the function
\begin{equation}	\label{yofsyscofssc}
	y_\sc(s) \equiv - \frac{\xoc }{ \yoc + (1+\lambda) \yonec } \
	\lim_{\tau \to \infty} \big[ \e(\tau) \big]^{- \frac{1+\l}{1-\l}}\
	y(s + \tfrac12 \beta \tau)
\end{equation}
is a finite function of the renormalised parameters, and gives the
scaling limit of the spectral curve near the tricritical point:
\begin{equation}	\label{solscalla}
\begin{aligned}
x_\sc(s) &= -2 M \ e^{-\lambda s} \sinh s ,
\\
y_\sc(s) &= 
	M^{\frac{1+\lambda}{1-\lambda}} \ e^{(1+\l ) s}
	- t \, M^{\frac{1-\lambda}{1+\lambda}} \ e^{- (1-\l) s} .
\end{aligned}
\end{equation}
This is the solution found in \cite{ Kostov:2025awi}.

\begin{figure}[t]
\centering
	\begin{minipage}{0.62\textwidth}
	\centering
	\includegraphics[width=0.58\textwidth]{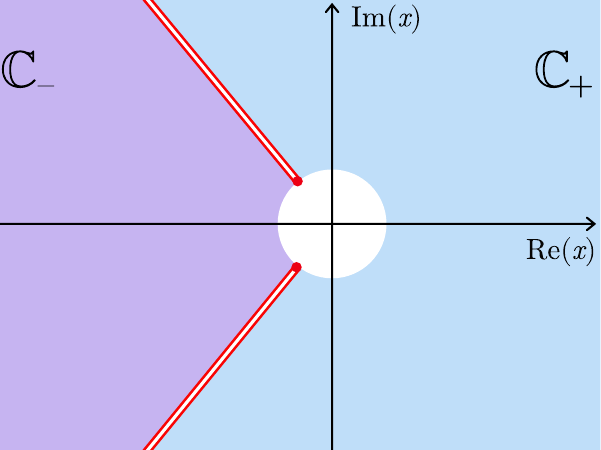} 
	\caption{%
	Complex $x$-plane in the scaling limit.  The red lines are the
	branch cuts of $Y(x)$, ending at $q^{\pm \frac12} {\mathrm{b} } $.
	The series \eqref{yofxseries2Main} converges in the blue region
	$\IC_+$.  The expansion in the purple region $\IC_-$ can be found
	using the symmetry $b \to 1/b$.  }
	\label{RegionsbbCpm}
	\end{minipage}
\end{figure}

In Appendix \ref{AppSeriesExps}, we use the parametrisation
\eqref{solscalla} to obtain the function $Y(x)$ in the scaling limit
as a power series around $x = \infty$.  After the renormalisations
\eqref{xofsxscofssc} and \eqref{yofsyscofssc},
\begin{equation}	\label{yofxseries2Main}
\begin{aligned}
	Y_\sc(x_\sc) &= \sum_{n=0}^{\infty} \frac{(-)^n}{n!} \textstyle
	\Bigg[ \left( \frac{1+\l}{1-\l} \right)^{\frac12} \left( (n-1)
	\frac{1+\l}{1-\l} \right) _n \left( -\frac{M}{x_\sc} \right)^{
	\frac{2}{1-\l} n} \left( -x_\sc\right) ^{\frac{1+\l}{1-\l} } \\
	 &\quad - t\textstyle \left( \frac{1-\l}{1+\l} \right)^{\frac12}
	 \left( (n-1) \frac{1-\l}{1+\l} \right) _n \ \left(-
	 \frac{M}{x_\sc} \right)^{ \frac{2}{1+\l} n} \left( - x_\sc\right)
	 ^{\frac{1-\l}{1+\l} } \Bigg] , \qquad x_\sc \in \IC_+ ,
\end{aligned}
\end{equation}
 where $(a)_n= \Gamma(a+n) / \Gamma(a) $ are Pochhammer symbols.  The
 coefficients of the two series are analytic (in fact polynomials) in
 the positive ratios $(1-\l)/(1+\l)$ and $(1+\l)/(1-\l)$, i.e.~they
 are analytic in $\l$ for $0 < \l < 1$.  As explained in Appendix
 \ref{AppSeriesExps}, the expansion holds for $\Re[ x_\sc ]> 0$, and
 the two series in the r.h.s.~of eq.\,\eqref{yofxseries2Main} converge
 for
\begin{equation}	\label{Converglarb}
| x | > {\mathrm{b} }  ,
\end{equation}
that is, outside a circle around the origin with radius ${\mathrm{b} }
$.  Together with the branch cuts, this circle splits the complex-$x$
plane into two domains which we call $\IC_\pm$, illustrated in
fig.\ref{RegionsbbCpm}.  In summary, the series
\eqref{yofxseries2Main} converges in $\IC_+$.

To obtain an expansion for $Y_\sc(x_\sc)$ that is valid in $\IC_-$, we
can use a symmetry of the spectral curve.  Recall that $q^{\frac12} =
e^{ \frac12 i \pi (1-\l)}$, hence the transformation $\l \mapsto -\l$
changes $q^{\frac12} \mapsto - \bar q^{\frac12}$.  It follows from
eq.\,\eqref{defPhi} that the spectral curve is invariant under $\l
\mapsto - \l$ and $x \mapsto - x$, therefore
\begin{equation}	\label{symmYofxla}
Y(-x) = Y(x) \big|_{\l \mapsto -\l} .
\end{equation}
This symmetry of the exact solution is almost present in the scaling
solution \eqref{solscalla}, but it is broken by the presence of $t$.
This is because with $\l \mapsto - \l$  the
expansion \eqref{yscb} must be reorganised, since the vicinity of the
critical point comes from the other term inside the square
brackets.\footnote{See the discussion below eq.\,\eqref{crittemp}.} The
final effect is that the transformation \eqref{symmYofxla}, in the
scaling limit, must be accompanied by   moving of $t$ to the other
term in the r.h.s.~of \eqref{solscalla}.  Taking this into account we
conclude that
\begin{equation}	\label{yofxseries2Main1}
\begin{aligned}
	  Y_\sc(x_\sc) = \sum_{n=0}^{\infty} & \frac{(-)^n}{n!} \textstyle
	  \Bigg[ t \, \left( \frac{1-\l}{1+\l} \right)^{\frac12} \left(
	  (n-1) \frac{1-\l}{1+\l} \right) _n \left( \frac{M}{x_\sc}
	  \right)^{ \frac{2n}{1+\l}} ( x_\sc)^{\frac{1-\l}{1+\l} } \\
		&\quad - \textstyle \left( \frac{1+\l}{1-\l} \right)^{\frac12}
		\left( (n-1) \frac{1+\l}{1-\l} \right) _n \left(
		\frac{M}{x_\sc} \right)^{ \frac{2n}{1-\l}} (
		x_\sc)^{\frac{1+\l}{1-\l} } \Bigg] , \qquad x_\sc \in \IC_- .
\end{aligned}
\end{equation}
Note that this formula holds again for $\l \in (0,1)$.

\subsubsection{Grand canonical  boundary entropy, 
susceptibility and free energy}

We would now like to derive the equation of state relating the
susceptibility and the boundary entropy.  It is more convenient, as in
\cite{ Kostov:2025awi}, to change from the canonical to the
grand-canonical ensemble because of the simpler analytic structure of
the free energy.  The integrand in \eqref{eigvalinx1bis} is an
$N\times N$ determinant, and can be formulated as the partition
function of a system of $N$ fermions.  On the other hand, the
grand-canonical partition function \footnote{More about the grand
partition function in \cite{1991hep.th....8019K,Ginsparg:1993is}.}
\begin{equation}	\label{Zgc}
\tilde \CZ(\tilka) \equiv \sum_{N=1}^\infty e^{- N \tilka / \hbar} \CZ_N ,
\end{equation}
 has nicer analytic properties as a function of the chemical potential
 $ \tilka$.  In particular, for a general potential it exhibits Toda
 integrability, see e.g. \cite{Kazakov:1998ji}.

We are interested in the thermodynamical limit $\hbar \to 0$, $\tilka
\to \infty$, with $\hbar\tilka$ and $\hbar\kappa$ fixed.  In this
limit the (genus-zero) canonical and grand-canonical free energies
\begin{equation}	\label{CZNisexpCF}
\CF_0(\kappa) = \hbar^2 \log \CZ_N  ,
\qquad
\tilde\CF_0(\tilka)  =  \hbar^2 \log \tilde \CZ(\tilka)  
\end{equation}
are related by Legendre transform.  The saddle-point approximation of
the sum \eqref{Zgc} gives
\begin{equation}	\label{CZmuistilfeFmu}
	\exp \frac1{\hbar^2} \tilde \CF_0(\tilka) = \sum_{N=1}^\infty \exp
	\frac1{\hbar^2} \big[ - \kappa \tilka + \CF_0(\kappa) \big]
	\approx \exp \frac1{\hbar^2} \big[ - \kappa\tilka + \CF_0(\kappa )
	\big]
\end{equation}
where on the r.h.s. $\k $ is a function of $ \tilka $ defined by the
stationary phase condition,
\begin{equation}	\label{mutok}
\p_\kappa \CF_0(\kappa) = \tilka .
\end{equation}
 Hence  
\begin{equation}	\label{cgrc}
\tilde \CF_0(\tilka) = - \kappa \, \tilka + \CF_0 (\kappa),
\end{equation}
which also gives 
\begin{equation}	\label{mutokDual}
\p_\tilka \tilde \CF_0(\tilka) = - \kappa .
\end{equation}
The dual pair of relations \eqref{mutok} and \eqref{mutokDual} 
show that the integrals of the spectral function $Y(x)$ on the 
two cycles $\CA$ and $\CB$, cf.~eq.\,\eqref{normn}, compute the first 
derivatives of the free energies in each ensemble.

We obtained in section \eqref{sec:unif} a simple formula,
eq.\,\eqref{frenb}, expressing the susceptibility in the canonical
ensemble $u(\kappa,t)\equiv \p_\kappa^2 \CF_0 $ as the reciprocal of
the modular parameter, and there is a similar formula here for the
susceptibility in the grand-canonical ensemble
which we normalise for convenience as
\begin{equation}	\label{DefSuscpa2CF}
\tscu (\tilka,t) \equiv  {  4 \pi^2} \p_\tilka^2 \tilde \CF_0 .
\end{equation}
Combining both eqs.\eqref{mutokDual} and \eqref{mutok}, we find that
the susceptibilities in both ensembles are related as
\begin{equation}
\begin{aligned}
	\p_\tilka^2 \tilde \CF_0 = - \frac{\p \kappa}{\p \tilka} = -
	\frac1{ \p \tilka / \p \kappa} = - \frac1{\p_\kappa^2 \CF_0},
\end{aligned}
\end{equation}
hence eq.\,\eqref{frenb} translates to
\begin{equation}	\label{frenbGC}
\begin{aligned}
\tscu(\tilka, t) = - {2 \pi \tau  } .
\end{aligned}
\end{equation}

In the scaling limit where $\tilka$ is close to the critical value
$\tilka_*$, eq.\,\eqref{eqC2} together with the scalings in
\eqref{yofsyscofssc} and \eqref{xofsxscofssc},  imply 
\begin{equation}	\label{RenormkappDu}
\begin{aligned}
\tilka - \tilka_* & = \text{constant} \times \frac{1}{2i} \int_{-
\frac{i \pi}{2} }^{ \frac{i\pi}{2} } ds \ x'(s ) \, y(s)
\\
& = \text{constant} \times \e^{\frac{2}{1-\l}} \times \frac{1}{2i}
\int_{- \frac{i \pi}{2} }^{ \frac{i\pi}{2} } ds \ x_\sc'(s ) \,
y_\sc(s) .
\end{aligned}
\end{equation}
This invites us to define  the renormalised chemical potential
$\mgc$ as
\begin{equation}	\label{RenormkappGC}
\tilka - \tilka_* = \text{constant} \times \e^\nu  \mgc ,
\quad
\nu = \frac{2}{1-\l} ,
\end{equation}
and we choose the constants in such a way that 
\begin{equation}	\label{Defnu2l}
\mgc = \frac{1}{2i} \int_{- \frac{i \pi}{2} }^{ \frac{i\pi}{2} } ds \
x_\sc'(s) \, y_\sc(s) .
\end{equation}
Given the renormalisation of $\tilka$ in \eqref{RenormkappGC}, we
introduce the renormalised grand-canonical free energy $\tilde F_0$ so
that the susceptibility does not contain power of $\e$,
\begin{equation}	\label{RenormF}
\tilde\CF_0 - \tilde\CF_{0*} = \text{constant} \times \e^{2\nu} \tilde
F_0 .
\end{equation}
The scaling with $\e$ is such that $\kappa - \kappa_* = - \p_\tilka
\tilde \CF_0 \sim \e^{\nu} \p_\mu \tilde F_0 , $ and thus we introduce
the renormalised cosmological constant $\mu $, such that
\begin{equation}	\label{Renormkapp}
\kappa - \kappa_* = \text{constant} \times \e^\nu  \mu  ,
\end{equation}
which has the same scaling as \eqref{RenormkappGC}.  Then we choose
the constants such that the scaling limit of \eqref{cgrc} reads
\begin{equation}	\label{cgrcs} 
\tilde F_0 (\mgc) = - \mgc \mu  + F_0(\mu ) ,
\end{equation}
while the renormalised cosmological constant and chemical potential
are related by
\begin{equation}	\label{cgrcs2} 
\mgc = \p_{\mu } F_0 , 
\qquad
\p_\mgc \tilde F_0 = - \mu .
\end{equation}

The integral in eq.\,\eqref{Defnu2l}  is  easily computed  and gives
\begin{equation}	\label{MLpak0}
\mgc = \frac{1 + \l}{2} M^{\frac{2}{1-\l}} - \frac{1 - \l}{2} \; t \;
M^{\frac{2}{1+\l}} .
\end{equation}
This determines the boundary entropy as a function of the chemical
potential and the temperature.  The transcendental equation
\eqref{MLpak0} is solved in Appendix \ref{appsolvtreqs} by Lagrange
inversion.  Introducing the dimensionless temperature
	\be\begin{aligned}  \hat t \equiv \left( \frac{2\mgc}{1 + \l}
	\right)^{-\frac{2\l}{1+\l}} \frac{1 - \l}{1 + \l} \; t ,
\end{aligned} \ee 
we write
\begin{equation}	\label{Mofhatt}
	M(\mgc,t) = - \frac{1-\l}{2} \left( \frac{ 2\mgc}{1+\l}
	\right)^{\frac{1-\l}{2} } \hat M(\hat t) ,
\end{equation}
where the scaling function $\hat M(\hat t)$ is a power series in $\hat
t$ with finite radius of convergence,
\begin{equation}	\label{Mofhatt2}
		\hat M(\hat t) = \sum_{n=0}^\infty \frac{ (-)^n \Gamma
		\big(\frac{2\l}{1+\l} n - \frac{1-\l}{2} \big) }{ n!  \Gamma
		\big( \frac{1+\l}{2} - \frac{1 - \l}{1+\l} n \big)} \ \hat t^n
		.
\end{equation}

Returning to eq.\,\eqref{MLpak0}, we can use it to derive the equation
of state for the susceptibility.  From eq.\,\eqref{Mvss} we have
\begin{equation}	\label{taufunctM}
\tau = - \frac{2}{\pi (1 - \l^2)} \log (\e M) ,
\end{equation}
and then eq.\,\eqref{frenbGC} immediately gives
\begin{equation}	\label{SusceSCGC}
\tscu =   \frac{4}{1-\l^2} \ \log (M \e)  .
\end{equation}
With the non-universal part of $\tscu$ subtracted, we thus have
\begin{equation}	\label{SusceSCGC1}
M = \exp \Big(  { 1-\l^2\over 4} \tscu \Big) ,
\end{equation}
and eq.\,\eqref{MLpak0} becomes the equation of state for the gravitational 
seven-vertex model claimed in \cite{ Kostov:2025awi},
	\footnote{ We are using a slightly different notation from \cite{
	Kostov:2025awi}: $\mgc$ here is $\mu$ there, and $\tscu$ here is
	$u$ there.  }
\begin{equation} \label{MLpak0SUC}
\mgc(\tscu,t) = \frac{1 + \l}{2}\; e^{ { 1+\l\over 2}\tscu} - \frac{1
- \l}{2} \; t \; e^{   {1-\l\over 2}\tscu} .
\end{equation}

 Furthermore the grand canonical free energy can be computed as a
 series by integrating twice \eqref{SusceSCGC} in $\tilde\mu$.  The
 normalisation is not important because it can be absorbed in the
 Planck constant.  The leading term
 takes the form
\begin{equation} \label{Suscpa2CFM1}
	  \tilde F_0 (\mgc) ={ \mgc^2\over 2} \;    \frac{ 4}{1-\l^2} \ \log (M \e) =
	  \begin{cases}
      & {2\over 1+\l} {\tilde\mu^2\over 2}\log\tilde\mu \qquad\qquad \quad  \text{ for \ }  t\to 0 , \\
      &   {2\over 1-\l} {(\tilde\mu/t)^2\over 2}\log|\tilde\mu /t|\qquad \text{ for \ }  t\to -\infty .
\end{cases}
\end{equation}
In the limit $t\to -\infty$ this
 is the genus-zero partition function of Liouville gravity with matter
 field a free boson compactified at radius \eqref{Rdense} obtained
 previously for the dense critical phase, namely $R|_{t\to -\infty }=
 {1-\l\over 2}$ (Strictly speaking, this is the free energy of the
 T-dual boson for which the volume of the target space is $2\pi/R$.)
 From the asymptotical form in the limit $t\to 0$ we can extract
 the compactification radius at the critical point (the dilute phase),
 \begin{equation}
 R\big|_{t=0}=  {1+\l\over 2}.
 \end{equation}

\subsection{Boundary length distribution in the scaling limit}
\label{SectBoundLengthDistro}

The disk partition function with fixed boundary length $\ell$ is given
by $\CU_L ={1\over L} \hbar \langle \Tr \X^ L\rangle,$ where $\X$ is
the original matrix in the partition function \eqref{SLMM} and $ L$ is
a positive integer.  After the shift in \eqref{RESCALING}, it becomes
\begin{equation}	\label{defCU}
\CU_L 
	={1\over  L} \hbar 
	\left\langle \Tr (1 + \X)^ L\right\rangle .
\end{equation}
In the scaling limit, the boundary length becomes macroscopic, $ L =
\ell / \e$, with $\ell $ finite.  Rescaling $\X\to \e \X$ we write, in
the thermodynamical limit,
\begin{equation}	\label{defCU2}
	\CU_L\to \CU(\ell) \approx {1\over \ell} \left\langle \hbar \Tr
	e^{\ell \X} \right\rangle \to {1\over \ell} \int dx \, \rho(x) \
	e^{\ell x} .
\end{equation}
 The disk partition function thus decays exponentially, with the
 exponent determined by the right edge of the eigenvalue distribution
 related to $M$ by \eqref{solforsbb1},
 \be\begin{aligned} \CU (\ell )\approx e^{- \ell\,{\mathrm{b} } /\e} =
 e^{-2 cM\ell}, \qquad c= (1+\l)^{-\frac{1+\l}{2}}
 (1-\l)^{-\frac{1-\l}{2}} .  \end{aligned} \ee

The disk partition function with a marked point on the boundary is the
inverse Laplace transform of the resolvent,
\begin{equation} \label{tildeWandWrho}
\begin{aligned}
\tilde W (\ell) &= \frac{1}{2\pi i} \int_\uparrow dx \ W(x) \, e^{
\ell x} =\int dx \ \rho(z) \, e^{ \ell x} =\ell \,\CU(\ell ).
\end{aligned}
\end{equation}
The arrow in the inverse Laplace transform integral denotes a vertical
contour that goes up in the imaginary direction, to the right of all
singularities and branch points of the integrand.  Conversely, the
resolvent is the Laplace transformation of the partition function with
a marked boundary,
\begin{equation}	\label{invLIW}
	\tilde W (\ell) = \frac{1}{2\pi i} \int_\uparrow dx \ W(x) \, e^{
	\ell x}, \qquad W(x) =\int_0^\infty \!  d\ell \ \tilde W(\ell) \,
	e^{- \ell x} .
\end{equation}
Thus $W(x)$ can be interpreted as the disk amplitude with a marked
point on the boundary and (complexified) boundary cosmological
constant $x$.

By deforming the contour of integration, the integral over the
vertical contour for the inverse Laplace transform \eqref{invLIW} is
transformed to the integral over a counterclockwise contour $\CC$
around the cut of $W(x)$,
\begin{equation}	\label{invLIWCC}
 \tilde W (\ell)
 	= \frac{1}{2\pi i} \oint_\CC  dx \ W(x) \, e^{ \ell x} .
\end{equation}
We can now  use   eq.\,\eqref{defPhi}   to express the resolvent   as
\begin{equation}	\label{invWofY}
	W (x) = \frac{i}{2} \Big[ q^{\frac12} Y(q^{\frac12} x) - \bar
	q^{\frac12} Y(\bar q^{\frac12} x) - i q W(q x) - i \bar q W(\bar q
	x) \Big] .
\end{equation}
The last two terms are analytic around the cut and do not contribute
to the integral \eqref{invLIWCC}.  The first two terms are not
analytic, because for $x$ on the cut of $W(x)$, $q^{\pm \frac12} x$ is
on either one of the cuts of $Y(x)$.  Thus the integral over $\CC$ can
be expressed as two integrals over contours $\CC_1$ and $\CC_2$,
around the upper and the lower branch cuts of $Y(x)$, respectively.
Integrating by parts, we obtain an integral over $Y$,
\begin{equation}	\label{invWofY34}
\begin{aligned}
	\tilde W (\ell) &= - \frac{1}{4\pi \ell} \left[ \oint_{\CC_1} \!
	e^{\bar q^{1/2} \ell x} \, d( q^{\frac12} Y(x)) - \oint_{\CC_2} \!
	e^{q^{1/2} \ell x} \, d ( \bar q^{\frac12} Y(x)) \right] \\
 &=\frac{1}{2\pi \ell} \Re \int_{-\infty}^{\infty} \!  e^{\bar q^{\frac12} \ell
	x_\sc(s + \frac12 i\pi) } \, d[q^{\frac12} y_\sc(s + \tfrac12 i
	\pi)] .	
\end{aligned}
\end{equation}

 We now introduce a new parameter $\omega$ which leads to somewhat
 simpler and more symmetric expressions,
\begin{equation}	\label{ChangeOfvarome}
\omega = \exp\left( - \frac{2s}{b + b^{-1}} \right) , \qquad b \equiv
\sqrt{ \frac{1-\l}{1+\l}} ,
\end{equation}
In the new parametrisation the spectral curve \eqref{solscalla} takes
the form
\begin{equation} \label{scalyxforMAp}
\begin{aligned}
x_\sc(s \pm \tfrac12 i\pi) &= q^{\pm 1/2} M \big( \omega^{1/b} +
\omega^{-b} \big) , \\
	y_\sc(s \pm \tfrac12 i\pi) &= - q^{\mp 1/2} \big( M^{1/b^2}
	\omega^{-1/b} + t M^{b^2} \omega^{b} \big) .
\end{aligned}
\end{equation}
Then  we write  \eqref{invWofY34}  as
\begin{equation}	\label{invWofY5}
	\tilde W (\ell) = \frac{1}{2\pi \ell} \int_0^\infty \!  d \omega \
	e^{- \ell M ( \omega^{1/b} + \omega^{-b} )} \, \Big( b t M^{b^2}
	\omega^{-1+b} - \frac1b M^{1/b^2} \omega^{-1-1/b} \Big) .
\end{equation}

The rhs of \eqref{invWofY34} can be expressed in terms of the special
function $K^{(b)}_\a (z)$ defined the integral representation
\begin{equation}	\label{defII} 
	K^{(b)}_\a (z) \equiv \frac12 \int_ 0^\infty \!  d\omega \; e^{-
	\frac12 z(\omega^{1/b} + \omega^{-b}) } \omega^{\a-1}, \qquad \Re
	z > 0.
\end{equation}
For $b = 1$, $K^{(1)}_{\a}(z) = K_{\a}(z)$ is the Macdonald function,
i.e.~the modified Bessel function of the second kind of order $\a$.
For $b\ne 1$ we will refer to $K^{(b)}_\a(z)$ as a `generalised
Macdonald function'.  For general $b > 0$,  it belong to
a family of functions known as `generalised Krätzel functions'
\cite{kratzel1975integral}.  Some relevant properties are derived in
Appendix \ref{AppKratzelFunctions}.  Using the symmetry
\begin{equation}	\label{symK}
K^{(b)}_{-\a}(z) = K^{(1/b)}_{\a}(z) ,
\end{equation}
 \eqref{invWofY5} can be cast into the form
\begin{equation}	\label{WIIb} 
	\tilde W(\ell) = \frac{1}{\pi \ell} \left( t \, b M^{ b^2}
	K^{(b)}_{ b}(2M\ell) - \frac{1}{b} M^{1/b^2} \; K^{(1/b)}_{1/b}
	(2M\ell) \right) .
\end{equation}

\section{Worldsheet CFT}
 \label{sect:WS}

In this section we consider the worldsheet QFT for the scaling limit
of the lattice theory.  We know that in the massive phases (regimes
$\rI$ and $\rIII$) the scaling limit is that of pure Liouville gravity
while in the dense phase (regime $\rII$) the scaling limit is in the
same universality class as the 6v model on planar graphs
\cite{Kostov:1999qx} whose continuum limit is that of Liouville
gravity with Gaussian matter field compactified at radius
\eqref{Rdense}.
 
The regime where we observe new physics is the vicinity of the
tricritical point $\{\Tstar , \k_*\}$ separating the phases of dense and
massive loops.  In \cite{ Kostov:2025awi} one of the authors claimed
that the scaling limit is described by sine-Liouville gravity, with
the sine-Liouville interaction constant identified as the renormalised
thermal coupling.  Below we review the analysis and the arguments
presented in \cite{ Kostov:2025awi} as well as the necessary
background in order to make the presentation self-contained.

\subsection{The 7v model on a flat lattice: mapping to a Coulomb gas.   }
 \label{sect:WSSG}

Let us articulate the statement in section \ref{sect:7vmWS} about the
continuum limit of the 7vM on the infinite honeycomb lattice.  It is
known \cite{Nienhuis:1984wm} that the equivalent height model with
Boltzmann weights \eqref{BweightsSOS} renormalises at large distances
to a Coulomb gas.  At the critical temperature $T=T_c$, eq.
\eqref{crT}, the height $h\in\mathbb{Z}$ renormalises at large
distances as a Gaussian field $\vp = \pi h$.  Since a vertex
configuration determines a height function up to a translation by
integer constant, the one should identify $\vp=\vp+ 2\pi$.  (The
symmetry $\vp\to -\vp$ turns the target space of the Gaussian field
into an orbifold with twice smaller compactification length $\pi$, see
e.g. \cite{Ginsparg:1987eb}).  The normalisation $g$ of the Gaussian
action
\be\begin{aligned}  \label{actFB} \CA_{_{\text{FB}}} [\vp ]&= {g\over 4\pi}
\int(\p_\mu \vp )^2  \label{Agauss} \end{aligned} \ee 
known as `Coulomb gas coupling' is determined by the vertex weights as
  \be\begin{aligned}  \label{gofp} g= {p+1\over p},
  \qquad  n\equiv  w^6+\bar w^6= 2\cos(\pi/p).  \end{aligned} \ee 
 In \cite{Nienhuis:1984wm}, the Gaussian field \eqref{actFB} is
interpreted as the UV fixed point of a two-dimensional Coulomb gas
where the fugacities of the Coulomb charges renormalise to zero.

Rescaling the gaussian field as $\vp\to \vp /R$ with $R=\sqrt{g}$, the
Coulomb coupling is set to one but the identification is now $\vp =
\vp+ \b$ where $\b= 2\pi R$ is the compactification length.  One also
says that the Gaussian field takes values in a circle of radius $R$.
The basic operators are then the electric, or vertex, operators
$\mathcal{O}_{n}=e^{in \varphi/R}$ and the magnetic, or vortex,
operators $\tilde{\mathcal{O}}_{m}=e^{i m R \tilde\vp}$ creating
discontinuity $2\pi mR $, with $m$ and $n$ both integers in our
normalisation.  The scaling dimensions of the electric and the
magnetic operators are respectively
  \be\begin{aligned} \label{dimsem} 
  x_{\pm n}=   {\textstyle{1\over 2}} (n/R)^2 \qquad \text{and}\qquad
  \tilde x_{\pm m}=  {\textstyle{1\over 2}} (mR)^2.
   \end{aligned} \ee 
The electric-magnetic duality, or T-duality, maps
\be\begin{aligned} 
n\leftrightarrow m, \qquad R\leftrightarrow 1/R.
\end{aligned} \ee 

To stress that the radius $R$ characterises the UV limit, we denote
$R=R_\uv $.  As argued by Nienhuis \cite{Nienhuis:1984wm}, the
perturbation with $t\sim T-T_c$ is triggered by a pair of electric
operators with charges $n=\pm 2$ with conformal dimensions
\be x_{\pm 2 }= {2\over R^2}= 2\ {p \over p+1 } \ee
which are relevant for $1<R_\uv <2$.  (In the 7v model there are no
magnetic operators because of the ice rule, except if they are
introduced by hand.)

Because of charge conservation, the bulk observables depend only on
the product \eqref{tmumu}.  When the fugacities $\mu_{\pm 2}$ of these
operators are small, the perturbation can be considered as triggered
by a neutral local `thermal' operator with coupling constant
 \be\begin{aligned}  \label{tmumu} t= \mu_{2}\, \mu_{-2}  \end{aligned} \ee 
and dimension
	\be\begin{aligned} \label{simSG} x_{_{\SG}} = x_2+x_{-2} -2 = 2 \,
	{p-1\over p+1} .  \end{aligned} \ee
If $ t>0$, the fugacities $\mu_{\pm 2} $ grow under renormalisation,
the effective compactification radius increases thus making these
fugacities more relevant and the system flows to a trivial massive
phase.  If, however, $ t<0$, the effective compactification radius
decreases and eventually renders the fugacities irrelevant.  As a
consequence, $\mu_2$ and $\mu_{-2}$ diminish and ultimately disappear
under renormalisation, and the system flows towards another gaussian
field with smaller compactification radius $ R_\ir <1$
\cite{Nienhuis:1984wm}.  We can assume that $\mu_2=\mu_{-2}$ because,
by charge conservation, the dynamics depends only on their product.

\subsection{Sine-Gordon QFT}

The local quantum field theory for the thermal perturbation is the
sine-Gordon QFT described in the UV limit by the action\footnote{The
connection with the traditional notation is $ \b_{ \mathrm{SG}} =
\sqrt{8\pi}/R $.}
\be\begin{aligned}  \label{actgbSG} \CA^{\text{orb}}_{_{SG}} [\vp ]
&=   \CA _{_{FB}} - \mu_{_{SG}} \int d^2 x \cos\left(  2   \vp/R \right)  
\end{aligned} \ee 
where $2\mu_{_{SG}}  =\mu_2=\mu_{-2}$.
 For $t = 4 \mu_{_{SG}}  ^2>0$ which corresponds to \emph{real}
  mass coupling $ \mu_{_{SG}} $, the
theory flows to a massive phase.  For $t< {0}$, which corresponds to
\emph{purely imaginary} mass coupling $\mu_{_{SG}} $, the sine-Gordon
theory experiences a non-unitary thermal flow which was later analysed
using integrability methods by Fendley, Saleur and Al.  Zamolodchikov
\cite{Fendley:1993wq,Fendley:1993xa,Zamolodchikov:1994za}.

One-loop RG calculations \cite{Jose:1977gm,Nienhuis:1984wm} support
the following relation between the compactification radii of the free
boson at the two extremities of the massless flow, $
R=R^{^{\mathrm{SG}}}_{^{\uv}} $ at $t=0$ and
$R=R^{^{\mathrm{SG}}}_{^{\ir}} $ at $t\to -\infty$,
\be\begin{aligned} (R^{^{\mathrm{SG}}}_{^{\uv}} )^2+
(R^{^{\mathrm{SG}}}_{^{\ir}} )^2 = 2, \label{RuvRir} \end{aligned} \ee
 or in terms of the parameter $p$,
 \be\begin{aligned}  \label{R0e0A} R^{^{\mathrm{SG}}}_{^{\uv}} = \sqrt{1+ 1/p} , \quad
 R^{^{\mathrm{SG}}}_{^{\ir}} = \sqrt{1-1/p}.  \end{aligned} \ee 
The limit $p\to\infty$ corresponds to the BKT radius $R_{\text{BKT}}
=1$ which in this case is also the self-dual radius.  

Geometrically, one can think of the oriented loops on the honeycomb
lattice as kinks interpolating between adjacent classical vacua of the
sine-Gordon potential.  In the massive phase $T>T_c$ ($t>0$), the
theory falls into one of the classical vacua.  Droplets of other
classical vacua can appear but they are exponentially suppressed.  At
$T=T_c$ ($t=0$), the kinks become massless and stay massless for
$T<T_c$ ($t<0$).  For large negative $t$ the kinks fill densely the
lattice and form a critical phase.

\subsection{The 7v model on a dynamical lattice and the sine-Liouville
gravity }

By general arguments summarised in the Introduction, it is expected
that the continuum limit of the 7v model on a dynamical lattice (in
the grand canonical ensemble) is described by the gravitational
sine-Gordon model, or equivalently, the sine-Liouville gravity
characterised by a Liouville interaction $\tilde \mu\, e^{2\phi}$ with
cosmological constant $\tilde \mu\sim \tilka _* - \tilka$ and
sine-Liouville perturbation generated by the gravitationally dressed
operators $\CO_{\pm 1}$ with couplings $ \mu_\pm$ such that
$\mu_+\mu_- \sim T-\Tstar $.  Again, the dynamics depends only on the
product of the two fugacities and we can assume $\mu_+=\mu_-= 2
\mu_{_{SL}}$.  The UV effective action of sine-Liouville gravity, in
our knowledge first studied by Moore \cite{Moore:1992ac}, is
 \begin{equation}\begin{aligned} \label{actionSG} \CA_{_\SL} [\phi,
 \vp ]&={1\over 4\pi} \int \limits_\CM d^2 z \sqrt{\hat g} \left[
 (\hat \nabla \phi)^2 + (\hat \nabla \vp)^2+ 2 \phi \hat R \right] +
 {\rm ghosts} \\
& +\tilde \mu \, \int d^2 z \sqrt{\hat g} \, e^{ 2 \phi } +
\mu_{_{SL}} \int d^2 z \sqrt{\hat g} \ e^{(2-{1\over R}) \phi}
\cos\left({\vp\over R}\right) .  \end{aligned} \end{equation}
A shift $\phi\to \phi + \log (r^2)$ is compensated by $\tilde
\mu\to\tilde \mu /r^2$ and $ \mu_{_{SL}} \to \mu_{_{SL}} /r^\a$ with
$\a= 2- 1/R$, so that the thermal coupling $t= 4\mu_{_{SL}}^2$ scales
as
  \begin{equation}\begin{aligned} \label{scalingSL} t\sim \tilde \mu^{
  2- 1/R} .  \end{aligned} \end{equation}

The exponents in the action \eqref{actionSG} are chosen so that the
exponential operators are strictly marginal.  Even if the theory is
conformal invariant, we can speak of correlation length set by the
cosmological constant $\tilde\mu$ in the sense that the typical size
of the world surface, that is the integrated Liouville term, is $\sim
1/\tilde \mu$.  As the area element is determined by the Liouville
field, the shift $\phi\to\phi + \log(r^2)$ above is the analogue of
the RG transformation in flat space.  The UV limit in this
interpretation corresponds to large $\tilde \mu$ or equivalently small
$ 2\mu_{_{SL}} = \sqrt{t}$.  In this limit the theory is that of a
Liouville gravity with a free compactified boson as matter field.

An important difference between the sine-Gordon action \eqref{actgbSG}
and the sine-Liouville action \eqref{actionSG} is that the
perturbation in the second case is triggered by the operators
$\CO_{\pm 1}$ with thrice smaller electric charge than in the SG
theory.  While on the flat lattice the phase factors of the loops are
$w^{\pm 6}$ (for the honeycomb) or $w^{\pm 4}$ (for the square
lattice), on the dynamical lattice they can be any integer power of
$w$ because of the curvature defects.  Therefore all spectrum of
electric charges \eqref{dimsem} is allowed and the lowest charges are
$n=\pm 1$ with dimension
  \be\begin{aligned} \label{dimsem1} 
  x_{\pm 1}=   {\textstyle{1\over 2}} R ^{-2}.
   \end{aligned} \ee 
 As a consequence, the BKT radius on dynamical lattice
 is twice smaller than on the flat lattice,
\be \label{RBKTD} R_{\mathrm{BKT}} =   {\textstyle{1\over 2}} \qquad \text{(dynamical
lattice)}.  \ee

  For   $\mu$ fixed, the  phase diagram of sine-Liouville gravity is characterised by three fixed points $t=0$ and $
t\to\pm\infty$ which is expected to describe the three critical phases of the
7v model on dynamical lattice. The flows between the  fixed points 
are in correspondence with the RG flows in the   sine-Gordon model in the plane.
   In particular, the massless flow in
the sine-Gordon with imaginary mass coupling has a counterpart in the
sine-Liouville gravity to which we refer to as \emph{gravitational
massless flow}.
 The results for the massless flow in the sine-Gordon model in the flat space
 suggest that the gravitational massless flow connects two theories of Liouville
gravity with compactified boson as matter field.  The  radius on the UV
side is by construction  $ \RSLc\equiv R$.
%
%
%
%
Sine-Liouville gravity is still an unsolved theory and it is not clear
how to  determine the IR  radius $R^{_{SL}}_{\ir}$ within the continuum theory.

\subsection{Matching the UV and IR parameters in sine-Liouville gravity}

The relation between $\RSLc$ and $R^{_{SL}}_{\ir}$ can be extracted
from the exact solution of the 7vMM, assuming the continuum limit is
described by sine-Liouville gravity.  The asymptotics of the free
energy at the two fixed points, eq.  \eqref{Suscpa2CFM1}, gives
\be\begin{aligned} \label{SLcSLd} R^{_{SL}}_{\ir}=1-\RSLc.
\end{aligned} \ee
This follows also from the scaling $t\sim \tilde \mu^{2\l\over 1+\l}$
we obtained in the matrix model near the tricritical point.  Comparing
with the scaling \eqref{scalingSL} we find
  \be
  \label{Ruv}
  2 - {1\over R} = {2\l\over 1+\l}\ \ \ \Rightarrow\ \ \ \RSLc=
  {1+\l\over 2} \ee
  while the limit $t\to -\infty$ is in the dense critical phase with
  \be\begin{aligned} \label{Rir}
  R^{_{SL}}_{\ir}= {1-\l\over 2}.
  \end{aligned} \ee 

Based on the solution of the discrete model, one can speculate that
the relation between the parameters of the sine-Gordon and the
sine-Liouville theories is a particular case of a universal map
relating the compactification radii before and after switching on
gravity.  Assume that such a map $f$ exists, we have $ \RSLc= f(
\RSGc)$ and $ R^{_{SL}}_{\ir}= f( \RSGd)$.  Comparing \eqref{SLcSLd}
and \eqref{RuvRir} we conclude that $f(x) ={1\over 2} x^2+ $ constant.
The constant should be zero because at the BKT point $ \RSGc=\RSGd=1 $
and $\RSLc=R^{_{SL}}_{\ir}=1/2$.  Thus we arrive at the relation
 \be\begin{aligned} \RSLc= {\textstyle{1\over 2}} ( \RSGc)^2 , \quad
 R^{_{SL}}_{\ir}= {\textstyle{1\over 2}} ( \RSGd)^2 .  \label{RSLRSG}
 \end{aligned} \ee
     Comparing \eqref{Ruv} with \eqref{R0e0A} and \eqref{RSLRSG},
  we can identify
  \be\begin{aligned}
  \label{relpla}
  p = {1\over \l}.  \end{aligned} \ee
 We see that although the lattice parameter $\l $ does not
 renormalise, the relation of the continuous parameter $p$ to the
 lattice parameter $w= e^{i\pi\l/2} $ depends on the lattice realisation.
 On the flat hexagonal lattice $1/p = 3\l$, on the flat square lattice
 $1/p = 2\l$, while on the dynamical lattice $1/p = \l$.

Finally, we can compare the equation of state from the discrete model
\eqref{MLpak0SUC} and the equation of state of sine-Liouville gravity
derived previously in the context of MQM perturbed by a condensate of
winding modes \cite{Kazakov:2000pm}.  The small $t$ expansion
reproduces the correlation functions of the sine-Liouville
perturbation conjectured by Moore \cite{Moore:1992ac}.  The match is
complete.  Interestingly, at the opposite end of the flow, where the
Gaussian field is perturbed by an irrelevant sine-Liouville operator,
the large $t$ expansion is automatically obtained from the small $t$
expansion using the symmetry of the equation of state under
	 \be\begin{aligned} \mu \to \mu_1= - {\mu\over t},\quad \l\to -\l,
	 \quad t\to - 1/t.  \label{muprim} \end{aligned} \ee
The symmetry of the equation is compatible with the phase structure of
the sine-Gordon model and with the expressions \eqref{Rir} and \eqref{Ruv}
of the UV and IR compactification radii being related by $\l\to -\l$.

\section{Discussion}

In this paper we derived the complete solution of a statistical model
of oriented self- and mutually avoiding loops on dynamical
triangulations, each loop weighted by a phase proportional to the
integrated curvature of the domain enclosed.  This loop model has a
local formulation in terms of the seven-vertex model on a dynamical
lattice, a one-parameter generalisation of the six-vertex model.

 We obtained the phase diagram and identified the critical phases.
 The phase diagram resembles qualitatively that of the $O(n)$ loop
 model, with three different phases characterised by massive, dense
 and dilute loops.  New physics emerges in the vicinity of the
 critical point separating the phases of massive and dense loops.  We
 claim that the universal behaviour in this regime is described by the
 sine-Liouville gravity which depends on two parameters, the
 cosmological constant $\tilde \mu\sim \k-\k_*$ and the   sine-Liouville
  coupling $ \mu_{_{SL}}\sim \sqrt{T-T_*}$.

 The UV effective action for sine-Liouville gravity,
 eq.\,\eqref{actionSG}, describes Liouville gravity with Gaussian
 matter field compactified at radius $R_\uv$ and perturbed by the
 sine-Liouville (SL) term.  The IR critical points appear in the limit
 of strong SL perturbation.  Knowing the fixed points of the matter
 field without gravity, SG theory, one expects that the limit of
 infinite large real SL coupling $\mu_{_{SL}}$ describes pure gravity
 while the limit of large imaginary coupling describes Liouville
 gravity with Gaussian matter field compactified at \emph{different
 radius} $R_\ir$.  Our solution of the discrete model confirms the
 claim made in \cite{Kostov:2025awi} about the exact value of $R_\ir$,
 eq.  \eqref{SLcSLd}.
   
The fact that the radius of the target space gets renormalised along
the massless flow for purely imaginary SL coupling may have important
consequences.  In particular, this fact was not taken into account in
the construction of a matrix model for the Euclidean 2D black hole in
\cite{Kazakov:2000pm}.  Therefore the claim of \cite{Kazakov:2000pm}
needs to be reconsidered.
    
The boundary observables that can be studied are of special interest
in our discrete model.  Concerning the physics in the bulk, the 7vMM
and the MQM perturbed by either momentum or vortex modes (for a review
see e.g. \cite{alexandrov2003matrixquantummechanicstwodimensional})
have the same continuum limit.  On the other hand, in presence of a
boundary the observables behave differently.

Indeed, since the scaling of the boundary length $\ell$ with the area
$A$ of the world sheet is the same as the scaling of $M$ with
$\tilde\mu$, eq.\,\eqref{MLpak0} implies that the length scales
anomalously as $\ell \sim A^{ 1-\l\over 2} $ for $t\to 0$.  Such
scaling cannot be realised in boundary Liouville gravity where
$\ell\sim \sqrt{A}$, that is, the boundary observables in the 7vMM are
\emph{not} described by FZZT branes.  In \cite{Kostov:2025awi}, it is
argued that the worldsheet action \eqref{actionSG} must have a
boundary term which is again of sine-Liouville type,
 \be\begin{aligned} \CA^B_{_\SL} [\phi, \vp ] =\mu_{^{SL}}^B \
 \int_{\mathrm{boundary}} e^{(1-R)\phi} \cos \left( R\tilde \vp\right)
 , \label{bdrytermSL} \end{aligned} \ee
where $\tilde\vp$ is the T-dual field of $\vp$.  The boundary
condition is Dirichlet for the height field $\vp$ and Neumann for the
T-dual field $\tilde\vp$ and $R=R_\uv$.  By charge neutrality the two
exponential fields $e^{\pm i R\tilde\varphi}$ appear only in pairs and
the effective boundary cosmological constant is $\mu_B \sim\left(
\mu_{^{SL}}^B\right)^2$ with the correct scaling $\mu_B \sim
\tilde\mu^{1-R}= \tilde\mu^{ 1-\l\over 2} $.  We are planning to
address the boundary theory in a future publication.

\acknowledgments

We thank Valentina Petkova and Galen Sotkov for their interest and
some important remarks.  The work of AAL  was financed in part by the Coordenação de Aperfeiçoamento de Pessoal de Nível Superior -- Brasil (CAPES) -- Finance Code 001. The work of MSJ  was financed in part  the Fundação de Amparo à Pesquisa e Inovação do Espírito Santo (FAPES) -- Finance Code 2025-4V3J4.

 \appendix

\section{The dual parametrisation}	\label{app_dualpar}

The spectral curve can be more naturally parameterised by theta
functions in terms of a dual uniformisation coordinate $u$ and a dual
modular parameter $\tilde \tau$ given by
\begin{equation} \label{svardef}
\tilde \tau = i / \tau ,
\qquad
u = s / \tau .
\end{equation}
Note that while $\tau$ is real, $\tilde \tau$ is purely imaginary.
The physical sheet \eqref{ParamRetangles} becomes the rectangle
\be \label{ParamRetangle} - {\textstyle{1\over 2}} \pi < \Re u <
{\textstyle{1\over 2}} \pi , \qquad -{1\over 2} \tilde\tau \pi < \Im u
< {\textstyle{1\over 2}} \tilde\tau \pi \ee
and the quasi-periodicity conditions \eqref{periodis}, translated to
the dual parametrisation, read
\begin{equation} \label{periodi}
\begin{aligned}
 x(u + \pi) &= x(u), &\qquad x(u+ \pi \tilde\tau ) &= q\, x(u) ,
 \\
y (u + \pi) &= y(u), &\qquad y (u + \pi\tilde\tau) &= \bar q\, y(u) .
\end{aligned}
\end{equation}
The solution to (\ref{periodi}) with the correct asymptotic conditions
for $Y(x)$ at $x = \infty$ are now given by
\begin{equation} \label{defxuAp}
x(u) = x_0 \frac{\tvt ( \frac12 \beta - u)}{\tvt(u)} ,
\qquad 
\begin{aligned}[t]
	y(u) &= y_0 \frac{\tvt ( \frac12 \beta + u)}{\tvt(u)} - i \tilde
	\tau y_1 \frac{d}{du} \left( \frac{\tvt ( \frac12 \beta +
	u)}{\tvt(u)} \right) \\
	&= - \frac{y_0}{x_0} x(-u) - \frac{i \tilde \tau}{x_0} y_1 x'(-u)
\end{aligned}
\end{equation}
where the    $\tvt(u)$ is  a Jacobi theta function:
\be \label{defvt}  
\tvt(u) \equiv \theta_1(u, e^{i\pi\tilde \tau}) 
\ee
  which satisfies the identities
\be\begin{aligned} 	\label{thetadiality} 
\tvt(u + \pi) &= - \tvt(u)\, ,
\\
\tvt(u + \pi\tilde\tau) &= - e^{\pi i \tilde\tau} e^{-2iu} \tvt(u)\, ,
\\
	\theta_1 ( u, e^{i\pi\tilde\tau} ) &= i \sqrt{
	\frac{i}{\tilde\tau}} \; \exp \left( - i u^2 / \pi \tilde\tau
	\right) \theta_1(u / \tilde\tau , e^{- i \pi / \tilde\tau}) .
	\end{aligned} \ee 
Expressing the r.h.s. of the third identity in terms of $s$ and
$\tau$, we obtain the theta function defined in \eqref{Defofvt},
\begin{equation}	\label{DefofvtAp}
\vt(s) = i \sqrt{\tau} \ \exp \left( - s^2 / \pi \tau \right)
\theta_1(- is, e^{-\pi \tau}) .
\end{equation}
In other words, we can translate relate the $u$ and $s$
parametrisations of the spectral curve by using the identities
\begin{equation}	\label{TranslThets} 
\tvt(u) = \vt(s) , 
\qquad
\partial_u^n \tvt(u) = \tau^n \partial_s^n \vt(s) ,
\qquad
\partial_s^n \vt(s) 
=
(- i \tilde \tau)^n \partial_u^n \tvt(u).
\end{equation}

The $u$-parametrisation of the spectral curve coordinate is more
natural in the context of enumeration of graphs and directly related
to the solution given in \cite{ELVEYPRICE2023105739} for the
six-vertex model.  The $s$-parametrisation used in the main text is
more appropriate for extracting the scaling limit where the real nome
$e^{-\pi \tau}$ becomes small.

\section{Fixing the parameters of the spectral curve} \label{app_solution}

We can fix the parameters $x_0$, $y_0$, $y_1$ of the solution
\eqref{solutionS} by imposing the asymptotic condition \eqref{asymJ}.
First, we expand $x(s)$ and $y(s)$ near the pole $s = 0$ where $x \to
\infty$, to find
\begin{align}
x(s) &= x_0 \left( \frac{\cT_1}{s} + \cT_2 + \cT_3 s \right) + O(s^2);
	\label{xsExps0}
\\
	y(s) &= \left( \frac{y_0}{s} - \frac{y_1}{ s^2} \right) \cT _1 -
	y_0 \cT _2 + y_1 \cT _3 + O(s), \label{ysExps0}
\end{align}
with $\cT_1$, $\cT _2$, $\cT _3$ given by \eqref{defCCC}.  Now we
substitute the expansion \eqref{xsExps0} into the asymptotic condition
\eqref{asymJ} and find
\begin{equation}	\label{YxsExps0}
	Y(x(s)) = - \frac{\cT _1 ^2 t_3 x_0^2 }{s^2} - \frac{2 \cT _1 \cT
	_2 t_3 x_0^2+ \cT _1 t_2 x_0 }{s} - \Big[ t_1 +x_0\cT _2 t_2 + t_3
	x_0^2 \left(2 \cT _1 \cT _3+ \cT _2 ^2\right) \Big] + O(s) .
\end{equation}
Matching the three first coefficients of \eqref{YxsExps0} with those
of \eqref{ysExps0} gives three algebraic equations for $y_0$, $y_1$
and $x_0$, whose solutions are
\begin{equation}	\label{x0y0t0ExactC}
\begin{aligned}
y_1 &= \cT _1 x_0^2 t_3 ,
\\
y_0 &= - t_2 x_0 - 2 \cT _2 t_3 x_0^2 ,
\\
x_0 &= - \frac{t_1/\cT _2}{ t_2+\sqrt{ t_2^2 - 3 (1 + \cT _1\cT _3 /
\cT _2^2) \; t_1 t_3 }} .
\end{aligned}
\end{equation}
There are, in fact, two solutions for $x_0$, corresponding to the two
branches of the square root.  We have chosen the branch such that
$T=0$ belongs to the main sheet.  The parameters depend on $T$ through
the couplings $t_1,t_2,t_3$ given in \eqref{t1t2t3}.  The expression
inside the square root in the formula for $x_0$ can be rewritten as
follows:
\begin{equation*}
\begin{aligned}
&t_2^2 - 3 (1 + \cT _1\cT _3 / \cT _2^2 ) t_1 t_3 
\\
	&\qquad = \frac14 \csc^2(\pi \lambda) \Big[ 1 + T(T-1) \Big( 4 + 3
	(1 + \cT _1 \cT _3 / \cT _2^2) \sec( \tfrac{\pi \l}{2} ) \sec(
	\tfrac{3\pi \l}{2} ) \sin^2(\pi \l) \Big) \Big] \\
&\qquad
	= 
	\frac14 \csc^2(\pi \lambda) \
	A \  (T^2 - T + 1/A)
\end{aligned}
\end{equation*}
where $A=A(\l,\t)$ denotes the expression in round brackets in the
second line.  The remaining quadratic polynomial factorizes as $(T^2 -
T + 1/A) = (T - T_-) (T - T_+)$, where the roots $T_\pm = \frac12 ( 1
\pm \sqrt{1 - 4/A})$ can be simplified to the form in
\eqref{Tpmlatau}.

\bigskip

An analogous computation determines the coefficients of the solution
at the critical point, eq.  \eqref{SolCritCurv}, whose expansion near
$s = 0$ is
\begin{subequations}
\begin{align}
	x_c(s) &= \frac{\xoc     }{2s} - \frac{ \lambda \xoc     }{2} +
	\frac{\xoc     ( 3 \lambda^2 - 1 )}{12} \; s + O(s^2),
\label{xsExps0Crit}
\\
	y_c(s) &= - \frac{ \yonec   }{2 s^2} + \frac{\yoc    }{2 s} + \frac{
	6 \lambda \, \yoc    + (3 \lambda^2 - 1) y_{1c } }{12} + O(s).
\label{ysExps0Crit}
\end{align}
\end{subequations}
Substituting \eqref{xsExps0Crit} into \eqref{asymJ} gives
\begin{equation}	\label{YxsExps0Crit}
	Y(x_c(s)) = - \frac{t_3 (\xoc    )^2 }{4 s^2} - \frac{ ( t_2 - t_3
	\xoc     ) \xoc    }{2s} - \Big[ t_1 - \tfrac12 \lambda \, t_2 \xoc    
	+ \tfrac1{12} ( 6 \lambda^2 - 1) \, t_3 \, (\xoc   ) ^2 \Big] + O(s) .
\end{equation}
Matching the coefficients of \eqref{YxsExps0Crit} and \eqref{ysExps0Crit}, we find%
\begin{equation}	\label{xcycyc1} 
\begin{aligned}
\yonec   = \tfrac12 \, t_3 (\xoc   ) ^2 , \quad \yoc    = - t_2 \xoc     +
\lambda \, t_3 \, (\xoc    )^2, \quad \xoc     = \frac{2 t_1 / \l }{ t_2 +
\sqrt{ t_2^2 + \frac{1 - 9 \lambda^2}{2 \lambda^2} \; t_1 \, t_3} } .
\end{aligned}
\end{equation}
Writing $t_1$, $t_2$, $t_3$ explicitly with \eqref{t1t2t3}, and
factorizing the polynomial inside the square root,
\begin{equation*}
\begin{aligned}
t_2^2 + \frac{1 - 9 \lambda^2}{2 \lambda^2} \; t_1 \, t_3
	&=
	\frac14 \csc^2 (\pi\l) \, \frac{(T - T^+_c)(T - T^-_c)}{T^+_c T^-_c}
\end{aligned}
\end{equation*}

\section{The limit $\tau \to \infty$ for the spectral curve}
\label{AppTauInftyLimit}

Here we describe the limit $\tau \to \infty$ for the exact solutions
of the spectral curve.  More precisely, we want to expand the theta
function defined in \eqref{Defofvt} in powers of $e^{- \pi \tau} \ll
1$.  We will need the expansions of $\vt(s)$ and its first three
derivatives:
\begin{subequations} \label{seriesfortvt}
\begin{align}
\vt (s) &
\begin{aligned}[t]
&\equiv i \sqrt{\tau} \ \exp( - s^2 / \pi \tau ) \theta_1(- i s , e^{-
\pi \tau }) .  \\
				&= 2 \sqrt{\tau} \; e^{- \frac{s^2}{\pi \tau}} \,
				e^{\frac14 \pi \tau} \Big[ \sinh s - e^{-2\pi \tau}
				\sinh (3s) + O(e^{-6 \pi \tau}) \Big]
\end{aligned}
\\
	\vt' (s) &= 2 \sqrt{\tau} \, e^{- \frac{s^2}{\pi \tau}} \,
	e^{\frac14 \pi \tau} \Bigg[ \Big( \cosh s - \frac{s \sinh s}{\pi
	\tau} \Big) + O(e^{-2\pi \tau} ) \Bigg]
\\
	\vt'' (s) &= 2 \sqrt{\tau} \, e^{- \frac{s^2}{\pi \tau}} \,
	e^{\frac14 \pi \tau} \Bigg[ \Big( \sinh s - \frac{2 \sinh s + 4 s
	\cosh s}{\pi \tau} + \frac{4 s^2 \sinh s}{\pi^2 \tau^2} \Big) +
	O(e^{-2\pi \tau} ) \Bigg] \\
	\vt''' (s) &= 2 \sqrt{\tau} \, e^{- \frac{s^2}{\pi \tau}} \,
	e^{\frac14 \pi \tau} \Bigg[ \Big( \cosh s + O(1/\tau) \Big) +
	O(e^{-2\pi \tau} )
	\Bigg].
\end{align}
\end{subequations}

The ratios of theta functions in eqs.\eqref{solutionS} require further
expansions because of the arguments at $\tfrac12 \beta \tau = \frac12
\pi(1-\l) \tau$,
\begin{subequations}
\begin{align}
\frac{ \vt (\frac{\beta \tau}{2} - s)}{\vt(s)} & \cong \frac{e^{-\l
s}}{2 \sinh s} \, \exp \big[ \tfrac14 \pi \tau (1 - \l^2) \big] ,
\label{ratiothe1}
\\
	\frac{ \vt' (\frac{\beta \tau}{2} + s)}{\vt(s)} & \cong
	\frac{e^{\l s}}{2 \l \sinh s} \, \exp \big[ \tfrac14 \pi \tau (1 -
	\l^2) \big] ,
\label{ratiothe2}
\\
\frac{ \vt'(s) \vt (\frac{\beta \tau}{2} + s)}{\vt^2(s)} & \cong
\frac{e^{\l s} \cosh s}{2 \sinh^2 s} \, \exp \big[ \tfrac14 \pi \tau
(1 - \l^2) \big] .
\label{ratiothe3}
\end{align}
\end{subequations}
Here and throughout this appendix, $\cong$ will mean ``equal at
leading order for $\tau \to \infty$''.  Since $\l < 1$, the
exponential above diverges, but the divergence is canceled by the
coefficients $x_0$, $y_0$, $y_1$.  To compute them, we need the
parameters $\cT _i$ in \eqref{defCCC} which are, to leading order,
\begin{subequations}		\label{CCClimit}
\begin{align}
\cT _1
&	\cong \frac12 \exp \big[ \tfrac14 \pi \tau (1 - \l^2) \big] ,
\label{CCClimit1}
\\
\cT _2
&	\cong - \frac{\l}{2} \exp \big[ \tfrac14 \pi  \tau  (1 - \l^2) \big] ,
\label{CCClimit2}
\\
\cT _3 & \cong - \frac{1 - 3\l^2}{12} \exp \big[ \tfrac14 \pi \tau (1
- \l^2) \big] .
\label{CCClimit3}
\end{align}
\end{subequations}
All of them are diverging.  The combination $\cT _1\cT _3 / \cT _2^2 =
(3\l^2 - 1) / 6 \l^2$ is finite, and the normalisation parameter $x_0$
in \eqref{x0y0t0ExactC} becomes
\begin{equation}	\label{x0limittauinf}
\begin{aligned}
		x_0 & \cong \frac{2 t_1/ \l }{ t_2 + \sqrt{ t_2^2 + \frac{1-
		9\l^2}{2\l^2} \; t_1 t_3 }} \exp \big[ - \tfrac14 \pi \tau (1
		- \l^2) \big] \\
&	= \xoc     \exp \big[ - \tfrac14 \pi  \tau  (1 - \l^2) \big] ,
\end{aligned}
\end{equation}
since the finite fraction in the first line is what we called $\xoc    $
in \eqref{xcycyc1}.  The vanishing exponential factor cancels the
diverging one in \eqref{ratiothe1}, so that we have a finite limit:
\begin{equation}
x(s) \cong 
	 \frac{\xoc     \, e^{-\l s}}{2 \sinh s} ,
\end{equation}
which is the hyperbolic solution on the critical line,
\eqref{SolCritCurvA}.  Similarly, it is immediate to check that the
factors of $e^{- \frac14 \pi \tau (1 - \l^2)}$ in
\eqref{ratiothe2}-\eqref{ratiothe3} cancel the ones in
\eqref{CCClimit2}-\eqref{CCClimit3}, so that the exact $y(s)$ becomes
the solution in \eqref{SolCritCurvB}.  In particular, $y_{1c}$ and
$y_{2c}$ given in \eqref{xcycyc1} are obtained from $y_1$ and $y_2$ in
\eqref{x0y0t0ExactC} by substituting the $\cT _i$ in \eqref{CCClimit}
and ignoring the exponential factors.

In  the scaling limit, we need to consider 
$x(\frac12 \beta \tau + s)$ for $\tau \to \infty$. 
Going back to \eqref{seriesfortvt}, we find
\begin{equation}
\begin{aligned}	\label{ratiothe1B}
	\frac{ \vt (- s)}{\vt(s + \frac{\beta \tau}{2})} \cong - 2 e^{- \l
	s} \sinh(s) \, \exp \big[ - \tfrac14 \pi \tau (1 - \l^2) \big] .
\end{aligned}
\end{equation}
Combining this with \eqref{x0limittauinf}, we conclude that
\begin{equation}	\label{xlimittauinfScaling}
	x(\tfrac12 \beta \tau + s) \cong - \xoc     \, e^{- \l s} \sinh(s)
	\exp \big[ - \tfrac12 \pi \tau (1 - \l^2) \big] .
\end{equation}
The exponential factors here do not cancel, the RHS of
eq.\,\eqref{xlimittauinfScaling} is the leading order contribution for
the vanishing of the function in the LHS, as $\tau \to \infty$.

To compute the asymptotics of $y(\frac12 \beta \tau + s)$ using
eq.\,\eqref{solutionSY} we need
\begin{equation}
\begin{aligned}	\label{ratiothe1C}
y_0 \; \frac{ \vt (s+\beta \tau )}{\vt(s + \frac{\beta \tau}{2})} &
\cong \yoc    \; e^{- (1-\l)s} \exp \big[ - \tfrac12 \pi \tau (1 -
\l)^2 \big] ,
\\
y_1 \; \frac{ \vt' (s+\beta \tau )}{\vt(s + \frac{\beta \tau}{2})} &
\cong - \yonec   (1 - 2\l) e^{- (1-\l)s} \exp \big[ - \tfrac12 \pi \tau
(1 - \l)^2 \big] ,
\\
- y_1 \; \frac{ \vt'(s + \frac{\tau\b}{2}) \vt(s+\tau\b ) }{ \vt(s +
\frac{\tau\b}{2})^2 } & \cong - \yonec   \l \; e^{- (1-\l)s} \exp \big[
- \tfrac12 \pi \tau (1 - \l)^2 \big] ,
\end{aligned}
\end{equation}
where we have computed the expansion of theta functions at leading order.
Therefore
\begin{equation}	\label{ylimittauinfScaling}
y(\tfrac12 \beta \tau + s) \cong
	\big[
	\yoc    - (1 - \l) \yonec   
	\big]
	e^{- (1-\l)s}
	\exp \big[ - \tfrac12 \pi \tau (1 - \l)^2 \big] .
\end{equation}
This holds, again, at leading order.  In the main text, we have
computed the result at next-to-leading order by using the symmetries
of the exact solution.

\section{The boundary entropy as a function of the chemical potential}
\label{appsolvtreqs}

Let us remind how a transcendental equation of the type
\begin{equation}	\label{Transeq}
p + r \, p^{1-\a} = q ,
\end{equation}
is solved by Lagrange inversion.  
 Consider the inverse
Laplace transformation of $p^\s$,
\begin{equation}
{\cal L}^{-1}_a[p^\s] = \int_\uparrow \frac{dq}{2\pi i} \ p^\s e^{a q}
= \frac1{2\pi i a} \int_\uparrow d[ e^{a q} ] \ p^\s \end{equation}
where the uparrow denotes a vertical contour as in
(\ref{tildeWandWrho}).  With an integration by parts, we arrive at
\begin{equation}
\begin{aligned}
	{\cal L}^{-1}_a[p^\s(q)] = - \frac{1}{2\pi i a} \int_\uparrow
	d[p^\s] e^{a q} = - \frac{\s}{2\pi i a} \int_\uparrow \!  dp \
	p^{\s-1} \ \exp [ a(p + r\, p^{1-\a}) ] .
\end{aligned}
\end{equation}
We have used eq.\,\eqref{Transeq} in the last equality, at which point
$p = p(q)$ inside the inverse Laplace transform becomes a solution of
the transcendental equation.  The idea is, then, to extract $p(q)$
from the integral above by performing a Laplace transform.  In order
to do so, we expand the exponential in a power series while defining
$ap = v$
\begin{equation}
\begin{aligned}
	{\cal L}^{-1}_a[p^\s] &= - \frac{\s}{a^{\s+1}} \sum_{n=0}^\infty
	\int_\uparrow \frac{dv}{2\pi i} \frac{(r\, a^\a)^n}{n!} \
	v^{\s-1+n(1-\a)} \ e^v \\
	&= - \frac{\s}{a^{\s+1}} \sum_{n=0}^\infty \frac{(r\, a^\a)^n}{n!
	\, \Gamma [(\a-1)n +1-\s]},
\end{aligned}
\end{equation}
where we used $\int_{\uparrow} e^{v} v^{-\a} = 2\pi i / \Gamma (\a)$.
Now we can Laplace-transform back all terms,
\begin{equation}
	p^\s(q) = \int_0^\infty \!  da \, e^{-a q} \, {\cal
	L}^{-1}_a[p^\s] = -\s \sum_{n=0}^\infty \frac{r^n}{n!  \Gamma
	[(n-1)n +1-\s]} \int_0^\infty \!  da \ e^{- a q} a^{-1-\s +n \a}
\end{equation}
and, since
\begin{equation}
\int_0^\infty \!  da \ a^{\a} e^{-a x} = \frac{\Gamma(\a+1)}{x^{a+1}}
,
\end{equation}
we finally arrive at the power series
\begin{equation}
	p^\s(q) = -\s q^\s \sum_{n=0}^\infty \frac{\Gamma (\a n -\s)}{n!
	\Gamma [(\a-1)n +1 - \s]} \left( \frac{r}{q^\a} \right)^n .
\end{equation}

Now we apply this formula to solve the equation for 
the boundary entropy \eqref{MLpak0}, we
write it as
\begin{equation}
M^\nu - r \;  M^{\nu (1 - \a)}
	=
	\frac{2 \mgc}{1 + \l},
\qquad
\a = \frac{2\l}{1+\l},
\quad
r = - \frac{1 - \l}{1 + \l} \; t .
\end{equation}
This has the form \eqref{Transeq}, with $p = M^\nu$ and $q = 2\mgc / (1+\l)$,
hence the solution is
\begin{equation}
		M = p^{1/\nu} = - \frac{1-\l}{2} \left( \frac{2\mgc}{1+\l}
		\right)^{\frac{1-\l}{2}} \sum_{n=0}^\infty \frac{ \Gamma
		\big(\frac{2\l}{1+\l} n - \frac{1-\l}{2} \big) }{ n!  \Gamma
		\big( \frac{1+\l}{2} - \frac{1 - \l}{1+\l} n \big)} \left[ -
		\left( \frac{2\mgc}{1 + \l} \right)^{-\frac{2\l}{1+\l}}
		\frac{1 - \l}{1 + \l} \; t \right]^n .
\end{equation}
This is the result given in eqs.\eqref{Mofhatt}-\eqref{Mofhatt2}.

\section{Properties of the generalised Macdonald function}
\label{AppKratzelFunctions}

In this apendix we list some of the properties of the generalised
Macdonald function $K_\a^{(b)}(z)$ defined in \eqref{defII},
\begin{equation}	\label{defIIAp} 
	K^{(b)}_\a (2z) = \frac12 \int_ 0^\infty \!  d\omega \; e^{-
	z(\omega^{1/b} + \omega^{-b}) } \omega^{\a-1}, \qquad \Re z > 0 .
\end{equation}
It is a special case of the \emph{generalised Krätzel functions}
\cite{kratzel1975integral},
\begin{equation}	\label{eq: GKF}
		K(\nu, \delta, \rho; x,y) = \int_0^\infty\!  d\omega \
		\omega^{\nu - 1} \exp \big[ -x \omega^{\delta} -
		y\omega^{-\rho} \big] , \quad \Re x, \ \Re y > 0, \quad
		\nu,\delta, \rho >0 ,
\end{equation}
see \cite{math8040526}, namely $K_{\a}^{(b)}(2z) = \tfrac{1}{2}
K(\a,1/b,b;z,z)$.  Following \cite{math8040526}, if we take a Mellin
transform with respect to $y$ of the integrand in \eqref{eq: GKF},
then integrate over $\omega$ and finally compute the inverse Mellin
transform, we obtain a new representation of the Krätzel function,
\begin{equation}	\label{eq: kratGamma}
	K(\a,\delta,\rho;x,y) = \frac{1}{x^{\a/\delta}} \int_{\uparrow}
	\frac{ds}{2\pi i} \, \Gamma(s) \Gamma \left(\frac{\a}{\delta} +
	\frac{\rho}{\delta}s\right) y^{-\rho s/\delta }x^{-s}.
\end{equation}
The integration is along a vertical contour that passes to the right
of all poles of the integrand.  From this, our $K$-function reduces to
\begin{equation}	\label{eq: KintGamma}
	K_{\a}^{(b)}(2z) = \frac{1}{2z^{b\a}} \int_\uparrow \frac{ds}{2\pi
	i} \, \Gamma(s) \Gamma(b \a + b^2s ) z^{-(b^2+1)s}
\end{equation}
and a series expansion can be found by computing the residues of
\eqref{eq: KintGamma}.  Closing the contour at infinity, the poles are
at those of the Gamma functions, at $s = -n$ and $b \a + b^2s = -n$,
with $n \in \mathbb N$,
\begin{equation}	\label{KintGamma2}
	K_{\a}^{(b)}(2z) = \frac12 \sum_{n=0}^{\infty} \frac{(-1)^n}{n!}
	\left[ \Gamma(b \a - b^2n) z^{(b^2+1)n - b \a} + \Gamma \Big(-
	\frac{\a}{b} - \frac{n}{b^2} \Big) z^{(1 + b^{-2}) n + \a / b}
	\right] .
\end{equation}
This can be written more suggestively as
\begin{equation}	\label{KbaIba}
K^{(b)}_\a (z) 
	=  
	I^{(b)}_{-\a} (z) + I^{(1/b)}_{\a}(z) , 
\end{equation}
where
\begin{equation}	\label{Ialpbseri}
\begin{aligned}
	I^{(b)}_\a (z) & \equiv \frac12 \sum_{n=0}^{\infty} \frac{(-1)^n
	\Gamma(- b \a - b^2n) }{n!} \left( \frac{z}{2} \right)^{(b^2+1)n +
	b \a} \\
	& = \frac12 \sum_{n=0}^{\infty} \frac{(-1)^{n+1} \pi }{ n!  \sin[
	\pi b( \a + bn)] \Gamma(1 + b \a + b^2n) } \left( \frac{z}{2}
	\right)^{(b^2+1)n + b \a} .
\end{aligned}
\end{equation}
For $b = 1$, $I^{(1)}_\a (z) = \frac\pi2 I_\a(z)$ reduces to the
modified Bessel function of the first kind, and eq.\,\eqref{KbaIba} is
therefore a generalised version of the usual relation between the
modified Bessel functions of the first and second kinds.
Eq.\,\eqref{KbaIba} makes the symmetry relation \eqref{symK} evident; it
can also be easily seen directly from the defining integral
\eqref{defIIAp}.

The $K$-functions satisfy  Bessel-like recurrence 
relations.  It  follows  from the integral representation
\eqref{defIIAp} that the pair of commuting linear differential
operators
\begin{equation}
 \mathbf{a}  _\a^- = - \frac{1}{b} \; z \p_z - \a, \qquad
  \mathbf{a}  ^+_\a = -b \; z\p_z + \a
\end{equation}
act on the K-functions as
\be\begin{aligned} 	\label{genrecr}
 \mathbf{a}  ^+_\a \; K^{(b)}_{\a} (2z) &= \frac12 \Big( b +
\frac{1}{b} \Big) z \, K^{(b)}_{\a+1/b}(2z) , \\
 \mathbf{a}  ^-_\a \; K^{(b)}_{\a} (2z) &=
\frac12 \Big( b + \frac{1}{b} \Big) z \, K^{(b)}_{\a-b } (2z) ,
\end{aligned} \ee 
which give a pair of recursion relations expressing
$\p_z K^{(b)}_\a$ as a sum of $K$-functions with different indices.
Explicitly, computing $(  \mathbf{a}   ^+_\a +  \mathbf{a}  ^-_\a)K^{(b)}_\a$ and $( b^{-1} {\mathrm{a} }^+_\a - b  \mathbf{a}   
^-_\a)K^{(b)}_\a$ gives

\begin{equation}	\label{RecrusxEqKK} 
\begin{aligned}
2 \p_z K^{(b)}_{\a} (2z) &= - K^{(b)}_{\a+1/b} (2z) - K^{(b)}_{\a-b}
(2z) ,
\\
2 \a K^{(b)}_{\a} (2z) &= \frac{1}{b} \; z K^{(b)}_{\a+1/b} (2z) - b
\, z\, K^{(b)}_{\a-b} (2z) .
\end{aligned}
\end{equation}
The product $  \mathbf{a}  ^+_\a  \mathbf{a}  ^-_\a \;
K^{(b)}_{\a}$ gives a second-order differential equation relating
functions with different indices:
\begin{equation}	\label{genBesseq}
(z \p_z + \a \, b) (z \p_z - \a/b) \; K^{(b)}_{\a} (2z)= \Big( b +
\frac{1}{b} \Big)^2 z^2 \, K^{(b)}_{\a - b+1/b}(2z) .
\end{equation}
For $b = 1$, the index of the function in the r.h.s.~is not shifted,
and the equation above reduces to the Bessel equation.

For large $z$, the integral \eqref{defIIAp} can be evaluated by the
saddle point method.  The saddle point is at $\omega = b^{\frac{2}{b +
1/b}}$,
\begin{equation}	\label{semiclas} 
	K^{(b)}_{\a} (2z) \sim \frac{1}{\sqrt{z}} \exp \big(-z \, Q b^{
	e_0/Q} \big) , \qquad Q = b + 1/b , \qquad e_0 = - b + 1/b .
\end{equation}
Translating $\l$ in terms of $b$, the combination that appears in the
exponent is, in fact, the position of the branch cut
\eqref{solforsbb}, i.e.~$Q b^{ e_0/Q} = {\mathrm{b} }  / M$.

\section{Series expansion for $Y(x)$ in the scaling limit}
\label{AppSeriesExps}

Here we derive a series expansion for the function $Y(x)$ at $x\to
+\infty$, in the scaling limit where the parametric representation is
\eqref{solscalla}.  We start by computing the inverse Laplace
transform
\begin{equation}	\label{LaplaceTrnsY}
\tilde Y_\sc(\ell) 
	= \frac{1}{2\pi i} \int_\uparrow dx \ e^{\ell x} \; Y_\sc(x) 
\end{equation}
by  deforming the original  integration contour  parallel to the imaginary axis 
into a pair of contours, $\CC_1$ and $\CC_2$,
surrounding the  two cuts of the integrand.  
 After integration by parts 
\be\begin{aligned} 
	\tilde Y_\sc(\ell) &= \frac{1}{2\pi i \ell} \int_\uparrow d( e^{\ell
	x}) \; Y_\sc(x) 
	\\
	&= \frac{1}{2\pi i \ell} \left( \int_{\CC_1} dY_\sc(x) \, e^{\ell
	x} + \int_{\CC_2}dY(x) e^{\ell x} \right) 
\\
 & = \frac{1}{\pi \ell} \Im \int_ {-\infty}^\infty dy(s + \tfrac12 i
 \pi) \, e^{\ell x(s + \frac12 i \pi) } \end{aligned} \ee

With the change of variables \eqref{ChangeOfvarome}, and using
eqs.\eqref{scalyxforMAp}, the integral \eqref{LaplaceTrnsY} becomes
\begin{equation}	\label{tiYellscZe}
\begin{aligned}
	\tilde Y_\sc(\ell) & = \frac{1}{\pi \ell} \Re \left[ i
	q^{-\frac12} \int_{0}^\infty d \Big( M^{1/b^2} \omega^{-1/b} + t
	M^{b^2} \omega^{b} \Big) \, \exp \left( - q^{\frac12} \ell M (
	\omega^{1/b} + \omega^{-b}) \right) \right].
\end{aligned}
\end{equation}
Hence, with the symmetry \eqref{symK}, 
\begin{equation}	\label{tiYellscZe2}
\begin{aligned}
	\tilde Y_\sc(\ell) = \frac{1}{\pi \ell} \Re \Bigg[ i q^{-\frac12}
	\left( t b M^{b^2} K_{b}^{(b)}(2 q^{\frac12} \ell M) -
	\frac{M^{1/b^2}}{b} K_{1/b}^{(1/b)}(2 q^{\frac12} \ell M) \right)
	\Bigg] .
\end{aligned}
\end{equation}

Using the series expansion \eqref{KintGamma2}, the first term in the
r.h.s.~of \eqref{tiYellscZe2} can be written as
\begin{equation}	\label{termitYell}
\begin{aligned}
	\Re \Big[ i q^{- \frac12} K_{b}^{(b)}(2 q^{\frac12} \ell M) \Big]
	&= - \sum_{n=0}^{\infty} \frac{(-1)^n}{n!} \Bigg[ \sin [ \pi (n-1)
	b^2] \Gamma \big[b^2(1 - n) \big] (\ell M)^{(b^2+1)n - b^2} \\
	&\qquad\qquad\qquad\qquad + \sin(n \pi) \Gamma \Big[- 1 -
	\frac{n}{b^2} \Big] (\ell M)^{(1 + b^{-2}) n + 1} \Bigg]
\end{aligned}
\end{equation}
We see that a whole series disappears because $\sin(n \pi) = 0$ and we
end up with
\begin{equation}
\begin{aligned}
	 \Re \Big[ i q^{- \frac12} K_{b}^{(b)}(2 q^{\frac12} \ell M) \Big]
	 &= - \sum_{n=0}^{\infty} \frac{(-1)^n}{n!  \ \Gamma \big[1 +
	 b^2(n-1) \big]} (\ell M)^{(b^2+1)n - b^2}.
\end{aligned}
\end{equation}
Therefore
\begin{equation}	\label{yofellAp}
	\tilde Y_\sc(\ell) = \sum_{n=0}^{\infty} \frac{(-1)^n}{n!} \Bigg[
	\frac{ b^{-1} \ (\ell M)^{(1+1/b^2)n} }{ \ell^{1+1/b^2} \Gamma
	\big[1 + (n-1)/b^2 \big]} - \frac{ t b \ (\ell M)^{(1+b^2)n} }{
	\ell^{1+b^2} \Gamma \big[1 + (n-1)b^2 \big]} \Bigg] .
\end{equation}

Once we know the expansion of the inverse Laplace transform, we can
obtain $Y_\sc(x)$ by performing the Laplace transformation term by term in
the expansion \eqref{yofellAp}.
Each of the integrals is of the form
\begin{equation}	\label{intGamma} 
\int_0^\infty \!  d\ell \ e^{-x \ell} \ \ell ^{s} = \frac{\Gamma (s+1)
}{x^{s+1}},
\end{equation}
resulting in a power series in $(M/x)^{1+b^2} = (M/x)^{ bQ} $,
\begin{subequations} \label{yofxseriesApp}
\begin{equation}
\begin{aligned}
	Y_\sc(x) = \sum_{n=0}^{\infty} \frac{(-1)^n}{n!} &\Bigg[
	\frac{1}{b} \frac{ \Gamma \big[n + (n-1)/b^2 \big] }{ \Gamma
	\big[1 + (n-1)/b^2 \big] } \left( \frac{M}{x} \right)^{(1+1/b^2)n}
	x^{1/b^2} \\
	&\quad - tb \frac{ \Gamma \big[n + (n-1)b^2 \big] }{ \Gamma \big[1
	+ (n-1)b^2 \big] } \left( \frac{M}{x} \right)^{(1+b^2)n} x^{b^2}
	\Bigg] .
\end{aligned}
\end{equation}
Note that the coefficients of both series are analytic functions of
$b$; in fact, they are Pochhammer polynomials in $b$ and $1/b$,
e.g.~$(a)_n = \Gamma(a+n) / \Gamma(a)$,
\begin{equation}	\label{yofxseries2}
\begin{aligned}
Y_\sc(x) = \sum_{n=0}^{\infty} \frac{(-1)^n}{n!} &\textstyle \Bigg[
\frac{ 1}{b} \left( {(n-1)\over b^2 }\right) _n\, x^{\frac1{b^2}}
\left( \frac{M}{x} \right)^{(1+1/b^2)n} - t\,b \big( (n-1)b^2 \big)_n
x^{b^2}\,\left( \frac{M}{x} \right)^{(1+b^2)n} \Bigg] .
\end{aligned}
\end{equation}
\end{subequations}

Since $\ell > 0$, the Laplace transform \eqref{intGamma} is only
defined for $\Re x > 0$, which must be therefore be assumed in the
asymptotic expansions \eqref{yofxseriesApp}.  To find the domain of
convergence of the series, consider the coefficients of the series
with $\sim x^{-(1+b^2)n}$; as $n \to \infty$,
\begin{equation}	\label{Coefcnninf}
\begin{aligned}\textstyle
	c_n = \frac1{n!} \frac{ \Gamma \big[n + (n-1)b^2 \big] }{ \Gamma
	\big[1 + (n-1)b^2 \big] } \approx \frac{1}{n!} [ (1 + b^2) n]^n
	\left( \frac{b + 1/b}{b} \right)^{b^2 n - \frac12} e^{-n}
\end{aligned}
\end{equation}
hence
$c_n / c_{n+1}
	\approx
	(1 + b^2)^{-1-b^2} b^{2b^2}
$,
and the series converges as long as
\begin{equation} \label{Coefcnninf1}
\begin{aligned}
|M / x |^{1+b^2} < (1 + b^2)^{-(1+b^2)} b^{2b^2} , \quad \text{hence}
\quad |x| > M (b + 1/b) b^{ \frac{1 - b^2}{1+b^2}} .
\end{aligned}
\end{equation}
The result is obviously symmetric under $b \to 1/b$, hence the radius
of convergence of the other series in \eqref{yofxseriesApp} is the
same as the above.  If we translate eq.\,\eqref{solforsbb} from $\l$
to $b$, the position of the branch point is precisely $ {\mathrm{b} }
= M (b + 1/b) b^{ \frac{1 - b^2}{1+b^2}} $.  Therefore we conclude
that the series \eqref{yofxseriesApp} for $Y(x)$ converges for $| x |
> {\mathrm{b} } $, as stated in \eqref{Converglarb}.

%
%

\begin{thebibliography}{10}

\bibitem{Kostov:2025awi}
I.~Kostov, {\it {Sine-Liouville gravity as a vertex model on planar graphs}},
  {\em JHEP} {\bf 06} (2026) 080,
  [\href{http://xxx.lanl.gov/abs/2512.1891}{{\tt arXiv:2512.1891}}].

\bibitem{1991hep.th....8019K}
I.~R. {Klebanov}, {\it {String Theory in Two Dimensions}},  {\em ArXiv High
  Energy Physics - Theory e-prints} (Aug., 1991)
  [\href{http://xxx.lanl.gov/abs/hep-th/91}{{\tt hep-th/91}}].

\bibitem{DiFrancesco:1993nw}
P.~Di~Francesco, P.~H. Ginsparg, and J.~Zinn-Justin, {\it {2-D Gravity and
  random matrices}},  {\em Phys. Rept.} {\bf 254} (1995) 1--133,
  [\href{http://xxx.lanl.gov/abs/hep-th/9306153}{{\tt hep-th/9306153}}].

\bibitem{Ginsparg:1993is}
P.~H. Ginsparg and G.~W. Moore, {\it {Lectures on 2-D gravity and 2-D string
  theory}},  \href{http://xxx.lanl.gov/abs/hep-th/9304011}{{\tt
  hep-th/9304011}}.

\bibitem{Ambjorn:1985az}
J.~Ambjorn, B.~Durhuus, and J.~Frohlich, {\it {Diseases of Triangulated Random
  Surface Models, and Possible Cures}},  {\em Nucl. Phys.} {\bf B257} (1985)
  433.

\bibitem{David:1985nj}
F.~David, {\it {A Model of Random Surfaces with Nontrivial Critical Behavior}},
   {\em Nucl. Phys.} {\bf B257} (1985) 543.

\bibitem{Boulatov:1986jd}
D.~Boulatov, V.~Kazakov, I.~Kostov, and A.~Migdal, {\it {Analytical and
  Numerical Study of the Model of Dynamically Triangulated Random Surfaces}},
  {\em Nucl. Phys.} {\bf B275} (1986) 641.

\bibitem{Duplantier:1988wc}
B.~Duplantier and I.~Kostov, {\it {Conformal spectra of polymers on a random
  surface}},  {\em Phys. Rev. Lett.} {\bf 61} (1988) 1433.

\bibitem{Duplantier:1989sx}
B.~Duplantier and I.~K. Kostov, {\it {Geometrical critical phenomena on a
  random surface of arbitrary genus}},  {\em Nucl. Phys.} {\bf B340} (1990)
  491--541.

\bibitem{Duplantier:1998fg}
B.~Duplantier, {\it {Random walks and quantum gravity in two dimensions}},
  {\em Phys. Rev. Lett.} {\bf 81} (1998) 5489--5492.

\bibitem{Duplantier:2003vx}
B.~Duplantier, {\it {Conformal fractal geometry and boundary quantum gravity}},
   \href{http://xxx.lanl.gov/abs/math-ph/0303034}{{\tt math-ph/0303034}}.

\bibitem{Boulatov:1986sb}
D.~Boulatov and V.~Kazakov, {\it {The Ising Model on Random Planar Lattice: The
  Structure of Phase Transition and the Exact Critical Exponents}},  {\em Phys.
  Lett.} {\bf 186B} (1987) 379.

\bibitem{AleshaZam-3pf}
A.~B. Zamolodchikov, {\it {Perturbed conformal field theory on fluctuating
  sphere}},  \href{http://xxx.lanl.gov/abs/hep-th/0508044}{{\tt
  hep-th/0508044}}.

\bibitem{Kostov:2006ry}
I.~K. Kostov, {\it {Thermal flow in the gravitational O(n) model}},  {\em Bulg.
  J. Phys.} {\bf 33} (2006), no.~s1 297--310,
  [\href{http://xxx.lanl.gov/abs/hep-th/0602075}{{\tt hep-th/0602075}}].

\bibitem{Ishimoto:2005ag}
Y.~Ishimoto and A.~B. Zamolodchikov, {\it {Massive Majorana fermion coupled to
  2D gravity and random lattice Ising model}},  {\em Theor. Math. Phys.} {\bf
  147} (2006) 755--776.

\bibitem{Zamolodchikov:2006xs}
A.~B. Zamolodchikov and A.~B. Zamolodchikov, {\it {Decay of Metastable Vacuum
  in Liouville Gravity}},  {\em Conf. Proc.} {\bf C060726} (2006) 1223--1228,
  [\href{http://xxx.lanl.gov/abs/hep-th/0608196}{{\tt hep-th/0608196}}].
  [,1223(2006)].

\bibitem{YL-On}
J.-E. Bourgine and I.~Kostov, {\it {On the Yang-Lee and Langer singularities in
  the O(n) loop model}},  {\em J.Stat.Mech.} {\bf 1201} (2012) P01024,
  [\href{http://xxx.lanl.gov/abs/1110.1108}{{\tt arXiv:1110.1108}}].

\bibitem{Fortuin:1971dw}
C.~M. Fortuin and P.~W. Kasteleyn, {\it {On the Random cluster model. 1.
  Introduction and relation to other models}},  {\em Physica} {\bf 57} (1972)
  536--564.

\bibitem{Kazakov:1986hu}
V.~Kazakov, {\it {Ising model on a dynamical planar random lattice: Exact
  solution}},  {\em Phys. Lett.} {\bf A119} (1986) 140--144.

\bibitem{Kostov:1988fy}
I.~Kostov, {\it {$O(n)$ vector model on a planar random surface: spectrum of
  anomalous dimensions}},  {\em Mod. Phys. Lett.} {\bf A4} (1989) 217.

\bibitem{Kostov:1989eg}
I.~Kostov, {\it {The ADE face models on a fluctuating planar lattice}},  {\em
  Nucl. Phys.} {\bf B326} (1989) 583-- 612.

\bibitem{Nienhuis:1984wm}
B.~Nienhuis, {\it {Critical behavior of two-dimensional spin models and charge
  asymmetry in the Coulomb gas}},  {\em J. Stat. Phys.} {\bf 34} (1984)
  731--761.

\bibitem{Ginsparg:1991bi}
P.~H. Ginsparg, {\it {Matrix models of 2-d gravity}},  1991.

\bibitem{Kostov:1999qx}
I.~Kostov, {\it {Exact solution of the six-vertex model on a random lattice}},
  {\em Nucl. Phys.} {\bf B575} (2000) 513--534,
  [\href{http://xxx.lanl.gov/abs/hep-th/9911023}{{\tt hep-th/9911023}}].

\bibitem{ZinnJustin:1999wt}
P.~Zinn-Justin, {\it {The six-vertex model on random lattices}},  {\em
  Europhys. Lett.} {\bf 50} (2000) 15--21,
  [\href{http://xxx.lanl.gov/abs/cond-mat/9909250}{{\tt cond-mat/9909250}}].

\bibitem{ELVEYPRICE2023105739}
A.~{Elvey Price} and P.~Zinn-Justin, {\it The six-vertex model on random planar
  maps revisited},  {\em Journal of Combinatorial Theory, Series A} {\bf 196}
  (2023) 105739.

\bibitem{Fendley:1993wq}
P.~Fendley, H.~Saleur, and A.~B. Zamolodchikov, {\it {Massless flows. 1. The
  Sine-Gordon and O(n) models}},  {\em Int. J. Mod. Phys.} {\bf A8} (1993)
  5717--5750, [\href{http://xxx.lanl.gov/abs/hep-th/9304050}{{\tt
  hep-th/9304050}}].

\bibitem{Fendley:1993xa}
P.~Fendley, H.~Saleur, and A.~B. Zamolodchikov, {\it {Massless flows, 2. The
  Exact S matrix approach}},  {\em Int. J. Mod. Phys.} {\bf A8} (1993)
  5751--5778, [\href{http://xxx.lanl.gov/abs/hep-th/9304051}{{\tt
  hep-th/9304051}}].

\bibitem{Zamolodchikov:1994za}
A.~B. Zamolodchikov, {\it {Thermodynamics of imaginary coupled sine-Gordon:
  Dense polymer finite size scaling function}},  {\em Phys. Lett.} {\bf B335}
  (1994) 436--443.

\bibitem{baxter1986q}
R.~Baxter, {\it q colourings of the triangular lattice},  {\em Journal of
  Physics A: Mathematical and General} {\bf 19} (1986), no.~14 2821.

\bibitem{t1974planar}
G.~t~Hooft, {\it A planar diagram theory for strong interactions},  {\em
  Nuclear Physics B} {\bf 72} (1974) 461--473.

\bibitem{Brezin:1977sv}
E.~Brezin, C.~Itzykson, G.~Parisi, and J.~B. Zuber, {\it {Planar Diagrams}},
  {\em Commun. Math. Phys.} {\bf 59} (1978) 35.

 \bibitem{Brezin:1979ba}
E.~Brezin, {\it {Planar diagrams}},  {\em Phys. Rept.} {\bf 49} (1979)
  221--227.

\bibitem{GR}
I.~Gradshteyn and I.~Ryzhik, {\it Table of integrals, series, and products}, .

\bibitem{Polyakov:1981rd}
A.~M. Polyakov, {\it {Quantum geometry of bosonic strings}},  {\em Phys. Lett.}
  {\bf B103} (1981) 207--210.

\bibitem{Kazakov:1998ji}
V.~Kazakov, I.~Kostov, and N.~A. Nekrasov, {\it {D-particles, matrix integrals
  and KP hierarchy}},  {\em Nucl. Phys.} {\bf B557} (1999) 413--442,
  [\href{http://xxx.lanl.gov/abs/hep-th/9810035}{{\tt hep-th/9810035}}].

\bibitem{kratzel1975integral}
E.~Kr{\"a}tzel, {\it Integral transformations of bessel type},  in {\em
  Generalized Functions and Operational Calculus, Proc. Conf. Varna},
  pp.~148--155, 1975.

\bibitem{Ginsparg:1987eb}
P.~H. Ginsparg, {\it {Curiosities at c = 1}},  {\em Nucl. Phys.} {\bf B295}
  (1988) 153--170.

\bibitem{Jose:1977gm}
J.~V. Jos\'e, L.~P. Kadanoff, S.~Kirkpatrick, and D.~R. Nelson, {\it
  Renormalization, vortices, and symmetry-breaking perturbations in the
  two-dimensional planar model},  {\em Phys. Rev. B} {\bf 16} (Aug, 1977)
  1217--1241.

\bibitem{Moore:1992ac}
G.~Moore, {\it Gravitational phase transitions and the sine-gordon model},
  \href{http://xxx.lanl.gov/abs/hep-th/9203061}{{\tt hep-th/9203061}}.

\bibitem{Kazakov:2000pm}
V.~Kazakov, I.~Kostov, and D.~Kutasov, {\it {A matrix model for the
  two-dimensional black hole}},  {\em Nucl. Phys.} {\bf B622} (2002) 141--188,
  [\href{http://xxx.lanl.gov/abs/hep-th/0101011}{{\tt hep-th/0101011}}].

\bibitem{alexandrov2003matrixquantummechanicstwodimensional}
S.~Alexandrov, {\em Matrix Quantum Mechanics and Two-dimensional String Theory
  in Non-trivial Backgrounds}.
\newblock PhD thesis, 2003.
\newblock \href{http://xxx.lanl.gov/abs/hep-th/0311273}{{\tt hep-th/0311273}}.

\bibitem{math8040526}
A.~M. Mathai and H.~J. Haubold, {\it Mathematical aspects of kr{\"a}tzel
  integral and kr{\"a}tzel transform},  {\em Mathematics} {\bf 8} (2020),
  no.~4.

\end{thebibliography}
%

\providecommand{\href}[2]{#2}\begingroup\raggedright\endgroup

 \end{document}